\documentclass[amsmath,amssymb,aps,pre,groupedaddress,superscriptaddress,twocolumn]{revtex4-2}
\usepackage[utf8]{inputenc}
\usepackage[english]{babel}
\usepackage[english]{varioref}
\usepackage{graphicx}
\usepackage{dcolumn}
\usepackage{bm}
\usepackage{amsmath}
\usepackage{amssymb}
\usepackage{xcolor}
\usepackage[font=small]{caption}
\usepackage{mathrsfs}
\usepackage{tabularx}
\usepackage[colorlinks=true,linkcolor=blue,urlcolor=blue,citecolor=blue,anchorcolor=blue]{hyperref}

\begin{document}

\title{Higher-order adaptive behaviors outperform pairwise strategies in mitigating contagion dynamics 
}

\author{Marco Mancastroppa}
\affiliation{Aix Marseille Univ, Université de Toulon, CNRS, CPT, Turing Center for Living Systems, 13009 Marseille, France}
\author{Márton Karsai}
\affiliation{Department of Network and Data Science, Central European University, 1100 Vienna, Austria}
\affiliation{National Laboratory for Health Security, HUN-REN Rényi Institute of Mathematics, 1053 Budapest, Hungary}
\author{Alain Barrat}
\affiliation{Aix Marseille Univ, Université de Toulon, CNRS, CPT, Turing Center for Living Systems, 13009 Marseille, France}

\begin{abstract}
When exposed to a contagion phenomenon, individuals may respond to the perceived risk of infection by adopting behavioral changes, aiming to reduce their exposure or their risk of infecting others. 
The social cost of such adaptive behaviors and their impact on the contagion dynamics have been investigated in pairwise networks, with binary interactions driving both contagion and risk perception. However, contagion and adaptive mechanisms can also be driven by group (higher-order) interactions. Here, we consider several adaptive behaviors triggered by awareness of risk perceived through higher-order and pairwise interactions, and we compare their impact on pairwise and higher-order contagion processes. By numerical simulations and a mean-field analytic approach, we show that adaptive behaviors driven by higher-order information are more effective in limiting the spread of a contagion, than similar mechanisms based on pairwise information. Meanwhile, they also entail a lower social cost, measured as the
reduction of the intensity of interactions in the population. Indeed, adaptive mechanisms based on higher-order information lead to a heterogeneous risk perception within the population, producing a higher alert on nodes with large hyperdegree (i.e., participating in many groups), on their neighborhoods, and on large groups. This in turn prevents the spreading process to exploit the properties
of these nodes and groups, which tend to drive and sustain the dynamics in the absence of adaptive behaviors.
\end{abstract}

\maketitle

\section{Introduction}
Several complex phenomena, including the spread of epidemics, opinions and information in a population, can be described as contagion processes on networks or on hypergraphs. In these structures nodes
represent the members of the population, and links (in a network) and
hyperlinks (in a hypergraph) represent the pairwise or group interactions along which
spreading can occur \cite{barrat2008dynamical,satorras2015epidemic,majhi2022,WANG20241}. Based on their perceived risk, individuals exposed to such processes may respond by adapting their social behavior to avoid infection (self-protection) and/or transmission (altruism) \cite{funk2010review,perra2011adaptive,verelst2016behavioural,de2020relationships,Degaetano2024modeling}. Adaptive behavioral patterns can emerge in diverse ways and can be broadly classified \cite{funk2010review,perra2011adaptive} by: 
(i) the \textit{information} determining how the risk is perceived and 
(ii) the \textit{behavioral changes} induced. 
For example, the awareness of an individual may derive from \textit{local} or \textit{global} information, driven respectively by the perception of the epidemic status of their neighborhood \cite{gross2006adaptive,bagnoli2007risk,marceau2010,moinet2018effect} or of the entire population (e.g. a country), facilitated by institutions or media \cite{berner2023adaptive,gross2008,perra2011adaptive}. 
Moreover, the behavioral changes induced by risk perception can take various forms
\cite{funk2010review,feniche2011adaptive,funk2009awearness,Ferguson2007,odor2025}.
They can influence the status of an individual to reduce
contagion risk, e.g. through vaccination, increased hygiene, 
physical distancing or the use of protective masks
\cite{degaetano2023,bagnoli2007risk,odor2025}, or even 
alter the structure of interactions \cite{gross2006adaptive,mancastroppa2020active,liu2025higher}, e.g., by rewiring potentially contagious links or through isolation.

These adaptive responses introduce a two-way coupling between the contagion dynamics and the behavior of individuals, impacting both the spreading process and the interactions 
between individuals in a population \cite{gross2009adaptive,berner2023adaptive,gross2008,SAYAMA20131645,mancastroppa2024,karsai2025}. 
Adaptive mechanisms yield in particular a reduction of the number or intensity
of interactions, hence suppressing the epidemic locally and influencing its global outcome. However, this reduction in interactions is possible only through a deterioration of the overall system social
activity and thus it comes with a social cost \cite{hollingsworth2011mitigation,Massaro2018,feniche2011adaptive,fenichel2013economic}. 
Different adaptive behaviors not only can have different effectiveness in mitigating an epidemic, but can also come with a different social cost (not to mention economic burden). It is important to take all these effects into account when designing interventions and mitigation strategies, to identify balanced solutions, which substantially lower the disease prevalence while keeping the social cost limited \cite{fraser2004,hollingsworth2011mitigation,Massaro2018,feniche2011adaptive,fenichel2013economic,colosi2022screening}.

Several frameworks have been developed to model adaptive behaviors in \textit{pairwise contagion processes}, where contagion occurs along pairwise interactions of a network \cite{gross2006adaptive,bagnoli2007risk,marceau2010,guo2013,karsai2025,odor2025,moinet2018effect,mancastroppa2020active,mancastroppa2024}. 
On the other hand, \textit{higher-order contagion processes}, in which contagion occurs in group interactions with non-linear reinforcement effects, exhibit a different phenomenology than pairwise processes, characterized by discontinuous transitions and multistability regimes \cite{Iacopini2019,FerrazdeArruda2023,FerrazdeArruda2021,WANG20241,Malizia2025,St-Onge2022,Mancastroppa2023,Battiston2021,BATTISTON20201,majhi2022}. 
The effects of adaptive behaviors on these higher-order processes can potentially be very different and are still largely unexplored \cite{mancastroppaCTside,Burgio2025,liu2025higher}. 
Moreover, just as higher-order contagion processes leverage group interactions, 
the adaptive mechanisms themselves can involve or be based on higher-order effects. For instance, risk perception can induce the splitting or merging of
groups \cite{Iacopini2024,liu2025higher}, leading to modifications of the hypergraph structure; risk awareness can also be perceived along group interactions, with an event-based perspective, instead of being perceived via pairwise interactions, with a contact-based perspective \cite{mancastroppa2025_2}. 
Building and studying models taking into account such higher-order mechanisms is therefore necessary to fully envision the interplay between
adaptation and spreading processes, and to determine which mechanisms lead to effective mitigation at lower social cost.

Here and in a companion paper \cite{mancastroppa2025_2}, we contribute to bridging this gap. We consider both pairwise and higher-order contagion processes in the presence of different adaptive behaviors, triggered by local information, which reduce the probability of contagion. 
In \cite{mancastroppa2025_2}, we focus on two of these adaptive mechanisms 
and on their impact on the epidemic phase transition. We show that they
do not change the critical behavior of pairwise processes, but profoundly 
alter the discontinuous transition of higher-order processes, shrinking 
the bistability region and even driving the transition to become continuous.
Here, we consider a broader range of adaptive strategies based on information of different nature and type. We generalize the results of \cite{mancastroppa2025_2}
and extend the range of analysis by systematically comparing the efficacy in epidemic containment and the social cost of these strategies.
In particular, we consider adaptive strategies where the information is based on the \textit{absolute} or \textit{relative} number of interactions perceived at risk.
These risky interactions can be either defined as (i) \textit{pairwise}, considering contacts with infectious individuals, (ii) \textit{higher-order}, based on groups containing infectious individuals, or (iii) \textit{hybrid}. 
Through numerical simulations and a mean-field analytical approach, we show that strategies based on absolute higher-order information are more efficient
in containing the spread, while having a lower social cost, compared to pairwise relative strategies. This performance stems from 
the exploitation of the heterogeneity and overlap of group interactions \cite{landryheterog2020,malizia2025disentangling,Malizia2025}: mechanisms based on absolute higher-order information lead to a 
heterogeneous risk perception across the population, with adaptation
focused on nodes with large hyperdegree and on large groups, which are
known to foster and sustain the spreading processes \cite{St-Onge2022,Mancastroppa2023}.
Moreover, this heterogeneous risk perception keeps the social cost low 
by focusing the reduction of interaction intensity on few hubs and groups.

The manuscript is organized as follows: in Sec.~\ref{sec:model}, we introduce
the spreading models and the adaptive mechanisms considered; in Sec.~\ref{sec:MF}, we present our mean-field approach; results of stochastic
numerical simulations and of the integration of the mean-field equations are 
described in Sec.~\ref{sec:results} and discussed in Sec.~\ref{sec:discussion}; Section \ref{sec:methods} contains details on the data sets
considered as substrate for the spreading processes, on the numerical simulations
and on the mean-field approach.

\section{The model}
\label{sec:model}

\begin{figure*}[ht!]
\includegraphics[width=\textwidth]{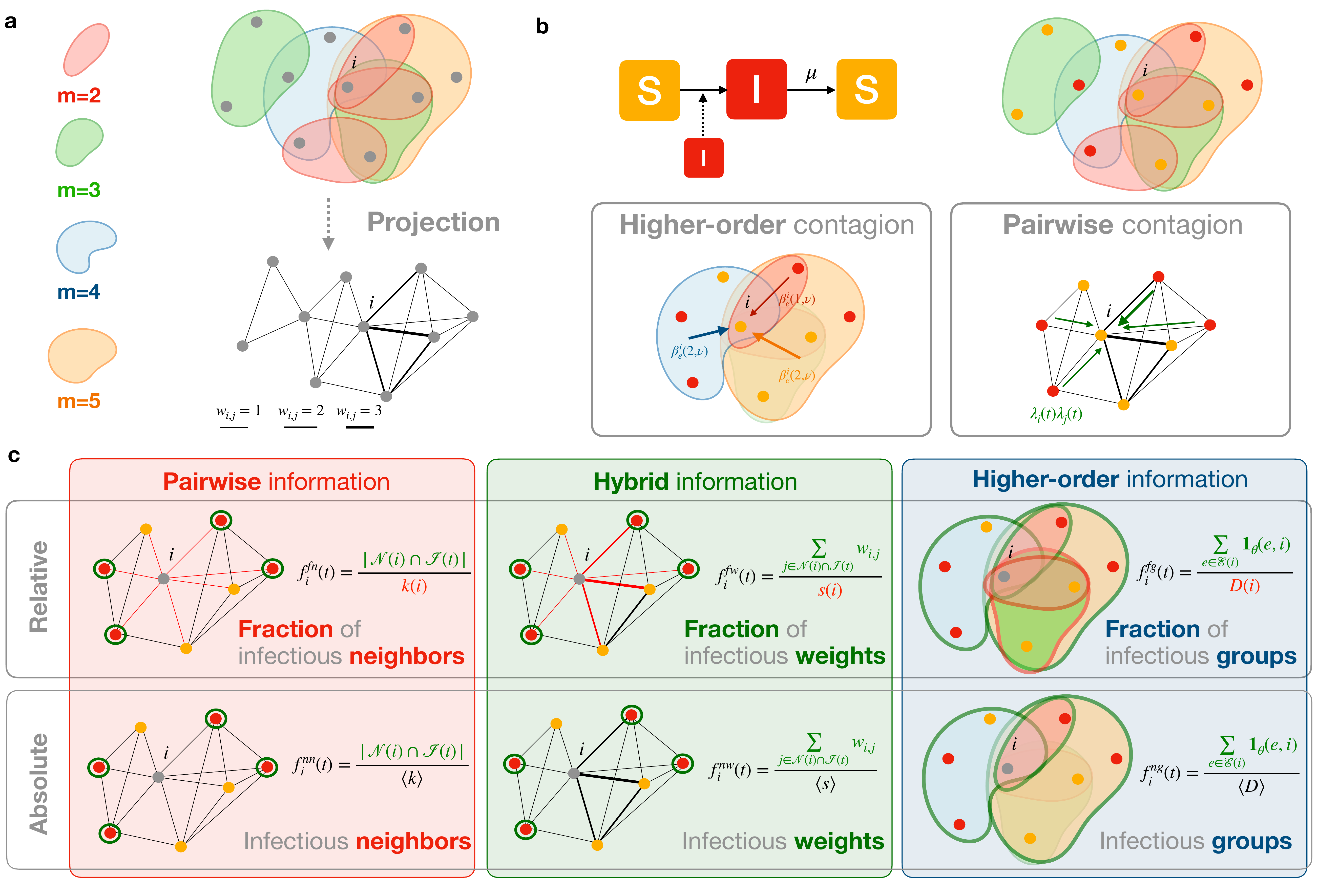}
\caption{\textbf{Schematic representation of the contagion and adaptive mechanisms.} \textbf{a}: an example of a hypergraph $\mathcal{H}$ and its weighted pairwise projection $\mathcal{G}$. 
\textbf{b}: schematic representation of the SIS contagion mechanisms for the node $i$, using higher-order contagion (left panel) or pairwise contagion (right panel). 
\textbf{c}: schematic representation of the six adaptive strategies considered for node $i$: the first and second rows show respectively the strategies based on relative (fraction) and absolute (number) information; the columns identify different nature of information: pairwise, higher-order and hybrid. 
In this case (considering $\theta=0.3$): $f_i^{fn}(t)=4/7$, $f_i^{nn}(t)=4/\langle k \rangle$, $f_i^{fw}(t)=5/11$, $f_i^{nw}(t)=5/\langle s \rangle$, $f_i^{fg}(t)=3/5$, $f_i^{ng}(t)=3/\langle D \rangle$.}
\label{fig:figure1}
\end{figure*}

\subsection{Contagion dynamics}
We consider a hypergraph $\mathcal{H}=(\mathcal{V},\mathcal{E})$, where $\mathcal{V}$ is the set of its $N$ nodes and $\mathcal{E}$ is the set of its $E$ hyperedges \cite{Battiston2021,BATTISTON20201}. We also consider the corresponding weighted projected graph $\mathcal{G}=(\mathcal{V},\mathcal{L})$, obtained by projecting each hyperedge onto the corresponding (fully connected) clique and assigning each link $l \in \mathcal{L}$ a weight $w_l$ equal to the number of hyperedges whose projection contains the link (see Fig. \ref{fig:figure1}\textbf{a}). The interactions described by $\mathcal{H}$ and $\mathcal{G}$ constitute both the substrate for contagion processes and for risk perception, which triggers adaptive behaviors. 

Here we consider the discrete-time Susceptible-Infected-Susceptible (SIS) model 
for the spread of infectious diseases (see Fig. \ref{fig:figure1}\textbf{b}):  
each node is at each time step either Susceptible ($S$) or Infectious ($I$). 
Infectious nodes recover spontaneously with probability $\mu$ per unit time, 
while susceptible nodes can become infectious if they interact with infectious nodes. 
We assign a parameter $\lambda_i(t)$ to each node $i \in \mathcal{V}$, and
we assume that the probability of contagion for a susceptible node depends on its own parameter $\lambda_i(t)$ and on the parameters of its infectious 
neighbors $\lambda_j(t) \forall j \in \mathcal{N}(i) \cap \mathcal{I}(t)$, 
where $\mathcal{I}(t)$ is the set of infected nodes at time $t$ and $\mathcal{N}(i)$ is the set of neighbors of $i$. Risk awareness of node $i$ is encoded in the potential time variation of $\lambda_i(t)$ \cite{odor2025,bagnoli2007risk}. 
We assume indeed that the parameter $\lambda_i(t)$ describes how the behavior of
node $i$ impacts both its risk of becoming infectious (when susceptible), or 
to infect others (when infectious).

We consider two possible infection mechanisms:
\begin{itemize}
    \item in the \textit{pairwise contagion} process, supported by $\mathcal{G}$, a susceptible node $i$ has probability $\lambda_i(t) \lambda_j(t)$ of being infected at time $t$ by each infectious neighbor $j$ \cite{satorras2015epidemic}. Furthermore, each link $\{i,j\} \in \mathcal{L}$ can be a source of infection with multiplicity equal to its weight $w_{i,j}$ (see Fig. \ref{fig:figure1}\textbf{b});
    \item in the \textit{higher-order contagion process}, taking place on $\mathcal{H}$, a susceptible node $i$ participating in a hyperedge $e$ with $i_e$ infectious nodes at time $t$ is infected within the group with probability $\beta_e^i(i_e,\nu)=\lambda_0 i_e^{\nu} \lambda_i(t) \langle \lambda_j(t)/\lambda_0 \rangle_{i_e}^{\nu}$ (see Fig. \ref{fig:figure1}\textbf{b}), where $\langle \cdot \rangle_{i_e}$ denotes the average over the $i_e$ infected nodes in $e$ and $\nu > 1$ defines the non-linearity of the process \cite{St-Onge2022}. In particular, this process models group effects, by considering a reinforcing effect when multiple infectious nodes are present in a group.
\end{itemize}

\subsection{Adaptive behaviors}
To take into account potential adaptive behaviors, we assume that each node at each
time has a local perception of the epidemiological state of its neighborhood.
This perception is encoded in an awareness function $f_i(t)$, which drives 
an exponential reduction in the parameter $\lambda_i$ \cite{odor2025,bagnoli2007risk,moinet2018effect}: 
\begin{equation}
    \lambda_i(t)= \lambda_0 e^{-f_i(t)}.
\end{equation}
The awareness of risk triggers therefore both (i) a self-protective behavior, if $i$ is susceptible, by reducing the probability of contagion for $i$
and (ii) an altruistic behavior, if $i$ is infectious, by reducing its probability of transmitting the disease to others. 
Such mechanisms model mild adaptive behaviors corresponding to an increased 
cautiousness (e.g., increased physical distancing, reduction of the frequency of
interactions, the use of protective masks, or increased hygiene), which decrease
contagion probabilities without altering the structure of interactions.

The baseline non-adaptive case (NAD), in which nodes do not change their behavior, is 
obtained for $f_i = 0, \forall i, \forall t$: $\lambda_i(t)=\lambda_0$ is constant and
uniform.
The contagion processes reduce then to the standard SIS model, either on a weighted graph, with infection probability per link $\lambda_0^2$ \cite{satorras2015epidemic}, or on a hypergraph with nonlinear contagion, with infection probability 
$\lambda_0^2 i_e^{\nu}$ for a susceptible 
node $i$ in a group having $i_e$ infectious nodes \cite{St-Onge2022}.

We consider several non-trivial forms of the awareness function $f_i(t)$,
depending on the \textit{nature} and \textit{type} of the information on $i$'s neighborhood, assumed to
determine $i$'s perception of risk. In particular, we consider that the risk assessment can be based on information of pairwise, group or hybrid \textit{nature}:
\begin{itemize}
    \item in the \textbf{pairwise} case, the information available to each node comes from its pairwise interactions in $\mathcal{G}$: a node's risk perception is based on how many infectious neighbors it has;
    \item in the \textbf{group (higher-order)} case, the perception of risk depends on the number of ``infectious'' groups (hyperedges) to which a node belongs, where group is considered as infectious if it includes a large enough fraction of infectious nodes. The relevant information is thus encoded in the hyperedges of $\mathcal{H}$;
    \item in the \textbf{hybrid} case, each node perceives a risk depending on its number of infectious neighbors in $\mathcal{G}$, but each neighbor $j$ contributes to $i$'s perception with the weight $w_{i,j}$, which retains higher-order information (as it gives the number of groups in $\mathcal{H}$ in which both $i$ and $j$ participate).   
\end{itemize}
Moreover, we distinguish \textit{two types} of information:
\begin{itemize}
    \item \textbf{absolute:} nodes' risk perception is based on their absolute number of infectious interactions (either pairwise or in groups);
    \item \textbf{relative:} nodes' risk perception depends on their fractions of infectious interactions, i.e., on the number of infectious interactions normalized by their total number of interactions.
\end{itemize}
Combining the various nature and type of information considered results in six different mechanisms (see Fig. \ref{fig:figure1}\textbf{c}), which can each be implemented
for the pairwise and for the higher-order contagion models.

For information driven by pairwise interactions, we obtain 
the \textit{``number of infectious neighbors''} ($nn$) strategy and the \textit{``fraction of infectious neighbors''} ($fn$) strategy (see Fig. \ref{fig:figure1}\textbf{c}). In the $nn$ strategy the absolute number of infectious neighbors determines the awareness level, while in the $fn$ strategy it is triggered by the fraction of infectious neighbors, yielding:
\begin{align}
    f_i^{nn}(t) &= |\mathcal{N}(i) \cap  \mathcal{I}(t)| /\langle k \rangle,\\
    f_i^{fn}(t) &= |\mathcal{N}(i) \cap  \mathcal{I}(t)| /k (i),
\end{align}
where $k (i)$ is the degree of node $i$ (its total number of neighbors) and $\langle k \rangle$ is the average degree in $\mathcal{G}$.

The two hybrid strategies are the \textit{``infectious weights''} ($nw$) strategy and the \textit{``fraction of infectious weights''} ($fw$) strategy (see Fig. \ref{fig:figure1}\textbf{c}). In the $nw$ strategy, the awareness function of a node $i$ is
proportional to the sum of its link weights to infectious neighbors. In the
$fw$ strategy, this sum is normalized by the total sum of its link weights (the node
strength $s(i)$ \cite{Barrat2004}):
\begin{align}
    f_i^{nw}(t) &= \sum\limits_{j \in \mathcal{N}(i) \cap \mathcal{I}(t)} w_{i,j} /\langle s \rangle,\\
    f_i^{fw}(t) &= \sum\limits_{j \in \mathcal{N}(i) \cap \mathcal{I}(t)} w_{i,j} /s (i),
\end{align}
where  $\langle s \rangle$ is the average node strength in $\mathcal{G}$.

Finally, information based on higher-order interactions yields  
the \textit{``number of infectious groups''} ($ng$) strategy and the 
\textit{``fraction of infectious groups''} ($fg$) strategy (see Fig. \ref{fig:figure1}\textbf{c}). 
In the $ng$ strategy the awareness of a node depends on the absolute number of ``infectious'' hyperedges in which it participates: a group is perceived as infectious by $i$ if the number $i_e$ of infected individuals within it perceived by $i$ (i.e., distinct from $i$) is $i_e> \theta (|e|-1)$, where $\theta \in [0,1]$ is a fixed threshold (parameter of the strategy that sets the level of alert). 
In the $fg$ strategy the awareness of a node depends on the fraction of 
infectious groups among the set $\mathcal{E}(i)$ of all hyperedges of $i$:
\begin{align}
    f_i^{ng}(t) &= \sum\limits_{e \in \mathcal{E}(i)} \bm{1}_{\theta}(e,i)/\langle D \rangle ,\\
    f_i^{fg}(t) &= \sum\limits_{e \in \mathcal{E}(i)} \bm{1}_{\theta}(e,i)/D(i),
\end{align}
where $D(i)$ is the hyperdegree of node $i$ (its number of hyperedges), $\langle D \rangle$ is the average hyperdegree in $\mathcal{H}$ and $\bm{1}_{\theta}(e,i)$ is the group indicator function, equal to $1$ if 
$i_e> \theta (|e|-1)$ (without counting $i$) and $0$ otherwise.

Note that, to make mechanisms based on absolute and relative information comparable, 
the awareness functions of the absolute strategies ($nn$, $nw$ and $ng$) are 
normalized by the average values of the degree, the strength and the 
hyperdegree, respectively.
In \cite{mancastroppa2025_2}, we restrict the analysis to the absolute pairwise $nn$ and absolute higher-order $ng$ cases; here, we consider all six mechanisms and compare their efficiencies and social costs.

\section{Mean-field approach}
\label{sec:MF}
We use an individual-based mean-field (IBMF) approach in continuous time \cite{satorras2015epidemic} to describe the epidemic dynamics in the presence of adaptive strategies for both types of contagion processes. We consider the evolution of the probability $P_i(t)$ that node $i$ is infected at time $t$,
by assuming that neighboring nodes are statistically independent, i.e.,
neglecting local correlations between nodes states. This mean-field approach makes it possible to \cite{satorras2015epidemic}: 
(i) account for the complete (higher-order) topological structure, which is essential for capturing the effect of adaptive behaviors; 
(ii) describe the epidemic dynamics at the level of individual nodes and groups. 

In the pairwise contagion, the equation for the evolution of $P_i(t)$ in the IBMF
framework is:
\begin{equation}
    \partial_t P_i(t)=-\mu P_i(t) + (1-P_i(t)) \sum\limits_{j \in \mathcal{V}} w_{i,j} \lambda_i(t) \lambda_j(t) P_j(t) ,
    \label{eq:MF_p}
\end{equation}
where the first term describes the recovery process when $i$ is infectious, and the second term the contagion process if $i$ is susceptible and has infectious neighbors (accounting for the link
weights in $\mathcal{G}$).

\begin{figure*}[htb!]
\includegraphics[width=\textwidth]{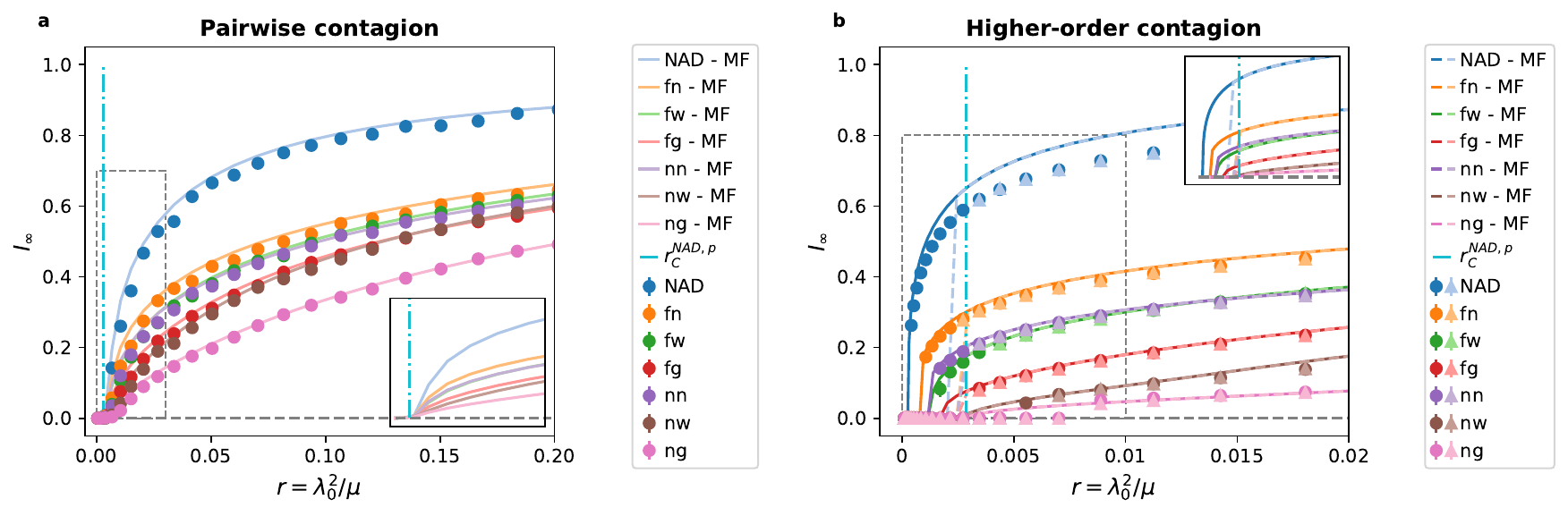}
\caption{\textbf{Phase diagram for pairwise and higher-order processes.} \textbf{a}, \textbf{b}: epidemic prevalence in the asymptotic steady state, $I_{\infty}$, as a function of the effective infection rate, $r=\lambda_0^2/\mu$, for the non-adaptive case (NAD) and for the six adaptation strategies, respectively for the pairwise and higher-order contagion process. We show the results of numerical simulations (markers) and of the integration of the mean-field equations (lines). For the higher-order contagion case, we consider simulations and numerical integration starting with a low (triangles and dashed lines) and a high (dots and solid lines) fraction of infected nodes. We consider the hospital dataset, the numerical results are averaged over $300$ simulations, $\nu=4$, $\theta=0.3$.
The insets represent a zoom on the mean-field results in the area marked by a dashed rectangle and the vertical line indicates the pairwise
epidemic threshold $r_C^{NAD,p}$. Note that the error-bars on numerical simulations are smaller than the corresponding markers (considering one standard deviation).}
\label{fig:figure2}
\end{figure*}

In the higher-order contagion, $P_i(t)$ is governed by:
\begin{widetext}
\begin{equation}
    \partial_t P_i(t)=-\mu P_i(t) + (1-P_i(t)) \sum\limits_{e \in \mathcal{E}(i)} \left[ \sum\limits_{i_e=1}^{|e|-1} Q_{e \setminus i}(i_e,t) \lambda_0 i_e^{\nu} \lambda_i(t) \overline{\langle \lambda_j(t)/\lambda_0 \rangle_{i_e}^{\nu}} \right] ,
    \label{eq:MF_HO}
\end{equation}
\end{widetext}
where the first term describes the recovery process and the second term the contagion process, due to the groups containing at least one infectious individual in which $i$ is involved. In particular, $Q_{e \setminus i}(i_e,t) $ is the probability that in the hyperedge $e$, where node $i$ is involved at time $t$, there are exactly $i_e$ infectious individuals (distinct from $i$); $\overline{\langle \lambda_j(t)/\lambda_0 \rangle_{i_e}^\nu}$ denotes a double average over these $i_e$ infectious individuals and over all possible configurations of $i_e$ infectious individuals in $e \backslash i$. 

In Sec. \ref{sec:methods}, we show that $\lambda_i(t)$, $Q_{e \setminus i}(i_e,t)$ and $\overline{\langle \lambda_i(t)/\lambda_0 \rangle_{i_e}^\nu}$ can be analytically obtained as functions of the probabilities $P_j(t)$ of all nodes $j$. This implies that
Eq.s \eqref{eq:MF_p} and Eq.s \eqref{eq:MF_HO} constitute two closed sets of $N$ coupled non-linear differential equations, which can thus be integrated numerically.

\section{Results}
\label{sec:results}
We perform stochastic numerical simulations of the dynamics of the pairwise
and higher-order contagion processes, both without adaptation and with 
each of the adaptation mechanisms. We consider as substrate of the contagion
processes several empirical hypergraphs describing face-to-face interactions between
individuals in a variety of contexts 
\cite{sociopatterns,Vanhems2013,genois2018,Genois2023,Stehle2011,Mastrandrea2015}
as well as synthetic hypergraphs with tunable properties (see Methods).
In each case, we also integrate numerically the mean-field equations 
(see Methods for more details) and compare the results. 
In the main text, we present the results obtained for processes on one of the empirical hypergraphs, built from data on interactions collected in a hospital \cite{Vanhems2013,sociopatterns}, and we refer to the Supplementary Material, SM, for analogous results on the other hypergraphs.
In particular, we first show the impact of the considered
adaptive strategies on the epidemic phase diagram and on the
epidemic phase transition, which separates the epidemic-free phase
from the phase with a finite density of infectious individuals
(Sec. \ref{sec:phase_diag}, see also \cite{mancastroppa2025_2}). 
We then focus on the active phase of the epidemic 
(Sec. \ref{sec:active}), considering the system's properties in the 
non-equilibrium endemic steady-state and in the early stages of the spread.

\subsection{The steady-state phase diagram}
\label{sec:phase_diag}
We show in Fig. \ref{fig:figure2} the epidemic prevalence in the asymptotic steady state $I_{\infty}$ as a function of the effective infection rate $r=\lambda_0^2/\mu$, for pairwise and higher-order contagion processes. 
Figure \ref{fig:figure2} shows the results for the non-adaptive case as well as for the six adaptive strategies considered. 
The IBMF approach presents very good agreement with the numerical simulations in all cases, showing in particular that this approach is
able to correctly capture the coupling between epidemic dynamics and adaptive behavior.

\begin{figure*}[ht!]
\includegraphics[width=\textwidth]{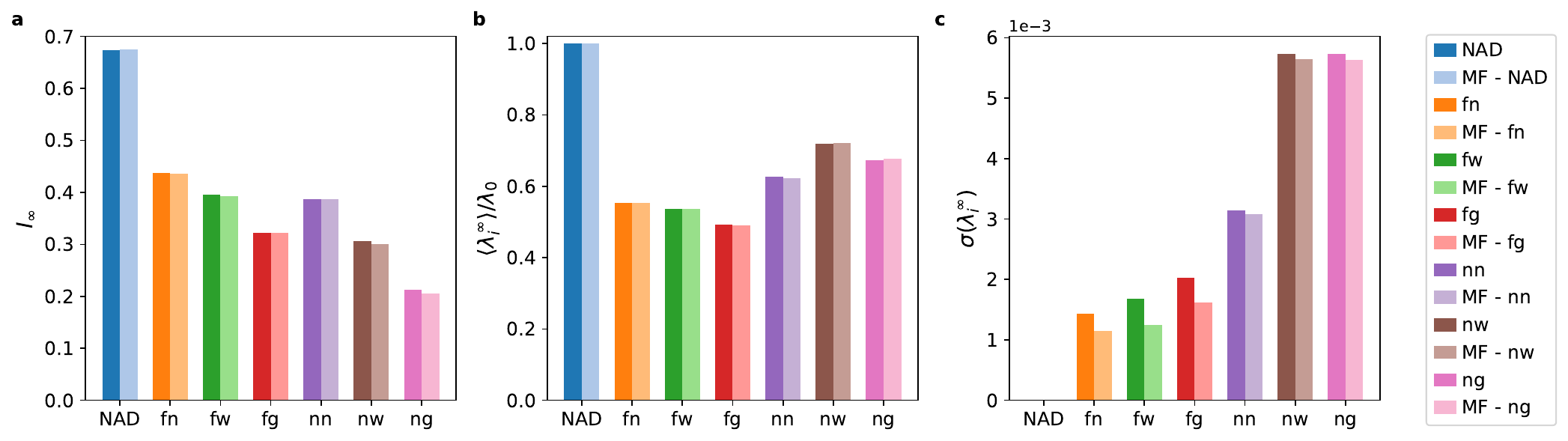}
\caption{\textbf{Impact of adaptive behaviors on the endemic state.} We consider the pairwise contagion process on the hospital dataset, with
$r=0.05$, $\theta=0.3$. 
For the NAD case and for each adaptive strategy, we show (through bar-plots): 
\textbf{a}: the endemic epidemic prevalence $I_{\infty}$; 
\textbf{b}: the average value of the transmission parameter $\langle \lambda_i^{\infty} \rangle/\lambda_0$ in the steady state; 
\textbf{c}: the standard deviation of the transmission parameter $\sigma(\lambda_i^{\infty})$ in the endemic state. 
In all panels, the left dark-colored bars are obtained by averaging over 300 numerical stochastic simulations, while the right light-colored bars are obtained through numerical integration of the mean-field equations (see legend).}
\label{fig:figure3}
\end{figure*}

As discussed in detail in \cite{mancastroppa2025_2}, the adaptive strategies can have very different impacts on the epidemic transition, depending
on which underlying contagion process is considered.
For the pairwise contagion, the usual continuous phase transition
is observed in the NAD case, with the epidemic threshold 
$r_C^{NAD,p}=1/\Lambda_w$ (where $\Lambda_w$ is the maximum eigenvalue of the weights matrix $\{w_{i,j}\}$, see Methods). Crucially, all
adaptive strategies present as well a continuous phase transition at
the same value of the epidemic threshold. This result can in fact 
be obtained within the IBMF approach, as shown in \cite{mancastroppa2025_2} for the $nn$ and $ng$ cases, and generalized in Sec. \ref{sec:methods}
for the other cases.
On the other hand, the higher-order contagion process in the non-adaptive case exhibits a discontinuous phase transition with a bistability region
\cite{Iacopini2019,St-Onge2022}. The
adaptive strategies impact this phase transition, by reducing both 
the range of the bistability region and the amplitude of the 
discontinuity. Some strategies suppress even completely this bistability
region, making the transition continuous (see Fig. \ref{fig:figure2}). We discuss this interesting phenomenon in detail in a companion paper \cite{mancastroppa2025_2}, showing how adaptive behaviors can neutralize bistability, by defusing group contagion and effectively
transforming the higher-order contagion process into a pairwise one.

Here, we focus on the active phase of the epidemic and we compare the performance of the different adaptive strategies in reducing the epidemic
prevalence, and their social cost, quantified by the reduction in contact
intensities. Figure \ref{fig:figure2} shows that 
all adaptive behaviors yield a reduction of the epidemic prevalence. 
However, a clear hierarchy is observed between strategies, which does
not depend on the epidemic parameter $r$ nor on the nature of the contagion
process. The nature and type of information triggering the adaptive behavior play both an important role: (i) strategies based on absolute information are more effective than those using relative information (for a given nature of information); 
(ii) strategies relying on higher-order information perform better than those using pairwise information (for a given type of information), with hybrid strategies having intermediate performance. In the following, we
study the differences between strategies in the active phase in more
detail.

\subsection{The active phase}
\label{sec:active}

As the hierarchy of strategies does not depend on the parameter $r>r_C^{NAD,p}$ nor on the contagion process, we now focus on the
pairwise contagion process at a fixed value of $r$. We investigate both the non-equilibrium endemic steady-state (asymptotic) and the epidemic transient (initial phases of the spread). We compare the different strategies with respect to their efficacy in mitigating the epidemic and with respect to their social cost, unveiling the adaptive mechanisms active at different time scales. We refer to the SM for similar results for the higher-order contagion
process.
 
\subsubsection{Endemic state}
\label{sec:endemic}
The performance of the adaptive strategies can be quantified, in the endemic state, along two dimensions. 
On the one hand, we estimate their efficacy as the reduction induced in the epidemic prevalence $I_{\infty}$ (compared to the NAD case).
On the other hand, we consider as a proxy of their social cost
the average reduction in the nodes parameters determining the contagion
probabilities, since we interpret this reduction as a tightening of social behaviors (e.g., limitation of durations or frequencies of interactions, mask-wearing, etc). We also consider in the SM another proxy of social cost (based on a link-centered perspective), with similar results.

\begin{figure*}[ht!]
\includegraphics[width=\textwidth]{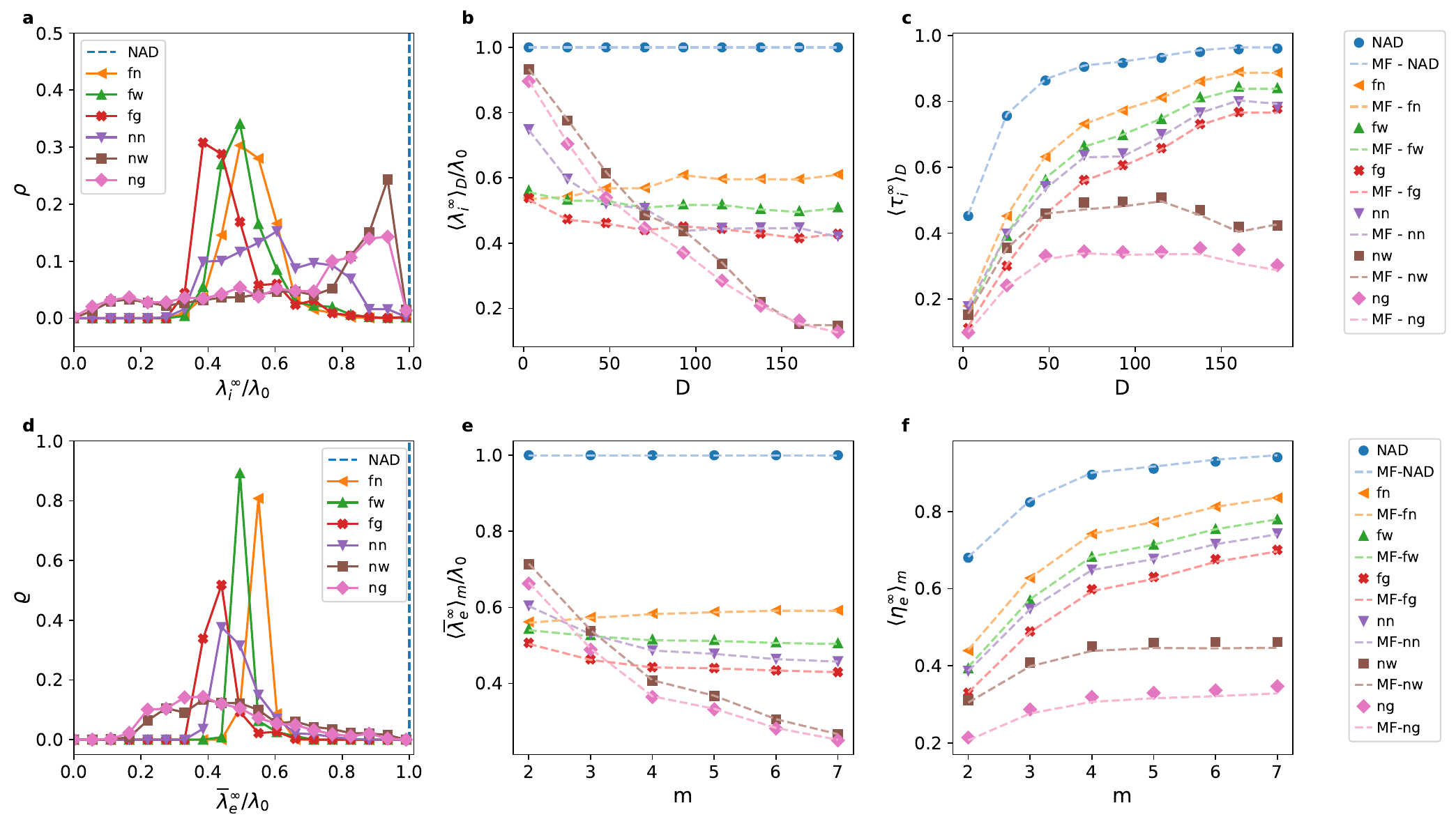}
\caption{\textbf{Heterogeneous risk perception of nodes and groups.} We consider the pairwise contagion process on the hospital dataset, $r=0.05$, $\theta=0.3$. We focus on the asymptotic steady state and we show the distribution $\rho(\lambda_i^{\infty}/\lambda_0)$ of the transmission parameter for nodes (panel \textbf{a}) and the distribution $\varrho(\overline{\lambda}_e^{\infty}/\lambda_0)$ of the average transmission parameter within a group (panel \textbf{d}), both obtained by averaging over 300 numerical simulations. We divide nodes into hyperdegree classes $D$ and for each class we estimate: \textbf{b}: the average reduction in the transmission parameter $\langle \lambda_i^{\infty} \rangle_D/\lambda_0$ in the steady state; \textbf{c}: the average fraction of time nodes spent infected $\langle \tau^{\infty}_i \rangle_D$ in the steady state. For each hyperedge size $m$ we estimate: \textbf{e}: the mean reduction of the transmission parameter within a group $\langle \overline{\lambda}_e^{\infty} \rangle_m/\lambda_0$ in the steady state; \textbf{f}: the average fraction of infected nodes within a group $\langle \eta^{\infty}_e \rangle_m$ in the steady state. All quantities are estimated for each node or hyperedge by averaging over 300 numerical simulations (markers) or by integrating numerically the mean-field equations (dashed curves). In all panels, we consider the NAD case and the six adaptive strategies.}
\label{fig:figure4}
\end{figure*}

In Fig. \ref{fig:figure3} we show
the average prevalence in the endemic steady-state (panel \textbf{a}) and the average parameter $\langle \lambda_i^{\infty} \rangle/\lambda_0$
(panel \textbf{b}, the value $1$ corresponding to the non-adaptive case), for
all strategies. The results highlight important differences 
among strategies; moreover, the hierarchy in terms of prevalence 
reduction is not exactly the same as the one obtained for social cost. For instance,
the $fg$ strategy yields a stronger prevalence reduction 
than the $fn$ strategy, but has also a higher social cost. 
Interestingly, the two best strategies for reducing the prevalence,
which are the ones based on absolute information integrating
a higher-order component ($ng$ and $nw$), are also the ones with the lowest
social cost (smallest reductions of $\langle \lambda_i^{\infty} \rangle/\lambda_0$). 
On the contrary, the two strategies based on relative and/or pairwise information ($fn$, $fw$) are the worst in epidemic containment and
among the ones with the highest costs.

Overall, the absolute and higher-order strategies are particularly
interesting as they manage to yield a low prevalence despite a lower awareness (lower cost) compared to other strategies. This suggests that their underlying adaptive mechanisms exploit the (higher-order) structure of the interactions differently from the other strategies.
We investigate this point in detail in the next subsection, but in Fig. \ref{fig:figure3}\textbf{c} we already provide some hints
in this respect. Indeed, Fig. \ref{fig:figure3}\textbf{b} describes only an average behavior, however heterogeneities have very important effects in spreading processes \cite{satorras2015epidemic}. The impact of adaptive strategies is also determined by the behavior of single nodes, which can be heterogeneous with different awareness levels. 
To compare the variability in nodes behavior in different strategies, we show in Fig. \ref{fig:figure3}\textbf{c} the standard deviation in the parameter
values, $\sigma(\lambda_i^{\infty})$: the best strategies $ng$ and $nw$ 
exhibit a high heterogeneity in awareness levels among nodes,
while relative and/or pairwise strategies yield 
more homogeneous alert levels within the population. 

\begin{figure*}[t!]
\includegraphics[width=\textwidth]{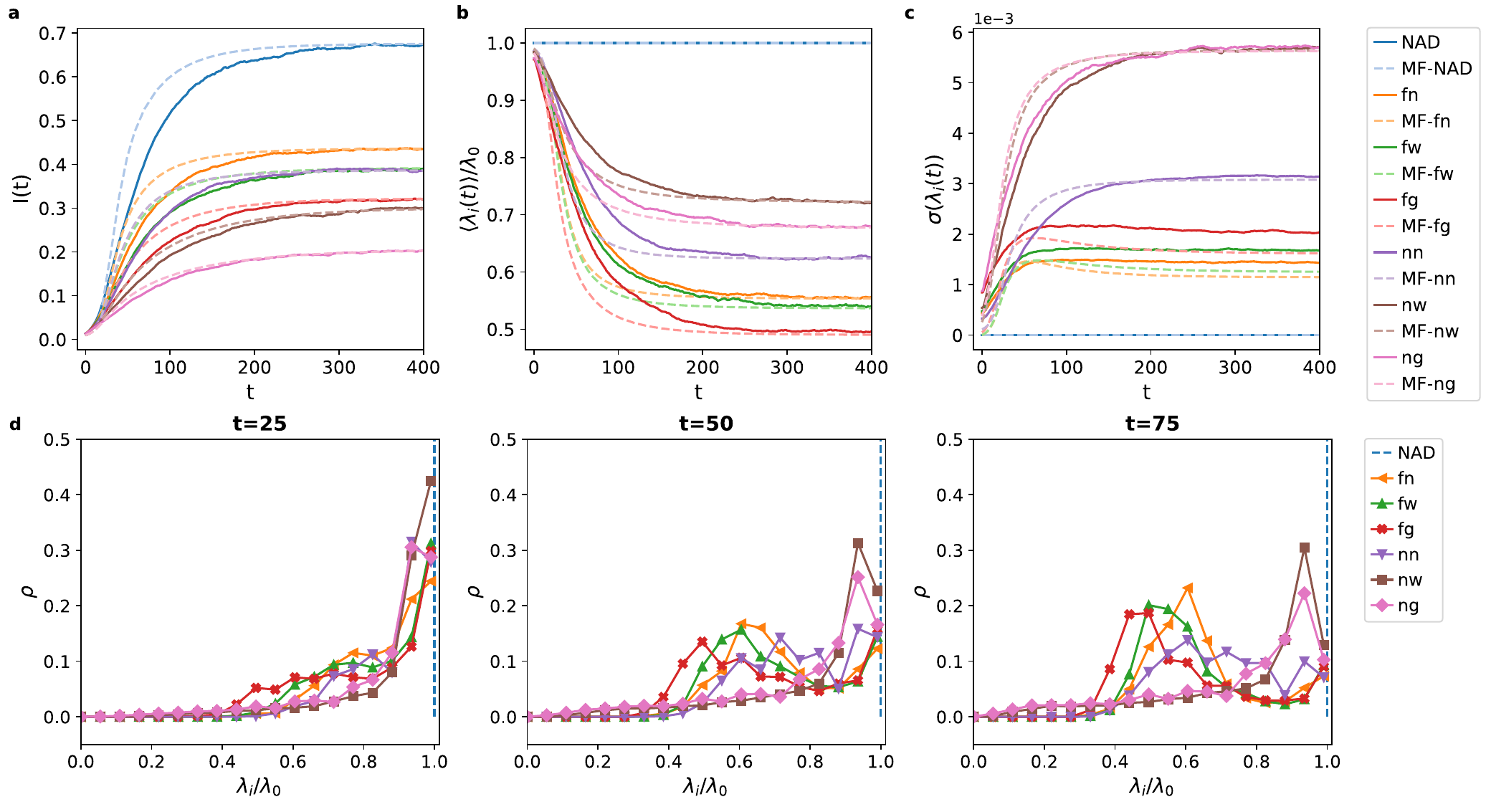}
\caption{\textbf{Impact of adaptive behaviors on the epidemic transient.} We consider the pairwise contagion process on the hospital dataset, with
$r=0.05$, $\theta=0.3$. 
\textbf{a}: temporal evolution of the fraction of infected nodes $I(t)$; 
\textbf{b}: temporal evolution of the mean transmission parameter $\langle \lambda_i(t) \rangle/\lambda_0$; 
\textbf{c}: temporal evolution of the standard deviation of the transmission parameter $\sigma(\lambda_i(t))$. 
Solid lines in \textbf{a}-\textbf{c} are obtained by averaging over 300 numerical simulations; dashed lines are obtained through numerical integration of the mean-field equations. 
\textbf{d}: distribution of the transmission parameter $\rho(\lambda_i(t)/\lambda_0)$ at different times $t$, obtained by averaging over 300 numerical simulations. In all panels we consider the NAD case and the six adaptive strategies.}
\label{fig:figure5}
\end{figure*}

\subsubsection{Heterogeneous awareness and local prevalence}
To unveil the microscopic mechanisms of adaptation and better understand the differences in efficacy and cost of the strategies, we now investigate how the various strategies lead to different distributions of local prevalence and risk awareness among nodes, and we study how these properties are distributed depending on nodes' and groups' properties.

Figure \ref{fig:figure4}\textbf{a} shows the distributions 
$\rho(\lambda_i^{\infty}/\lambda_0)$ of the transmission parameters
of the various nodes in the steady state, for the various strategies.
The three strategies based on relative information ($fn$, $fw$, $fg$)
present all a homogeneous distribution, well peaked around 
their average value $\left\langle \lambda_i/\lambda_0 \right\rangle$. The
social cost of these strategies is thus similar for all nodes.
On the contrary, the strategies based on absolute and higher-order information ($nw$ and $ng$) present a broader distribution, with large fluctuations and an asymmetric shape: most nodes carry a small social
cost, with $\lambda_i/\lambda_0 \sim 1$, but the distribution extends
to very small values, with a few nodes having 
$\lambda_i/\lambda_0 \sim 0$ and thus a high social cost.
Hence, the best strategies produce a heterogeneous response in the system, with large variability in the risk perception; 
the worst strategies instead produce a homogeneous reduction of
contact intensity within the population. 

A similar mechanism is observed at the group level: we can define for each group $e$ the average of its transmission parameters in the steady state, 
$\overline{\lambda}_e^{\infty}/\lambda_0$. The reduction
of this average with respect to the NAD case can be seen as an estimate of 
the awareness level within the group. Figure \ref{fig:figure4}\textbf{d} shows the resulting distribution $\varrho(\overline{\lambda}_e^{\infty}/\lambda_0)$: 
also in this case, the higher-order and absolute strategies present a broad distribution, unveiling a heterogeneous distribution of awareness among groups, ranging from low to very high levels of alert; on the contrary, pairwise and/or relative strategies present homogeneous distributions, with all groups having a similar level of alert.

Let us now analyse how the awareness is distributed among nodes and groups,
as a function of their structural properties. Hubs and large groups are indeed
well known to have an important role in sustaining the spreading, both
for pairwise and higher-order processes \cite{barrat2008dynamical,St-Onge2022,Mancastroppa2023}. To this aim, we show in Fig. \ref{fig:figure4}\textbf{b},\textbf{e} the average value of transmission parameters of nodes and groups,
$\lambda_i^{\infty}/\lambda_0$ and $\overline{\lambda}_e^{\infty}/\lambda_0$, as a function respectively of the nodes' hyperdegree and of the groups' size. 
As expected from Fig. \ref{fig:figure4}\textbf{a},\textbf{d}, the reduction in transmission parameters is independent from these properties for the strategies based on relative information. The $nn$ strategy, which yields intermediate results, has a smaller alert (awareness of risk) only for nodes with small hyperdegree and for small groups. On the other hand, the $nw$ and $ng$ strategies yield an alert level that is monotonously increasing with the hyperdegree and the group size: they lead to a strong reduction of the transmission parameter for high-hyperdegree nodes and within large groups, while the reduction is very limited for low-hyperdegree nodes and in small groups. 
This has a twofold effect on the spreading: (i) it prevents high-hyperdegree nodes and nodes in large groups from becoming infected when they are susceptible; 
(ii) whenever these nodes or groups become infectious, 
their risk of transmitting further the infection is strongly reduced.
The fact that the reduction in transmission parameters is focused on those nodes
also strongly lowers the social cost of adaptation: high levels of alert are focused only on a small part of the population (high-hyperdegree nodes, with 
$\lambda_i^{\infty}/\lambda_0 \ll 1$) while the rest of the population behaves normally (with $\lambda_i^{\infty}/\lambda_0 \sim 1$), thus preserving the activity of many nodes. 

Obtaining a high level of alert on nodes and groups usually at high-risk of infection and transmission is in fact akin to targeted immunization strategies \cite{satorras2015epidemic,mancastroppa2024},
which are known to significantly limit the spread by hindering the tendency of hubs to become a reservoir of the infection. 
We illustrate this effect by computing, in the endemic steady state, 
the fraction of time spent in the infectious state of each node $i$, 
$\tau_i^{\infty}$, and the average fraction of infectious nodes contained in each group $e$, $\eta_e^{\infty}$.
Figure \ref{fig:figure4}\textbf{c},\textbf{f} shows how these quantities vary with the hyperdegree of nodes and with the size of groups. 
In the NAD case, as expected 
the epidemic tends to be localized in high-hyperdegree nodes and in large groups, where the probability of infection is high. A similar localization of the spread is obtained for the relative and/or pairwise strategies ($fn$, $fw$, $fg$, $nn$).
Conversely, the absolute higher-order strategies ($nw$, $ng$) yield a much less
localized spread: the probability of infection of high-hyperdegree nodes and large groups becomes similar to that of other nodes and groups. 
Hence, these strategies drastically reduce the impact of the epidemic by decoupling it from the most potentially risky nodes and groups, which normally sustain the epidemic acting as reservoir of the infection.

Note that these results are obtained both through numerical simulations and integration of the IBMF equations (see Methods), with a good agreement between 
these two approaches, even regarding the social cost and the behavior of single nodes and groups (see Fig.s \ref{fig:figure3}, \ref{fig:figure4}). 

\begin{figure}[htb]
\includegraphics[width=\columnwidth]{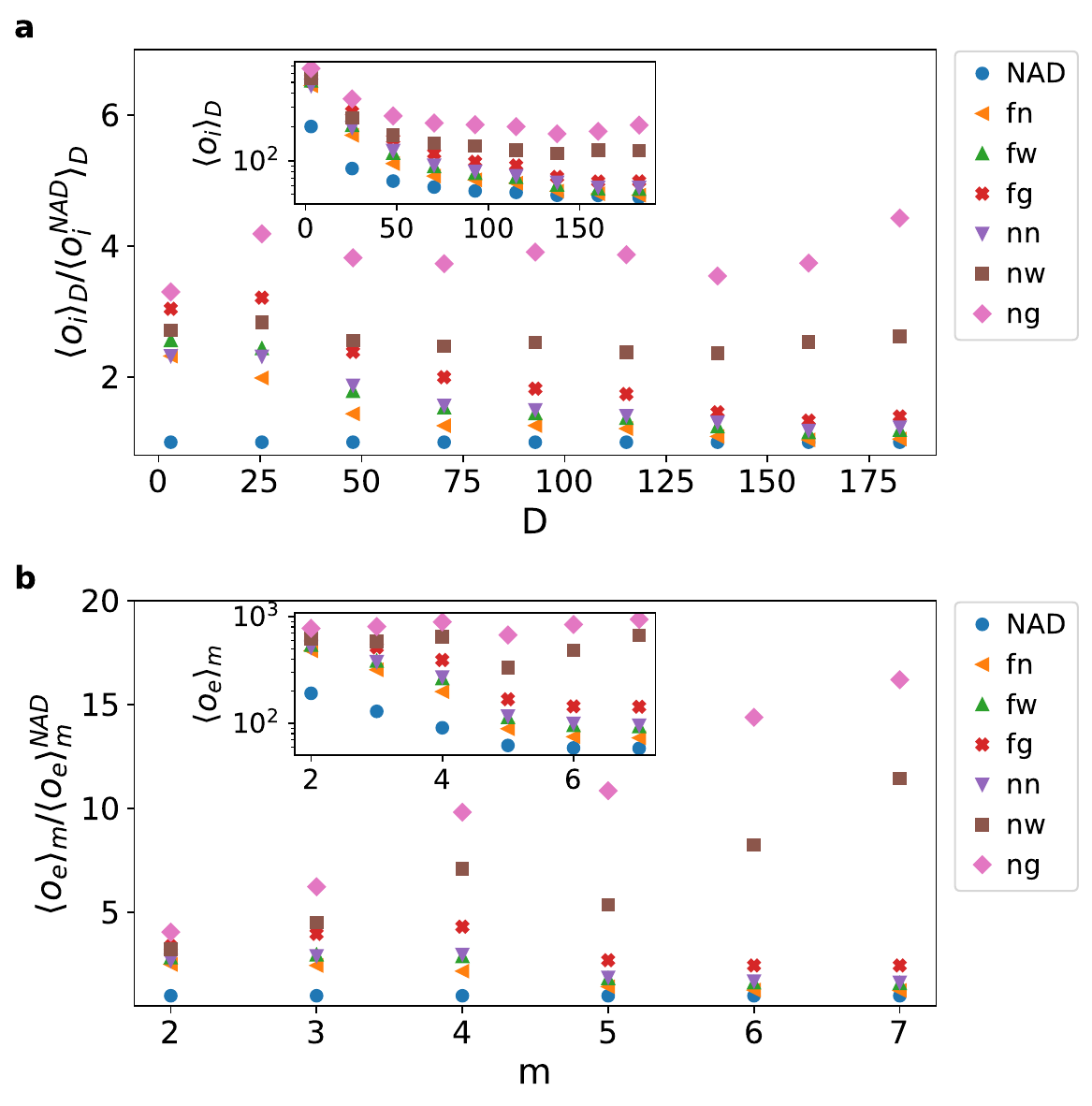}
\caption{\textbf{Adaptation in the early stages of the epidemic.} We consider the pairwise contagion process on the hospital dataset, with $r=0.05$ and $\theta=0.3$. 
\textbf{a}: we divide nodes into classes according to their hyperdegree  $D$ and for each hyperdegree class $D$ we compute the average time of first infection $\langle o_i \rangle_D/\langle o_i \rangle^{NAD}_D$ compared to the NAD case (in the inset we show the absolute value). 
\textbf{b}: for each hyperedge size $m$ we compute the average time of first infection $\langle o_e \rangle_m/\langle o_e \rangle^{NAD}_m$ compared to the NAD case (in the inset we show the absolute value). The time of first infection is estimated for each node or hyperedge by averaging over 300 numerical simulations. In all panels we consider the NAD case and the six adaptive strategies.}
\label{fig:figure6}
\end{figure}

\subsubsection{Early phases of the epidemic}
\label{sec:init}
While the previous results describe the result of adaptation strategies in the steady state, it is also of interest to study how these diverse situations are reached, and
thus to focus on the early stages of the spread.
Figure \ref{fig:figure5}\textbf{a}-\textbf{c} report the temporal dynamics of the epidemic prevalence and of the social behavior, through 
the evolution of the fraction of infectious nodes $I(t)$, of the average transmission parameter  $\langle \lambda_i(t)\rangle/\lambda_0$ and its fluctuations 
$\sigma(\lambda_i(t))$ (see the SM for results on the very first steps of the propagation). 
Both stochastic simulations and IBMF equations show that the hierarchy between the various strategies, both in terms of prevalence and of social costs, emerge 
very rapidly and remain stable during the epidemic transient.
Figure \ref{fig:figure5}\textbf{d} indicates that the differences in the
distributions of awareness  $\rho(\lambda_i(t)/\lambda_0)$ also emerge rapidly.
Initially, all nodes feature the same $\lambda_i=\lambda_0$, 
hence $\rho(\lambda_i/\lambda_0) \sim \delta(\lambda_i/\lambda_0-1)$, 
but the spread rapidly triggers some adaptive behaviors: 
nodes with $\lambda_i/\lambda_0 \sim 0$ emerge rapidly for 
the strategies based on absolute and higher-order information, while 
the majority of nodes remains with $\lambda_i/\lambda_0 \sim 1$, yielding a heterogeneous distribution early in the spread. 
On the contrary, the distributions for the other strategies (based on relative and/or pairwise information) keep a more homogeneous shape with their average decreasing over time. Indeed, 
the relations between prevalence or social cost and node or group properties shown in Fig. \ref{fig:figure4} for the steady state are already present in these early stages of the spread (see SM).

\begin{figure*}[ht!]
\includegraphics[width=\textwidth]{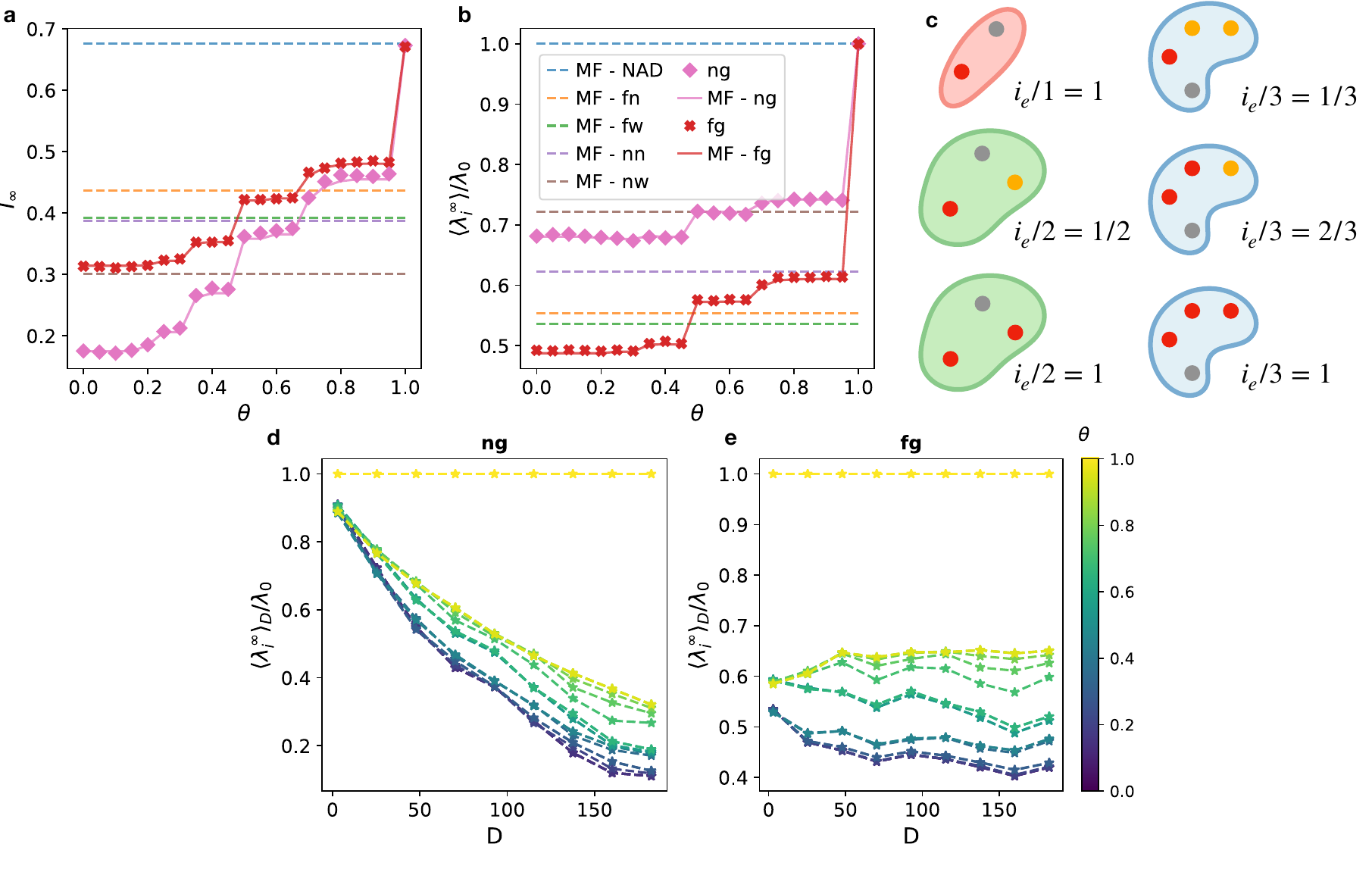}
\caption{\textbf{Impact of the level of alert $\theta$.} 
We consider the pairwise contagion process on the hospital dataset with $r=0.05$. 
In panels \textbf{a}, \textbf{b} we show respectively the epidemic prevalence $I_{\infty}$ and the average transmission parameter $\langle \lambda_i^{\infty} \rangle/\lambda_0$ in the asymptotic steady state as a function of $\theta$ for the $ng$ and $fg$ strategies. 
The markers indicate the results averaged over 300 numerical simulations and the curves represent the results of numerical integration of the mean-field equations (also for the other strategies - horizontal dashed lines). 
In panel \textbf{c} we sketch the epidemic configurations a node (gray) can perceive in hyperedges of size $2,3,4$, and for each of them we indicate the perceived fraction of infected nodes $i_e/(|e|-1)$. 
In panel \textbf{d}, \textbf{e} we divide nodes into hyperdegree classes $D$ and for each class we estimate the mean transmission 
parameter $\langle \lambda_i^{\infty} \rangle_D/\lambda_0$ in the steady state, obtained by integrating the mean-field equations for the $ng$, $fg$ strategies with different $\theta$ values (see colorbar).}
\label{fig:figure7}
\end{figure*}

We moreover investigate how the adaptive mechanisms impact the way in which the contagion reaches various nodes and groups in these early stages. We
compute for each node $i$ and each group $e$ their time of first infection,
respectively $o_i$ and $o_e$ (see Methods). Figure 
\ref{fig:figure6}\textbf{a},\textbf{b} show how these quantities 
vary with the hyperdegree of nodes and the size of groups. 
In the NAD case the first nodes to be infected are those with high hyperdegree,
with a progressive cascade towards nodes of decreasing  hyperdegree; 
analogously, the first groups to be infected are those of large size, followed progressively by those of smaller size. 
The relative and/or pairwise strategies do not change this hierarchy of first infection, neither for the nodes nor for the groups, and in fact they 
even reinforce it: indeed, they delay the infection of small groups and low-hyperdegree nodes, but do not change much the time of first infection of hubs and large groups. 
On the contrary, the absolute and higher-order strategies strongly delay the infection of all nodes, and in particular of the high-hyperdegree nodes and of the large groups, making the hierarchy in the order of first infection much less
pronounced.

\subsubsection{Impact of the level of alert}
All the previous results have been obtained for a given value of the parameter
$\theta$, which defines the level of alert for the strategies based on group information, namely, the minimum fraction of infectious nodes that a node $i$ needs to 
detect in a group to which it belongs to deem it infectious. Lower values of $\theta$ means that it is easier to consider a group as a risk,
which can be interpreted as an increase in the cautiousness of the nodes
(potentially with a higher social cost).
We thus now investigate the impact of $\theta$ 
on efficacy and social cost of the strategies $fg$ and $ng$. 

Figure \ref{fig:figure7}\textbf{a},\textbf{b} show the epidemic prevalence $I_{\infty}$ and the average transmission parameter 
$\langle \lambda_i^{\infty} \rangle/\lambda_0$ in the asymptotic state as a function of $\theta$ (for both 
the IBMF approach and the numerical simulations). 
For $\theta=1$ the strategies both coincide with the NAD case, since no group is considered infectious, regardless of its epidemiological status. 
As soon as $\theta$ is reduced below 1, all groups $e$ with $(|e|-1)$ infected nodes perceived by the node $i$ are considered as infectious 
(as from the point of view of $i$, all nodes in the group are infectious):
the adaptive strategies are then triggered, and we observe a drop in both
prevalence and average transmission parameter. As $\theta$ decreases, successive plateaux
and drops are observed: these occur at values of $\theta$ corresponding to thresholds leading to additional configurations of groups to be considered as infectious. 
Indeed, as the groups in the dataset have only finite sizes, the perceived fraction of infected nodes in a group can only assume specific values.
For example, for $\theta \in [0.5,1)$ a node $i$ can consider a group of size 3 to be infectious only if the other two nodes are infected; as soon as $\theta<0.5$, it
is enough that one of them is infectious to consider the group to be a risk (see Fig. \ref{fig:figure7}\textbf{c}). 
We note that the impact of $\theta$ is qualitatively similar on the $ng$ and $fg$ strategies. The difference between them increases as $\theta$ decreases, 
with the $ng$ strategy becoming better than the $fg$ both in terms of prevalence and of social cost.

As both prevalence and social cost depend on $\theta$, so does the comparison with the other strategies.
In the present dataset, for $\theta > 2/3$,  only groups almost fully infected are considered as infectious.
Hence it is difficult to trigger the adaptive mechanism, and the prevalence is not reduced strongly:
the higher-order strategies are then less efficient than the other strategies. 
However, as soon as $\theta \lesssim 2/3$ (resp. $\lesssim 0.5$),  
the $ng$ strategy becomes the second most efficient (resp. the most efficient) in containing the spread.
The $fg$ strategy becomes moreover almost as efficient as the $nw$ one.
Thus, even if $\theta$ impacts the performance of the higher-order strategies, 
their performance remains the best on a broad range of parameter values. 
Note that $\theta=0.5$ is not an exceedingly cautious or unrealistic behavior, since it means that a group is considered as infectious if half of its members -apart from the observing node- are infectious.

The impact of the alert level $\theta$ on the social cost is more limited (see Fig. \ref{fig:figure7}\textbf{b}). The absolute higher-order strategy, $ng$, presents consistently a low-cost even for low $\theta$, showing a reduction in $\lambda_i$ of at most 30\%, smaller than the other
strategies (except $nw$). Figure \ref{fig:figure7}\textbf{d} indeed shows that the $ng$ strategy
exploits the structural heterogeneities, as described previously, in the same way for all $\theta < 1$,
with a higher reduction in transmission parameters for nodes with higher hyperdegree.
Similarly, the $fg$ strategy leads to a reduction almost independent from the hyperdegree for
all values of $\theta$ (see Fig. \ref{fig:figure7}\textbf{e}). 

Overall, the absolute higher-order strategies can balance social cost and impact on the epidemic, for a wide range of values of the alert level. In particular, realistic alert levels still guarantee the effectiveness in epidemic control with a low social cost.

\subsection{Mechanisms driving adaptation}
The results described above provide insights into the underlying adaptive mechanisms of the various strategies. Moreover, they identify the key features affecting the social cost and performance of such adaptive strategies: The strategies performance depends strongly on the nature and type of information triggering the adaptive behaviors, and on how they correlate with nodes structural properties. 

Regarding the role of the \textit{type} of information, 
let us first consider strategies where the awareness of a node is based on its \textit{relative} number of infectious interactions. To reach a given value of its awareness function $f_i$, a hub needs then to have a higher number of infectious interactions than a low-degree node. This means that high-degree nodes will not
reach a significant level of awareness if the local prevalence around them is not already large. Their
known vulnerability to a rapid infection and their risk of being a source of superspreading events
are thus not hindered. Moreover, the risk perception and subsequent decrease in risk transmission parameters
are spread out on all nodes, independently from their degree or hyperdegree.
On the other hand, for strategies in which the awareness depends on the \textit{absolute} number of infectious
interactions, hubs reach high levels of awareness more easily than low-degree nodes. These
strategies thus exploit the structural heterogeneity, similarly to targeted immunization strategies \cite{satorras2015epidemic,mancastroppa2024}, by
effectively focusing awareness efforts on the hubs. However, the absolute strategies obtain this as an effective result of their basic mechanisms, without
explicitly targeting hubs, thus without the need for global knowledge of the network structure \cite{cohen2003efficient}. This results in a twofold effect: susceptible hubs with infectious neighbors decrease their probability of becoming infectious and, if they become infectious, their spreading power is reduced.

The \textit{nature} of the information defining the awareness plays an important role as well. Strategies based on pairwise information simply count the number of infectious neighbors, regardless of 
the number of times they are encountered (i.e., of the number of groups in which they co-participate). This leads
to an underestimation of the role of important contagion paths \cite{cencetti24_complex}.
For strategies including higher-order information instead, the multiplicity of interactions is taken into
account to determine the awareness level. In particular, not only hubs but also nodes with a limited
number of neighbors, belonging to overlapping hyperedges, can trigger high levels of awareness in their neighborhood when they are infectious.
These considerations apply also to hybrid strategies, as they retain 
the information on the number of interactions between pairs of nodes. 

Overall, the absolute strategies based on higher-order information, $nw$ and $ng$, provide the best performance by fully exploiting the heterogeneity and overlap of interactions. They effectively target the risk alert on nodes with large hyperdegree and within large groups. This reduces their probability of being infected and of transmitting the infection, hence defusing superspreading events that, in the non-adaptive case, drive the epidemic spreading. This heterogeneous impact strongly reduces the prevalence while maintaining a low social cost, with a limited reduction in the activity of most nodes.

We finally note that the structure of the underlying hyper-network plays a crucial role: in the SM, we show how different strategies become equivalent, canceling out their differences, if the underlying hypergraph does not present heterogeneity or higher-order interactions. Moreover, we show that both hyperdegree heterogeneity and hyperedge overlap amplify these differences.

\section{Discussion}
\label{sec:discussion}

Here and in a companion paper \cite{mancastroppa2025_2}, we have developed and studied a series of adaptive mechanisms driven by the perception of risk during an ongoing pairwise or higher-order spreading process on a (hyper)network.
We have considered adaptive behaviors driven by local awareness of the epidemic: nodes change their behavior based on pairwise
or higher-order information about the health status of their neighborhood. This behavioral change reduces their probability of becoming infectious and of transmitting the infection.
In particular, we defined six adaptive strategies with different type and nature of the information triggering such behavioral changes: the level of awareness can be driven by the absolute or relative number of infectious interactions perceived by a node, and the corresponding information can be based on higher-order, pairwise, or hybrid interactions. 
Using a mean-field analytic approach and numerical simulations, we analyzed the strategies' 
effects on contagion processes. 
In \cite{mancastroppa2025_2}, we have focused on the epidemic phase transition. We have shown
that adaptation does not modify the epidemic phase transition in pairwise processes, maintaining its continuous nature and the same epidemic threshold. Meanwhile it reduces and can even suppress the bistability regime typical of higher-order processes, neutralizing the explosiveness of the higher-order transition. 
In the present paper, we have instead compared their effectiveness in containing contagion and (proxies of)
their social cost.
Adaptive behaviors guided by absolute higher-order information yield the best performances, producing a high level of epidemic control while having a low social cost. Instead, relative pairwise strategies produce a limited control despite a high social cost. 
The superior performance of the absolute higher-order strategies lies in their ability to exploit the heterogeneity and overlap of group interactions. Indeed, they produce a heterogeneous risk perception in the population, with high level of alert on high-hyperdegree nodes, on their neighborhood and within large groups. This hinders the tendency of the spread to localize on high-hyperdegree nodes and large groups 
already in the early stages of the epidemic and also in the endemic state. Hence, this heterogeneous risk perception neutralizes superspreading events, which would otherwise drive and sustain high levels of contagion, and it also keeps the social cost low since the high alert is dynamically limited to hubs.

Our work provides insights into the effects of adaptive behaviors on contagion processes on hypergraphs, showing how to systematically compare the performance and cost of different adaptive strategies. We highlight that adaptive mechanisms driven by higher-order interactions can lead to a different phenomenology than when risk perception is triggered by pairwise information. 
Our results provide also a methodological and conceptual foundation for new research directions. For example, the development of other higher-order adaptive mechanisms, involving topological modifications such as groups merging and splitting \cite{Iacopini2024,liu2025higher}; or the introduction of dynamic adaptive behaviors on temporal hypergraphs \cite{Iacopini2024,MancastroppaEPJD,EATH_2025}, taking into account the temporal nature of interactions \cite{moinet2018effect,mancastroppa2020active,mancastroppa2024}.

\section{Methods}
\label{sec:methods}
\subsection{Empirical and synthetic hypergraphs} 
We built hypergraphs from empirical datasets describing time-resolved face-to-face interactions in several environments \cite{sociopatterns,Vanhems2013,genois2018,Genois2023,Stehle2011,Mastrandrea2015}: a hospital \cite{Vanhems2013,sociopatterns}, conferences \cite{Genois2023} and schools \cite{Stehle2011,sociopatterns,Mastrandrea2015} (see SM for details on each dataset and preprocessing). These hypergraphs cover a wide spectrum of properties (see SM), ranging from 76 to 327 nodes and from 315 to 4795 hyperedges of sizes up to 7: for example, the hospital dataset considered in the main text features $76$ nodes and $1102$ hyperedges.

We also consider synthetic hypergraphs obtained: (i) applying different reshuffling methods to the empirical ones \cite{agostinelli2025higher,Mancastroppa2023}; (ii) setting specific features and mechanisms for hyperedges formation (such as specific hyperdegree distributions or overlap level). We refer to the SM for more details.

\subsection{Numerical simulations}
We performed numerical simulations of the epidemic processes in discrete-time: infected individuals recover with probability $\mu$ per unit of time; a susceptible individual $i$ becomes infectious at time $t$ with probability $p_i^p(t)$ when considering the pairwise process or with probability $p_i^{ho}(t)$ when considering the higher-order process. In particular: 
\begin{align}
    p_i^p(t)&=1-\prod\limits_{j \in \mathcal{V}} (1-\lambda_i(t) \lambda_j(t))^{w_{i,j}},\\
    p_i^{ho}(t)&=1-\prod\limits_{e \in \mathcal{E}(i)} (1- \beta_e^i(i_e,\nu)),
\end{align}
where all possible contagion channels are considered.

\textit{When building the phase diagram:} the simulation is stopped when the system reaches the absorbing state. If the system does not reach the absorbing state, the simulation is run for a minimum of $T$ time-steps and then the system is assumed to have reached the stationary state at the time-step $t>T$ if: (i) $\sigma(\bm{I}_{0.8t,t})<10^{-2}$ and (ii) $|\overline{\bm{I}_{0.9t,t}}-\overline{\bm{I}_{0.8t,0.9t}}|<10^{-2}$, where $\bm{I}_{a,b}=[|\mathcal{I}|(a),|\mathcal{I}|(a+1),...,|\mathcal{I}|(b-1),|\mathcal{I}|(b)]$ is the epidemic prevalence from time $a$ to $b$, $\sigma(\bm{I}_{a,b})$ is its standard deviation and $\overline{\bm{I}_{a,b}}$ is its average. The pairwise contagion process is simulated starting from a single infectious individual (selected uniformly at random) in a completely susceptible population, with $\mu=10^{-2}$, $T=1000$; the results are averaged over different initial conditions and over all outcomes of the epidemic (as long as at least one transmission event has occurred during the simulation). The higher-order process is simulated starting with either 2\% or 80\% of the population in the infectious state (selected uniformly at random), in order to explore both branches of the phase diagram in the bistability region. In this case $\mu=10^{-1}$, $T=500$, and the results are averaged over different initial conditions and only over the realizations where the epidemic reaches an endemic state. For both type of processes, the asymptotic epidemic prevalence is then calculated as $I_{\infty}=\overline{\bm{I}_{0.8t,t}}$ and the results are averaged over 300 numerical simulations. 

\textit{When focusing on a specific $r>r_C^{NAD,p}$:} if the system does not reach the absorbing state, the simulation is run for $T$ times-steps, where $T$ is set to be large enough to allow reaching the steady state for the non-adaptive case and for any adaptive strategy. The pairwise contagion process is simulated starting from a single infectious individual (selected uniformly at random) in a completely susceptible population, with $\mu=10^{-2}$, $T=1000$; the higher-order process is simulated starting with 2\% of the population in the infectious state (selected uniformly at random), with $\mu=10^{-1}$, $T=500$. In each simulation, we evaluate at each time $t$ the fraction of infectious nodes $I(t)$; the parameters $\lambda_i(t) \forall i$; the time of first infection of nodes $o_i$ and of groups $o_e$ (assuming that the first infection of a group occurs the first time it has more than 75\% of infectious nodes). Focusing then on the asymptotic state, i.e. on the time interval $\Delta T=[0.8T,T]$, we measure for each node $\lambda_i^{\infty}$ as the average of $\lambda_i(t)$ over $\Delta T$; $\tau_i^{\infty}$ as the fraction of time $i$ spent infected in $\Delta T$; $\eta_e^{\infty}$ as the average fraction of infectious nodes in $e$ in $\Delta T$; and $\overline{\lambda}_e^{\infty}= \sum_{j \in e} \lambda_j^{\infty}/|e|$. For both type of processes, the results are averaged over 300 numerical simulations with different initial conditions (as long as at least one transmission event has occurred during the simulation).

\subsection{Integration of mean-field equations}
The mean-field equations are integrated numerically, considering the same epidemiological parameters as the numerical simulations. 

\textit{When building the phase diagram:} the pairwise process is integrated starting from $P_i(0)=0.01 \, \forall i$. The higher-order process is integrated either from $P_i(0)=0.01 \, \forall i$ or from $P_i(0)=0.8 \, \forall i$, to explore the two branches of the phase diagram. In this case $I_{\infty}= \sum_{i \in \mathcal{V}} P_i(t \to \infty) /N$.

\textit{When focusing on a specific $r>r_C^{NAD,p}$:} both the pairwise and higher-order processes are integrated starting from $P_i(0)=0.01 \, \forall i$. In each numerical integration we obtain at each time $t$ the probability $P_i(t) \forall i$: thus we can calculate the fraction of infected nodes as the average probability of being infected $I(t)=\sum_{i \in \mathcal{V}} P_i(t) /N$; the parameter $\lambda_i(t)$ for each node (as detailed in the next subsection); the fraction of infected nodes in a group $e$ as $\eta_e(t)=\sum_{j \in e} P_i(t)/|e|$; the average parameter in a group $e$ as $\overline{\lambda}_e(t)= \sum_{j \in e} \lambda_j(t)/|e|$. Focusing on the asymptotic state, i.e. considering $P_i(t \to \infty)$, we can similarly calculate $\lambda_i^{\infty}$, $\tau_i^{\infty}=P_i(t \to \infty)$, $\eta_e^{\infty}$ and $\overline{\lambda}_e^{\infty}$.

\subsection{Individual-based mean-field approach} 
Here we show that we obtain two closed set of 
mean-field equations for the pairwise and higher-order contagion, by expressing all quantities as functions of the $P_i(t)$. This generalizes to all six strategies the results of \cite{mancastroppa2025_2}.

We first evaluate the probability $Q_{e \setminus i}(i_e,t)$ that in the hyperedge $e$ at time $t$ there are exactly $i_e$ infected individuals (other than $i$): 
\begin{equation}
    Q_{e \setminus i}(i_e,t)= \sum\limits_{A \in \mathcal{C}_{e \setminus i}(i_e)} \prod\limits_{j \in A} P_j(t) \prod\limits_{k \in e \setminus \{i,A\}} (1-P_k(t)),
    \label{eq:Q}
\end{equation}
where $\mathcal{C}_{e \setminus i}(i_e)$ is the set of all possible sets of $i_e$ nodes in $e \setminus i$. 

We then need to evaluate the term  
$\overline{\langle \lambda_j(t)/\lambda_0 \rangle_{i_e}^\nu}$, 
which involves a first average over the $i_e$ infected individuals in $e$, and then a second average over all possible configurations of $i_e$ infected individuals in $e \setminus i$:
\begin{equation}
    \overline{\langle \lambda_j(t)/\lambda_0 \rangle_{i_e}^\nu}= 
    \sum\limits_{A \in \mathcal{C}_{e \setminus i}(i_e)} H_A(t) 
    \left(\frac{1}{i_e} \sum\limits_{j \in A} e^{- f_j(t)} \right)^\nu,
\end{equation}
where:
\begin{equation}
    H_A(t)=\frac{\prod\limits_{j \in A} P_j(t) \prod\limits_{k \in e \setminus \{i,A\}} (1-P_k(t))}{\sum\limits_{B \in \mathcal{C}_{e \setminus i}(i_e)} \prod\limits_{j \in B} P_j(t) \prod\limits_{k \in e \setminus \{i,B\}} (1-P_k(t))}.
    \label{eq:HA}
\end{equation}

We moreover need to evaluate the awareness function $f_i(t)$ for the different strategies. 

The "fraction of infectious neighbors" strategy ($fn$) and the "infectious neighbors" strategy ($nn$) are: 
\begin{align}
    f_i^{fn}(t) &= \sum\limits_{n=0}^{k(i)} n Q_{\mathcal{N}(i)}(n,t)/k(i) ,\\
    f_i^{nn}(t) &= \sum\limits_{n=0}^{k(i)} n Q_{\mathcal{N}(i)}(n,t)/\langle k \rangle ,
\end{align}
where $Q_{\mathcal{N}(i)}(n,t) $ is the probability that $i$ has exactly $n$ infected neighbors at time $t$, which is calculated analogously to Eq. \eqref{eq:Q}, $k(i)=|\mathcal{N}(i)|$ and $\langle k \rangle$ is the average degree in $\mathcal{G}$.

The "fraction of infectious weights" strategy ($fw$) and the "infectious weights" strategy ($nw$) are: 
\begin{align}
    f_i^{fw}(t) &= \sum\limits_{n=0}^{k(i)} Q_{\mathcal{N}(i)}(n,t) W_i(n,t)/s(i) ,\\
    f_i^{nw}(t) &= \sum\limits_{n=0}^{k(i)} Q_{\mathcal{N}(i)}(n,t) W_i(n,t)/\langle s \rangle ,
\end{align}
where $s(i)$ is the strength of $i$ in $\mathcal{G}$ and $W_i(n,t)$ is the average sum of weights of the $n$ infected nodes in the neighborhood of $i$ at time $t$. In particular: 
\begin{equation}
    W_i(n,t)= \sum\limits_{A \in \mathcal{C}_{\mathcal{N}(i)}(n)} H_A(t) \sum\limits_{j \in A} w_{i,j},
\end{equation}
where $\mathcal{C}_{\mathcal{N}(i)}(n)$ is the set of all possible sets of $n$ nodes in $\mathcal{N}(i)$ and $H_A(t)$ is calculated analogously to Eq. \eqref{eq:HA}. At each time $t$ for each node $i$ it is necessary to consider all possible configurations of $n$ nodes in $\mathcal{N}(i)$, for every value of $n \in [0,k(i)]$: this evaluation is generally not computationally feasible since the size of $\mathcal{N}(i)$ can be potentially large, therefore we consider an approximation. We assume that:
\begin{equation}
    W_i(n,t) \sim n \overline{w_{i,j}}^I(t),
\end{equation}
where $\overline{w_{i,j}}^I(t)$ is the average weight of nodes in the neighborhood of $i$, where each node contributes to the average proportionally to its probability of being infected at time $t$. In particular: 
\begin{equation}
    \overline{w_{i,j}}^I(t)=\sum\limits_{j \in \mathcal{N}(i)} w_{i,j} \frac{P_j(t)}{\sum\limits_{k \in \mathcal{N}(i)} P_k(t)}.
\end{equation}
This makes the evaluation computationally feasible, and the approximation leads to:
\begin{align}
    f_i^{fw}(t) &= \sum\limits_{n=0}^{k(i)} n Q_{\mathcal{N}(i)}(n,t) \overline{w_{i,j}}^I(t)/s(i) ,\\
    f_i^{nw}(t) &= \sum\limits_{n=0}^{k(i)} n Q_{\mathcal{N}(i)}(n,t) \overline{w_{i,j}}^I(t)/\langle s \rangle.
\end{align}

The "fraction of infectious groups" strategy ($fg$) and the "infectious groups" strategy ($ng$) are: 
\begin{align}
    f_i^{fg}(t) &= \sum\limits_{g=0}^{D(i)} g Z_{\mathcal{E}(i)}(g,t)/D(i) ,\\
    f_i^{ng}(t) &= \sum\limits_{g=0}^{D(i)} g Z_{\mathcal{E}(i)}(g,t)/\langle D \rangle ,
\end{align}
where $D(i)$ is the hyperdegree of node $i$, $\mathcal{E}(i)$ is the set of hyperedges in which $i$ is involved and $Z_{\mathcal{E}(i)}(g,t)$ is the probability of having $g$ infectious hyperedges at time $t$ among the hyperedges in $\mathcal{E}(i)$. 
We can write:
\begin{equation}
    Z_{\mathcal{E}(i)}(g,t)=\sum\limits_{A \in \mathcal{C}_{\mathcal{E}(i)}(g)} \prod\limits_{e \in A} \xi_{e,i}(t) \prod\limits_{h \in \mathcal{E}(i) \setminus A} (1-\xi_{h,i}(t)),
\end{equation}
where $\mathcal{C}_{\mathcal{E}(i)}(g)$ is the set of all possible sets of $g$ hyperedges
in $\mathcal{E}(i)$ and $\xi_{e,i}(t)$ is the probability that the hyperedge $e$ is infectious
at time $t$ for node $i$. In particular:
\begin{equation}
    \xi_{e,i}(t)=\sum\limits_{i_e=\phi_e(\theta)}^{|e|-1} Q_{e \setminus i}(i_e,t),
\end{equation}
where $\phi_e(\theta)=\lfloor \theta (|e|-1) \rfloor +1$ is the minimum number of infectious individuals in $e$ (distinct from $i$) so that $i$ perceives $e$ as at-risk.

\subsection{Pairwise epidemic threshold} 
By applying a linear stability analysis to Eq. \eqref{eq:MF_p}, we obtain a mean-field estimation of the epidemic threshold for the pairwise contagion \cite{satorras2015epidemic}. 
In the NAD case, Eq. \eqref{eq:MF_p} reduces to: 
\begin{equation}
    \partial_t P_i(t)= -\mu P_i(t) + \lambda_0^2 (1-P_i(t)) \sum\limits_{j \in \mathcal{V}}  w_{i,j} P_j(t).
    \label{eq:MF_p_NAD}
\end{equation}
Linearizing Eq. \eqref{eq:MF_p_NAD} around the absorbing state $\bm{P}=[0,0,...,0]$ leads to the Jacobian matrix with element $J_{i,j}=-\mu \delta_{i,j} + \lambda_0^2 w_{i,j}$. The absorbing state is thus stable if and only if the largest eigenvalue of $\{J_{i,j}\}$ is negative, i.e. if:
\begin{equation}
   r=\lambda_0^2/\mu<1/\Lambda_w \equiv r_C^{NAD,p},
   \label{eq:conditions}
\end{equation}
where $\Lambda_w$ is the largest eigenvalue of $\{w_{i,j}\}$.

The adaptive strategies we considered feature the same Jacobian matrix as the NAD case. Indeed, expanding around the absorbing state and keeping only the linear leading order
yields:
$Q_{\mathcal{N}(i)}(n,t) \sim \delta_{n,1} \sum_{j \in \mathcal{N}(i)} P_j(t)$
and
$Q_{e \setminus i}(i_e,t) \sim \delta_{i_e,1} \sum_{j \in e \setminus i} P_j(t)$,
hence:
\begin{align}
    f_i^{fn}(t) &\sim \sum\limits_{j \in \mathcal{N}(i)} P_j(t)/k(i),\\
    f_i^{nn}(t) &\sim \sum\limits_{j \in \mathcal{N}(i)} P_j(t)/\langle k \rangle,\\
    f_i^{fw}(t) &\sim \sum\limits_{j \in \mathcal{N}(i)} \overline{w_{i,j}}^I(t) P_j(t)/s(i),\\
    f_i^{nw}(t) &\sim \sum\limits_{j \in \mathcal{N}(i)} \overline{w_{i,j}}^I(t) P_j(t)/\langle s \rangle,\\
    f_i^{fg}(t) &\sim \sum\limits_{e \in \mathcal{E}(i) | \phi_e(\theta) \leq 1} \sum\limits_{j \in e \setminus i} P_j(t)/D(i),\\
    f_i^{ng}(t) &\sim \sum\limits_{e \in \mathcal{E}(i) | \phi_e(\theta) \leq 1} \sum\limits_{j \in e \setminus i} P_j(t)/\langle D \rangle.
\end{align}
Therefore, for all strategies $\lambda_i(t) \sim \lambda_0 \forall i, \forall t$ and, inserting in Eq. \eqref{eq:MF_p}, we recover the NAD case of Eq. \eqref{eq:MF_p_NAD}, which implies that the epidemic thresholds for the pairwise contagion in all the six adaptive cases are the same as for the NAD case. This generalizes to all six strategies the results of \cite{mancastroppa2025_2}.

\section*{Data availability}
The data that support the findings of this study are publicly available. The SocioPatterns data sets are available at \cite{sociopatterns, conf_data}.

\begin{acknowledgments}
M.M., M.K. and A.B. acknowledge support from the Agence Nationale de la Recherche (ANR) project DATAREDUX (ANR-19-CE46-0008). M.K. received support from the National Laboratory for Health Security (RRF-2.3.1-21-2022-00006); the MOMA WWTF project; and the BEQUAL NKFI-ADVANCED grant (153172).
\end{acknowledgments}

\bibliographystyle{naturemag}
\bibliography{references_HO_adaptive_epi}

\end{document}


\title{Supplementary Material for "Higher-order adaptive behaviors outperform pairwise strategies in mitigating contagion dynamics"
}

\author{Marco Mancastroppa}
\affiliation{Aix Marseille Univ, Université de Toulon, CNRS, CPT, Turing Center for Living Systems, 13009 Marseille, France}
\author{Márton Karsai}
\affiliation{Department of Network and Data Science, Central European University, 1100 Vienna, Austria}
\affiliation{National Laboratory for Health Security, HUN-REN Rényi Institute of Mathematics, 1053 Budapest, Hungary}
\author{Alain Barrat}
\affiliation{Aix Marseille Univ, Université de Toulon, CNRS, CPT, Turing Center for Living Systems, 13009 Marseille, France}

\maketitle



In this Supplementary Material we present additional information on the considered (empirical and synthetic) hypergraphs, as well as additional results. In Section \ref{sez:section1} we present in detail all the empirical datasets considered, their preprocessing procedure, and their basic properties. In Section \ref{sez:section2}, for all these datasets, we report results on the impact of the adaptive behaviors on both the contagion dynamics and the societal cost of adaptation, considering both pairwise and higher-order contagion
(since in the main text we present for simplicity results restricted to one dataset and to pairwise contagion). We also report on (i) the local impact of these behaviors when there are few infected individuals, 
(ii) an alternative estimation of the adaptation social cost and 
(iii) additional results on the epidemic transient, for the dataset used in the main text. 
In Section \ref{sez:section3}, we present the considered synthetic hypergraphs, their basic features and the procedures to generate them. Finally, in Section \ref{sez:section4}, we present results on the impact of adaptive strategies on these synthetic hypergraphs and we discuss the effects of hyperdegree heterogeneity and hyperedge overlap on the efficacy and social cost of the strategies.

\section{Empirical hypergraphs}
\label{sez:section1}
We consider empirical datasets describing time-resolved face-to-face interactions in several settings: a hospital (LH10 - \cite{Vanhems2013,sociopatterns}), two schools (Thiers13, LyonSchool - \cite{Stehle2011,sociopatterns,Mastrandrea2015}) and three conferences (WS16, ECIR19, ICCSS17 - \cite{Genois2023}). The datasets consist in time-resolved pairwise interactions: we preprocess the data to obtain static hypergraphs, by using a standard preprocessing method \cite{Mancastroppa2023,iacopini2022group}. The aggregate hypergraph $\mathcal{H}$ is built as follows: (i) all pairwise interactions are aggregated on time windows of duration $\Delta t$; (ii) all maximal cliques (not contained in other cliques) are promoted to hyperedges; (iii) only group interactions observed at least $\delta$ times over the entire time span are retained as hyperedges of $\mathcal{H}$. 

In Supplementary Table \ref{tab:table1}, we report for each dataset 
the setting in which the interactions were collected, the parameters $\Delta t$ and $\delta$ of the preprocessing and the main higher-order topological properties of the obtained empirical hypergraphs $\mathcal{H}$: the number of nodes $N$, the number of hyperedges $E$, the average hyperedge size $\langle m \rangle$, the average hyperdegree $\langle D \rangle$, the average degree $\langle k \rangle$, the average link weight $\langle w_l \rangle$ in the weighted projected graph $\mathcal{G}$, and the heterogeneity of their distributions. In Supplementary Fig. \ref{fig:figure1} we show, for each dataset, the hyperedge size distribution $\Psi(m)$, the distribution $P(k/\langle k \rangle)$ of the degree and $P(w_l/\langle w_l \rangle)$ of the link weights in $\mathcal{G}$, the distribution $P(D/\langle D \rangle)$ of the hyperdegree in $\mathcal{H}$.

Supplementary Table \ref{tab:table1} and Supplementary Fig. \ref{fig:figure1} show that
the empirical datasets considered describe interactions collected in a wide range of different settings, and present different statistical properties.

\begin{table}[h!]
    \begin{tabular}{|c|c|c|c|c|c|c|c|c|c|c|c|c|c|}
    \hline
    \hline
          & Setting & $\Delta t$ & $\delta$ & N & E & $\langle m \rangle$ & $\langle D \rangle$ & $\langle D^2 \rangle/\langle D \rangle^2$ & $\langle k \rangle$ & $\langle k^2 \rangle/\langle k \rangle^2$ & $\langle w_l \rangle$ & $\langle w_l^2 \rangle/\langle w_l \rangle^2$ \\ \hline
         LH10 & Hospital & 15 min &  1  &  76  &  1102  &  3.4 &  50.0  &  2.0  &  30.4  &  1.3  &  4.6  &  3.4\\
         Thiers13 & High-school & 15 min &  1  &  327  &  4795  &  3.1 &  45.3  &  1.3  &  35.6  &  1.1  &  3.0  &  2.2\\
         LyonSchool & Primary school & 15 min &  3  &  242  &  1188  &  2.4 &  11.8  &  1.2  &  13.9  &  1.2  &  1.3  &  1.3\\
         WS16 & Conference & 5 min &  2  &  137  &  315  &  3.4 &  7.8  &  1.3  &  16.1  &  1.3  &  1.3  &  1.3\\
         ECIR19 & Conference & 5 min &  2  &  169  &  708  &  2.7  &  11.3  &  1.3  &  17.3  &  1.3  &  1.3  &  1.4 \\
         ICCSS17 & Conference & 5 min &  2  &  271  &  1473  &  2.6 &  13.9  &  1.3  &  20.4  &  1.4  &  1.2  &  1.3\\
    \hline
    \hline
    \end{tabular}
    \caption{\textbf{Properties of empirical datasets.} For each empirical dataset we report: the setting in which the face-to-face interactions have been collected, the preprocessing parameters $\Delta t$ and $\delta$, the number of nodes $N$, the number of hyperedges $E$, their average size $\langle m \rangle$, the average hyperdegree $\langle D \rangle$ and the heterogeneity $\langle D^2 \rangle/\langle D \rangle^2$ of its distribution, the average degree $\langle k \rangle$ and the heterogeneity $\langle k^2 \rangle/\langle k \rangle^2$ of its distribution, the average weight $\langle w_l \rangle$ in the projected graph $\mathcal{G}$ and the heterogeneity $\langle w_l^2 \rangle/\langle w_l \rangle^2$ of its distribution. Note that the statistics of the weights in $\mathcal{G}$ considers only existing links, so $\langle w_l \rangle \geq 1$.} 
    \label{tab:table1}
\end{table}

\clearpage
\begin{figure*}[ht!]
\includegraphics[width=0.97\textwidth]{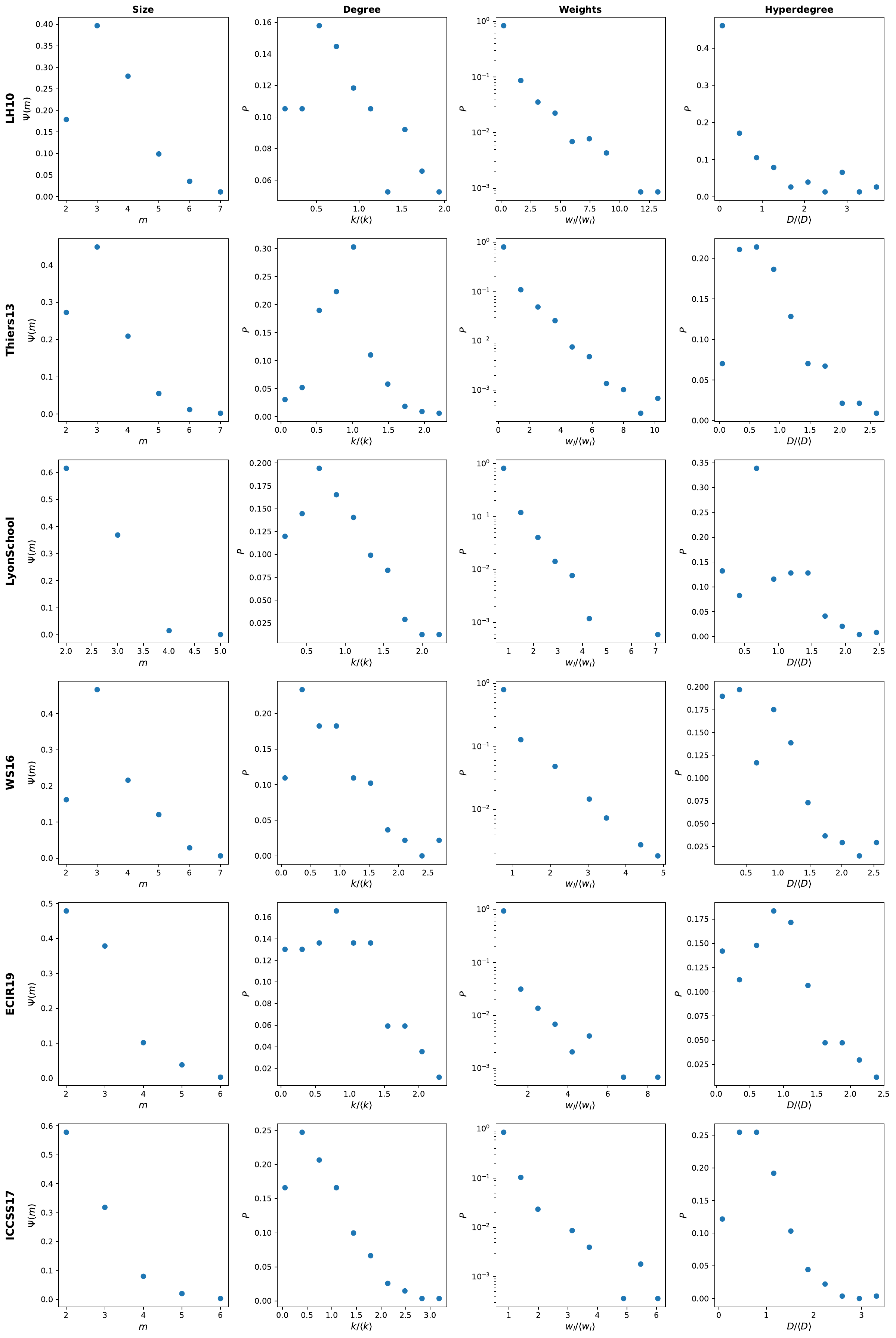}
\caption{\textbf{Properties of empirical datasets.} Each row corresponds to a different dataset (see labels): for each of them the first column shows the hyperedge size distribution $\Psi(m)$; the second column the distribution $P(k/\langle k \rangle)$ of the degree in $\mathcal{G}$; the third column the distribution $P(w_l/\langle w_l \rangle)$ of the link weights in $\mathcal{G}$; the fourth column the distribution $P(D/\langle D \rangle)$ of the hyperdegree in $\mathcal{H}$.}
\label{fig:figure1}
\end{figure*}

\clearpage
\section{Results on empirical hypergraphs}
\label{sez:section2}
Here we focus on the impact of all the adaptive strategies on pairwise and higher-order contagion processes, considering the empirical datasets as substrates for the spreading: we report results on the efficacy of the adaptive behaviors and on their social cost, comparing the non-adaptive (NAD) case with the different strategies. 
In particular, in Supplementary Fig.s \ref{fig:figure3}, \ref{fig:figure7} we report the temporal evolution of the fraction of infected nodes $I(t)$ and of the average risk parameter $\langle \lambda_i(t) \rangle/\lambda_0$, and we focus on the endemic state, showing the epidemic prevalence $I_{\infty}$ and the average parameter $\langle \lambda_i^{\infty} \rangle/\lambda_0$ in the asymptotic steady state.
Supplementary Fig.s \ref{fig:figure4}, \ref{fig:figure8} show the temporal evolution of the distribution $\rho(\lambda_i(t)/\lambda_0)$ of the risk parameter. 
In Supplementary Fig.s \ref{fig:figure5}, \ref{fig:figure9} we focus on how different nodes behave in the asymptotic steady state and in the epidemic transient: we divide nodes in classes of hyperdegree $D$ and for each class we estimate the average time of first infection $\langle o_i \rangle_D$, the average time spent infected in the asymptotic state $\langle \tau_i^{\infty} \rangle_D$ and the average asymptotic risk parameter 
$\langle \lambda_i^{\infty} \rangle_D/\lambda_0$.
Supplementary Fig.s \ref{fig:figure6}, \ref{fig:figure10} focus on how different groups behave in the asymptotic steady state and in the epidemic transient: we divide groups in classes of size $m$ and for each class we estimate the average time of first infection $\langle o_e \rangle_m$, the average fraction of infected nodes within them in the asymptotic state $\langle \eta_e^{\infty} \rangle_m$ and the average asymptotic risk parameter within them $\langle \overline{\lambda}_e^{\infty} \rangle_m/\lambda_0$. 
Supplementary Fig. \ref{fig:figure2} shows the local impact of the adaptive behaviors when there are only a few infected individuals: we estimate it through the local reduction of the infection probability for links (in the pairwise contagion) or groups (in the higher-order contagion). \\

The different datasets exhibit the same phenomenology described in the main text (where we focused only on the hospital dataset and on pairwise contagion), producing the same hierarchy in the adaptive strategies efficacy and social cost. Furthermore, the same adaptive mechanisms and the same phenomenology are active for both pairwise and higher-order contagion. Note that the amplitude of the differences between the various strategies, and the level of agreement of the IBMF approach with the numerical results, can vary for different datasets: this is due to the specific topology of the empirical hypergraphs, including the hyperdegree heterogeneity and hyperedge overlap (as we will see in the next Sections) and the presence of meso-structures increasing the relevance of dynamical correlations between neighbouring nodes, which are neglected by the IBMF approach.

These results show the robustness of our findings, which hold even for higher-order contagion processes and even when considering empirical systems very different between each other, in terms of settings and statistical properties (see Section \ref{sez:section1}).\\

In Supplementary Fig. \ref{fig:figureSM03} we report an alternative estimation of the social cost of adaptive behaviors. The average $\langle \lambda_i^{\infty}\rangle/\lambda_0$ considered in the main text estimates indeed the average behavior change of nodes, hence considering the social cost with an individual-centered point of view. Here, instead, we estimate the social cost as the reduction $\langle \lambda_i^{\infty} \lambda_j^{\infty} \rangle_{i,j}/\lambda_0^2$, i.e. as the change in the social intensity of the connection $i,j$ averaged over all the connections $i,j$, thus having a connection-centered point of view of the social cost. This definition, by focusing on links, takes into account that changes in the behavior of high-degree nodes can have a greater impact on the social activity of the population than those of low-degree nodes, due to their higher participation in the society activity. 
The results show that the two definitions produce the same qualitative phenomenology: the hierarchy in the social cost of the adaptive strategies is the same, with the higher-order and absolute strategies maintaining a cost that is lower (or equal) to the other strategies. Moreover, considering the connection-centered definition, the differences in the social costs of the adaptive strategies are slightly reduced compared to the individual-centered definition. \\

Finally, in Supplementary Fig.s \ref{fig:figureSM01}, \ref{fig:figureSM02} we consider the hospital dataset (LH10, shown in the main text) and we report the features describing the behavior of single nodes and groups (shown in Supplementary Fig.s \ref{fig:figure5}, \ref{fig:figure6}, \ref{fig:figure9}, \ref{fig:figure10}) at specific times $t$ in the epidemic transient. These results show how adaptive mechanisms and differences between the various strategies emerge already in the early stages of the epidemic, with higher-order and absolute strategies already triggering heterogeneous levels of alert.

\clearpage
\begin{figure*}[ht!]
\includegraphics[width=\textwidth]{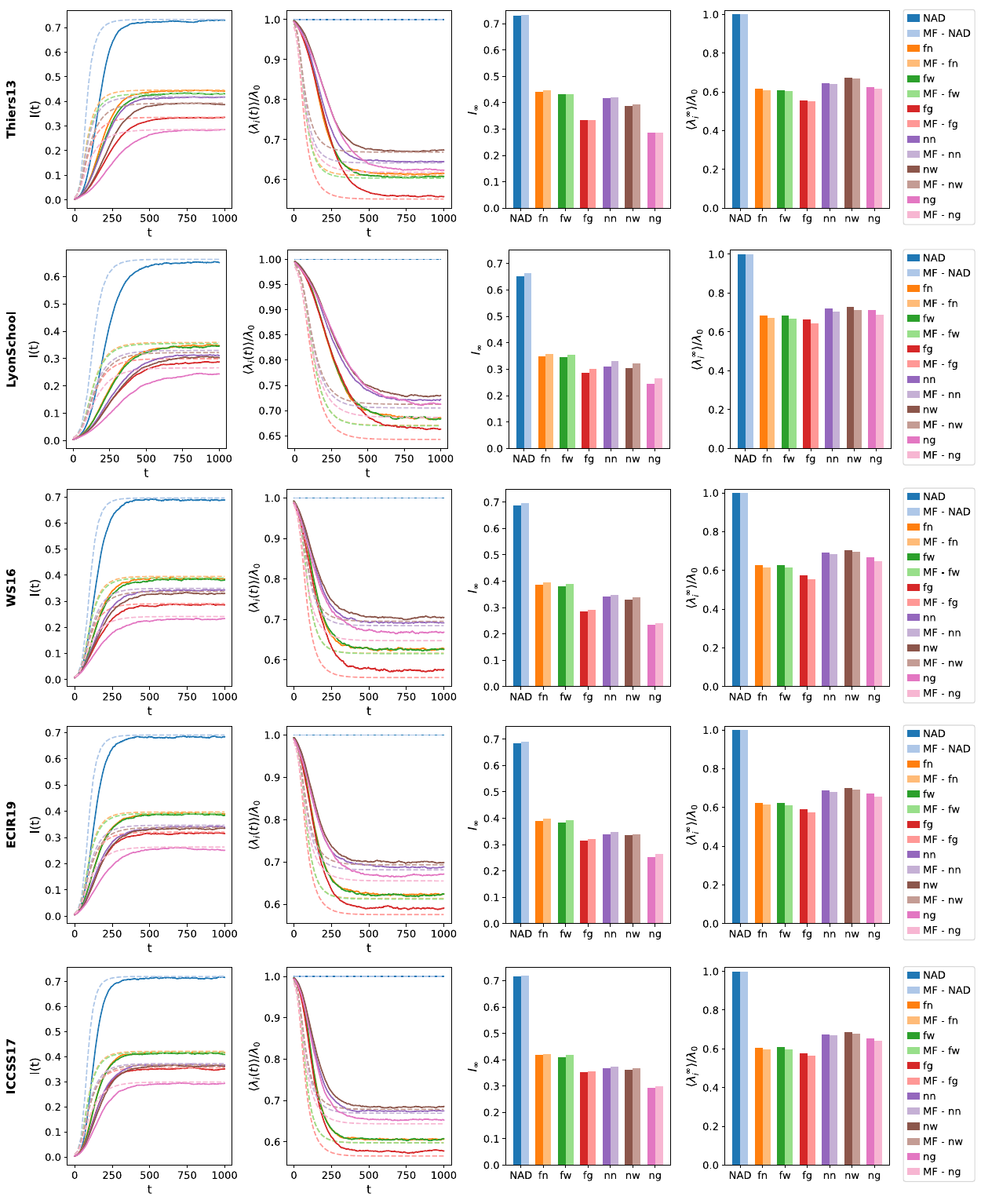}
\caption{\textbf{Containment efficacy and social cost - Empirical - Pairwise contagion.} Each row corresponds to a different dataset (see labels): for each of them, (i) the first column shows the temporal evolution of the fraction $I(t)$ of infected nodes; (ii) the second column shows the temporal evolution of the mean reduction in the risk parameter $\langle \lambda_i(t) \rangle/\lambda_0$. Solid lines are obtained by averaging over 300 numerical simulations; dashed lines are obtained through numerical integration of the mean-field equations. In the third and fourth columns we show (through a bar-plot) respectively (iii) the epidemic prevalence $I_{\infty}$ and (iv) the average  $\langle \lambda_i^{\infty} \rangle /\lambda_0$ in the asymptotic steady state, for the NAD case and for all the adaptive strategies. The dark-colored bars correspond to the average of 300 numerical simulations while light-colored bars are obtained through numerical integration of the mean-field equations (see legend). We consider the pairwise contagion with $\theta=0.3$, with $r=0.05$ for the Thiers13 dataset and $r=0.2$ for all the others. In all panels we consider the NAD case and the six adaptive strategies.}
\label{fig:figure3}
\end{figure*}

\clearpage
\begin{figure*}[ht!]
\includegraphics[width=\textwidth]{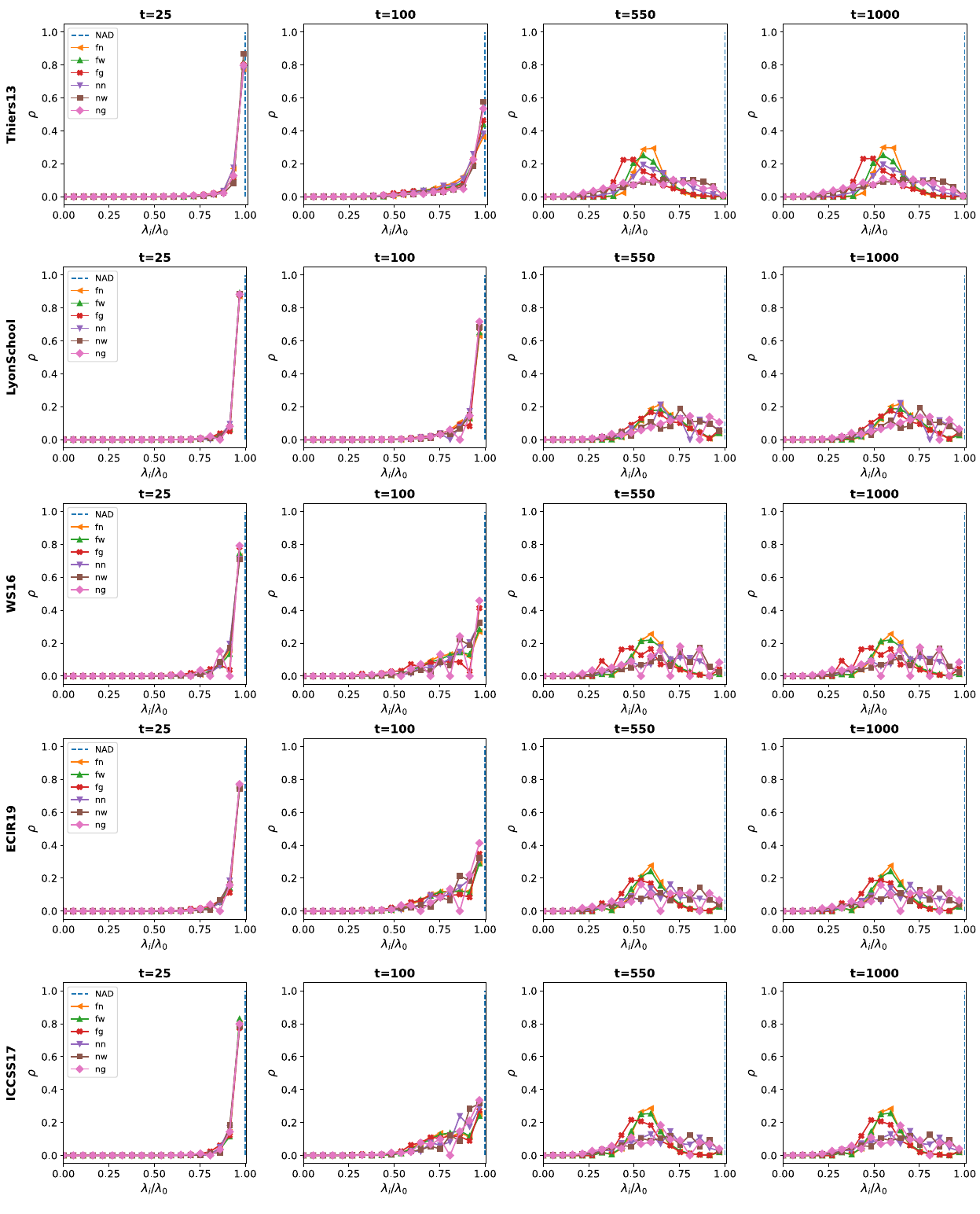}
\caption{\textbf{Evolution of risk parameter distribution - Empirical - Pairwise contagion.} Each row corresponds to a different dataset (see labels): for each of them, we show the distribution of the parameter $\rho(\lambda_i(t)/\lambda_0)$ at different times $t$, obtained by averaging over 300 numerical simulations. We consider the pairwise contagion with $\theta=0.3$, with $r=0.05$ for the Thiers13 dataset and $r=0.2$ for all the others. In all panels we consider the NAD case and the six adaptive strategies.}
\label{fig:figure4}
\end{figure*}

\clearpage
\begin{figure*}[ht!]
\includegraphics[width=0.9\textwidth]{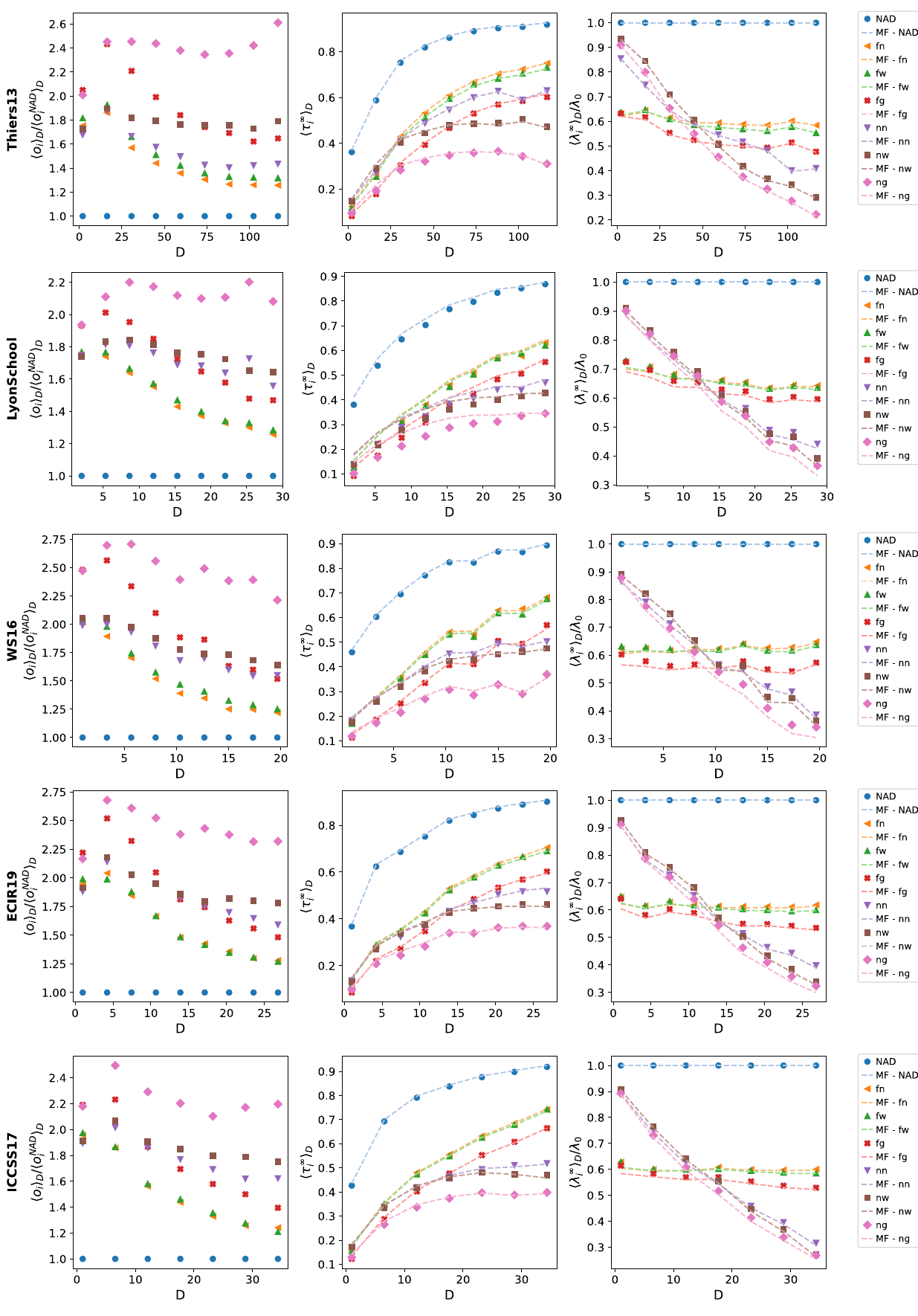}
\caption{\textbf{Behavior of single nodes - Empirical - Pairwise contagion.} Each row corresponds to a different dataset (see labels). For each of them, we divide nodes into hyperdegree classes $D$ and for each class we estimate: (i) the average time of first infection $\langle o_i \rangle_D/\langle o_i \rangle_D^{NAD}$ compared to the NAD case; (ii) the average fraction of time nodes spent infected $\langle \tau^{\infty}_i \rangle_D$ in the steady state; (iii) the average reduction in the risk parameter $\langle \lambda_i^{\infty} \rangle_D/\lambda_0$ in the steady state. All quantities are estimated for each node by averaging over 300 numerical simulations (markers) or by integrating numerically the mean-field equations (dashed curves). We consider the pairwise contagion with $\theta=0.3$, with $r=0.05$ for the Thiers13 dataset and $r=0.2$ for all the others. In all panels we consider the NAD case and the six adaptive strategies.}
\label{fig:figure5}
\end{figure*}

\clearpage
\begin{figure*}[ht!]
\includegraphics[width=0.9\textwidth]{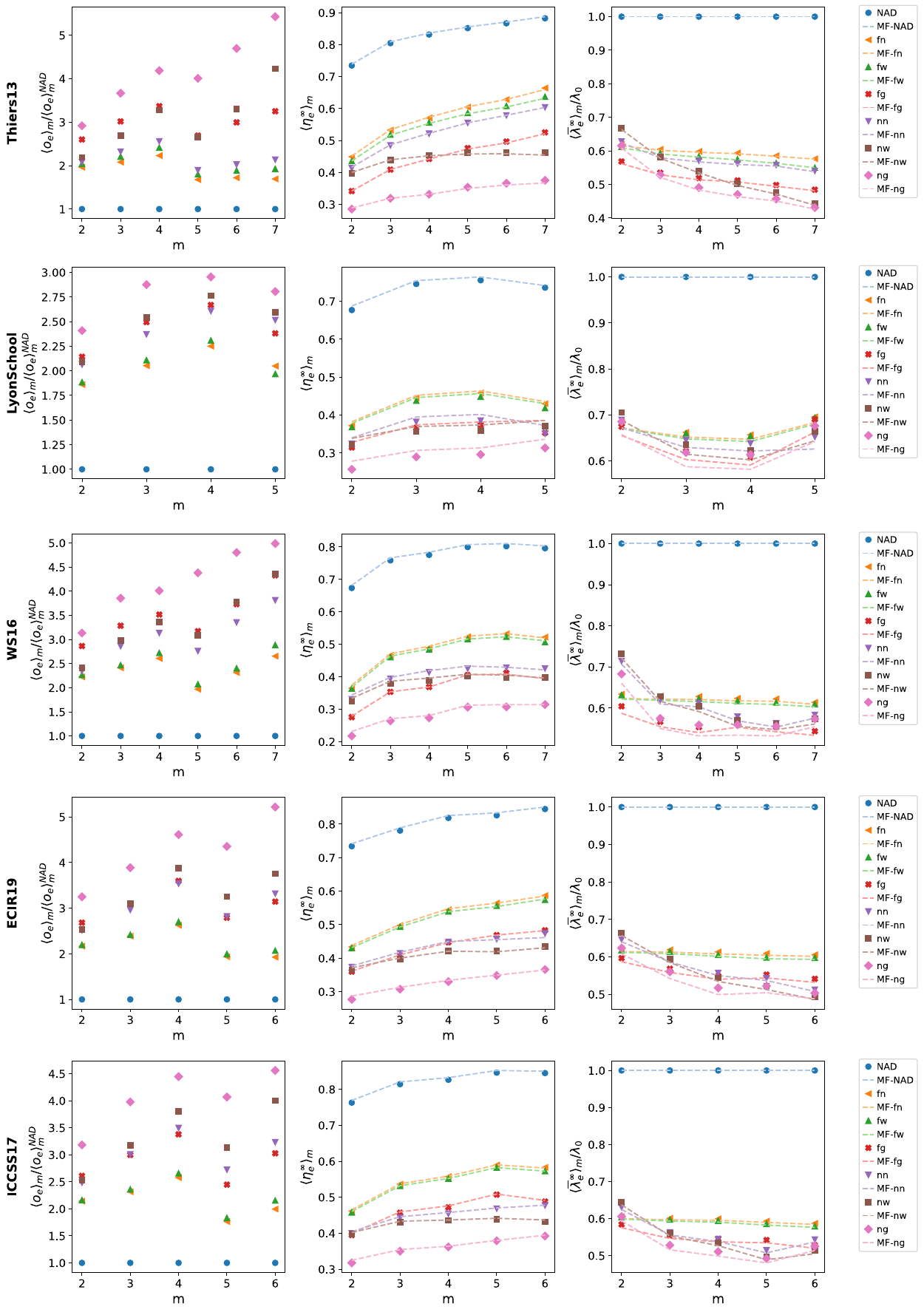}
\caption{\textbf{Behavior of single groups - Empirical - Pairwise contagion.} Each row corresponds to a different dataset (see labels). For each of them, for all hyperedge sizes $m$ we estimate: (i) the average time of first infection $\langle o_e \rangle_m/\langle o_e \rangle_m^{NAD}$ compared to the NAD case; (ii) the average fraction of infected nodes within a group in the steady state $\langle \eta^{\infty}_e \rangle_m$; (iii) the mean reduction of the risk parameter within a group $\langle \overline{\lambda}_e^{\infty} \rangle_m/\lambda_0$ in the steady state. All quantities are estimated for each group by averaging over 300 numerical simulations (markers) or by integrating numerically the mean-field equations (dashed curves). We consider the pairwise contagion with $\theta=0.3$, with $r=0.05$ for the Thiers13 dataset and $r=0.2$ for all the others. In all panels we consider the NAD case and the six adaptive strategies.}
\label{fig:figure6}
\end{figure*}

\clearpage
\begin{figure*}[ht!]
\includegraphics[width=0.95\textwidth]{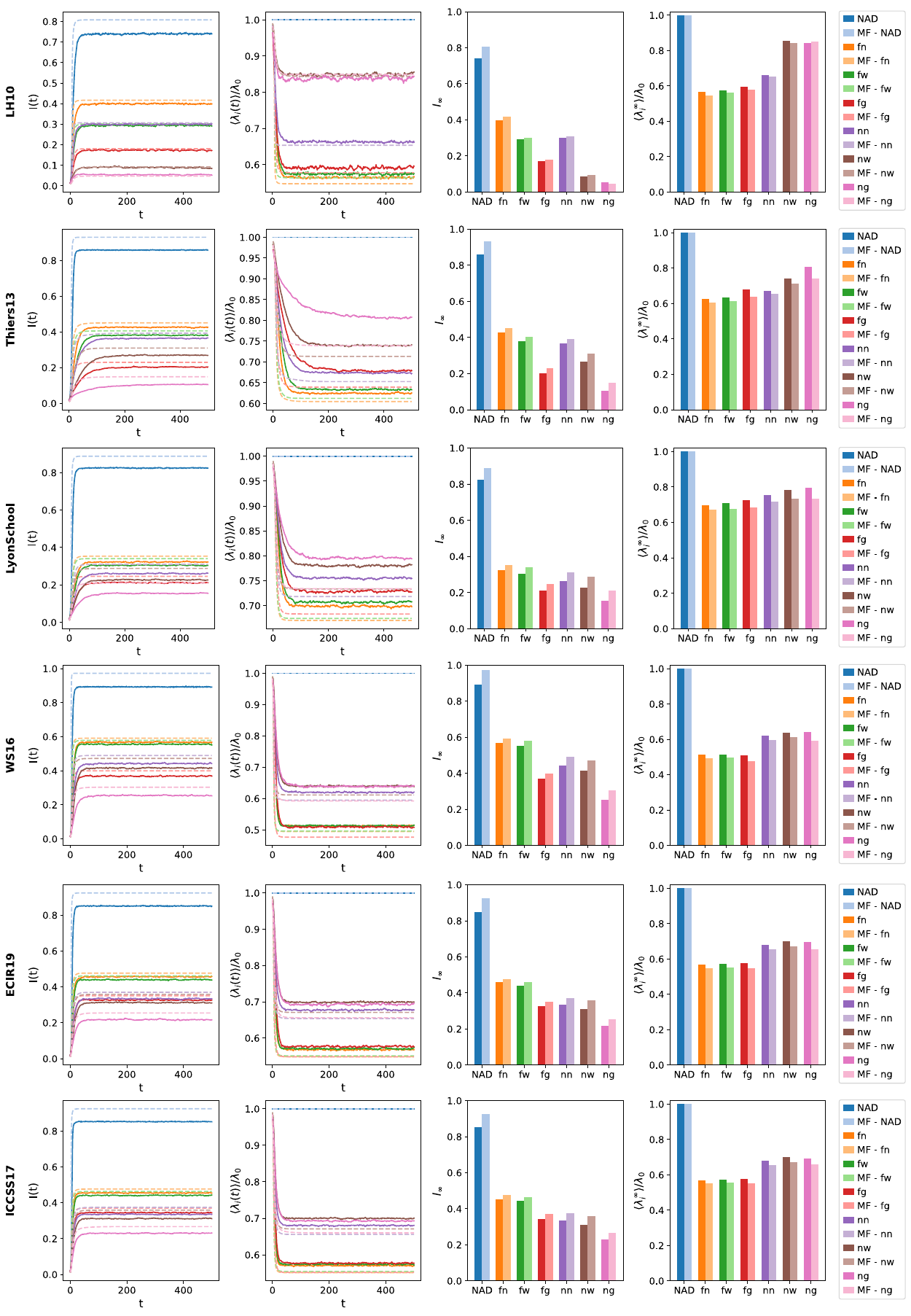}
\caption{\textbf{Containment efficacy and social cost - Empirical - Higher-order contagion.} Analogous to Supplementary Fig. \ref{fig:figure3}, but here we consider the higher-order contagion with $\theta=0.3$, $r=0.01$ for the LH10 dataset, $r=0.02$ for the Thiers13 dataset and $r=0.2$ for all the others. In all panels we consider the NAD case and the six adaptive strategies.}
\label{fig:figure7}
\end{figure*}

\clearpage
\begin{figure*}[ht!]
\includegraphics[width=\textwidth]{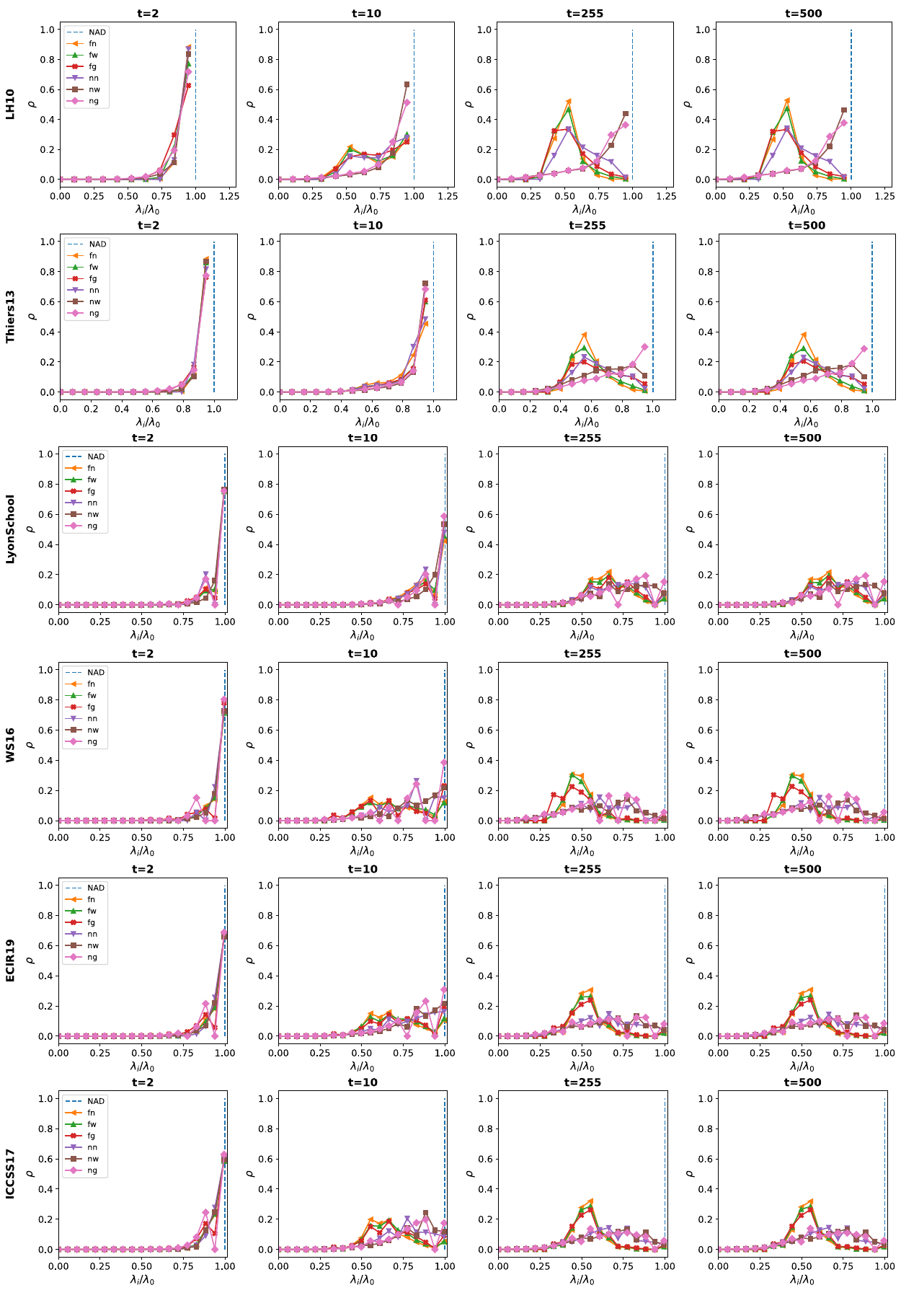}
\caption{\textbf{Evolution of risk parameter distribution - Empirical - Higher-order contagion.} Analogous to Supplementary Fig. \ref{fig:figure4}, but here we consider the higher-order contagion with $\theta=0.3$, $r=0.01$ for the LH10 dataset, $r=0.02$ for the Thiers13 dataset and $r=0.2$ for all the others. In all panels we consider the NAD case and the six adaptive strategies.}
\label{fig:figure8}
\end{figure*}

\clearpage
\begin{figure*}[ht!]
\includegraphics[width=0.8\textwidth]{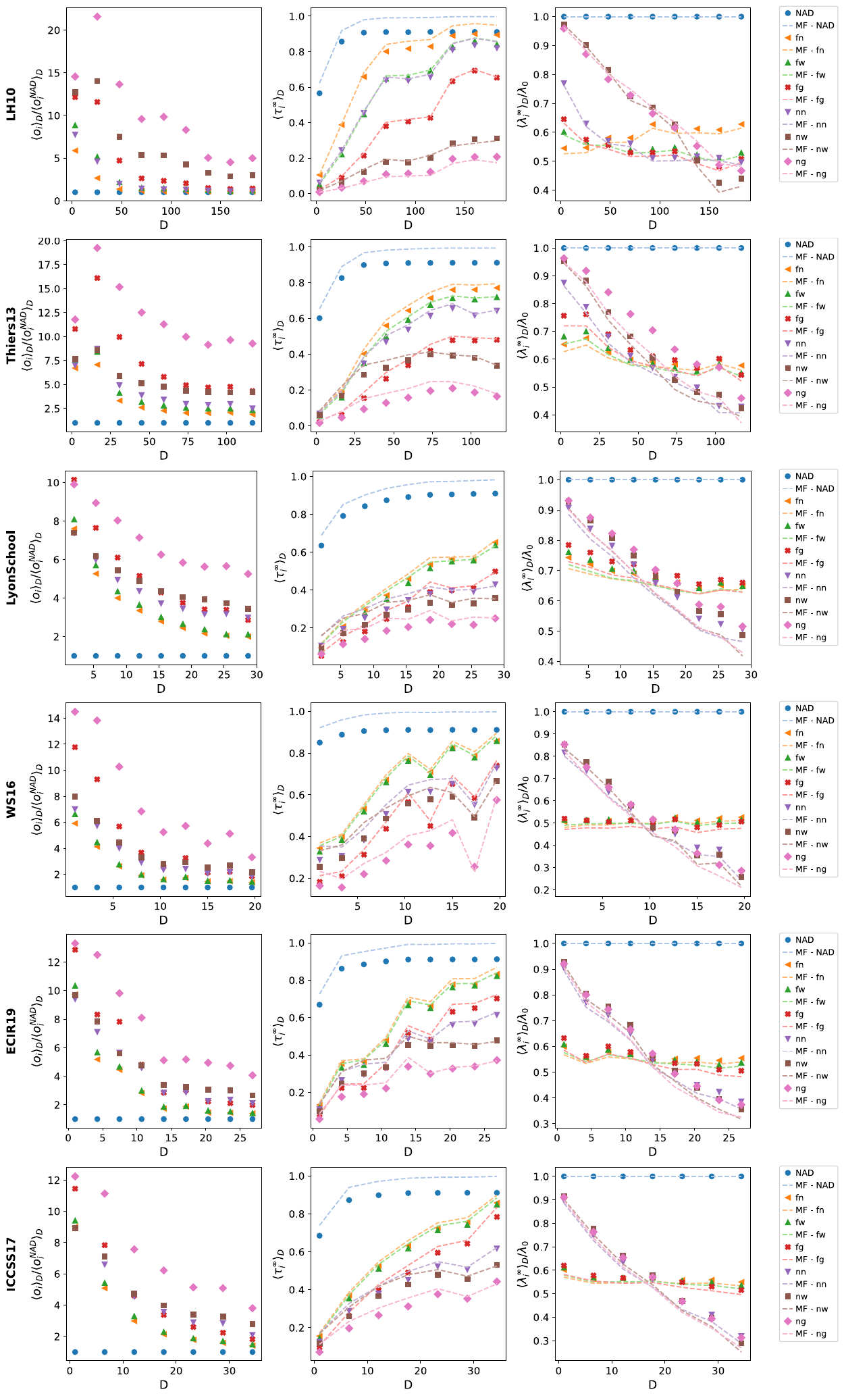}
\caption{\textbf{Behavior of single nodes - Empirical - Higher-order contagion.} Analogous to Supplementary Fig. \ref{fig:figure5}, but here we consider the higher-order contagion with $\theta=0.3$, $r=0.01$ for the LH10 dataset, $r=0.02$ for the Thiers13 dataset and $r=0.2$ for all the others. In all panels we consider the NAD case and the six adaptive strategies.}
\label{fig:figure9}
\end{figure*}

\clearpage
\begin{figure*}[ht!]
\includegraphics[width=0.8\textwidth]{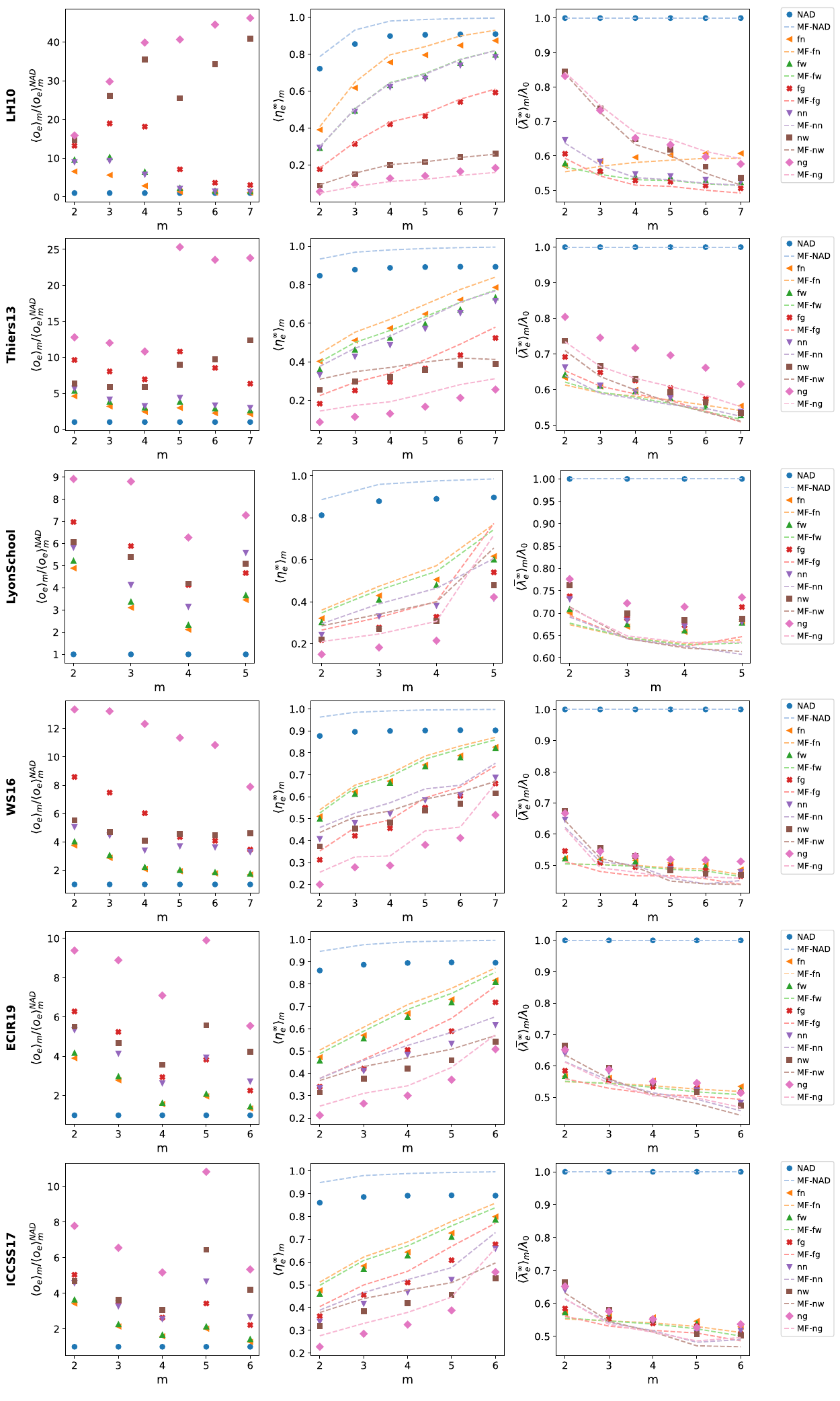}
\caption{\textbf{Behavior of single groups - Empirical - Higher-order contagion.} Analogous to Supplementary Fig. \ref{fig:figure6}, but here we consider the higher-order contagion with $\theta=0.3$, $r=0.01$ for the LH10 dataset, $r=0.02$ for the Thiers13 dataset and $r=0.2$ for all the others. In all panels we consider the NAD case and the six adaptive strategies.}
\label{fig:figure10}
\end{figure*}

\clearpage
\begin{figure*}[ht!]
\includegraphics[width=0.85\textwidth]{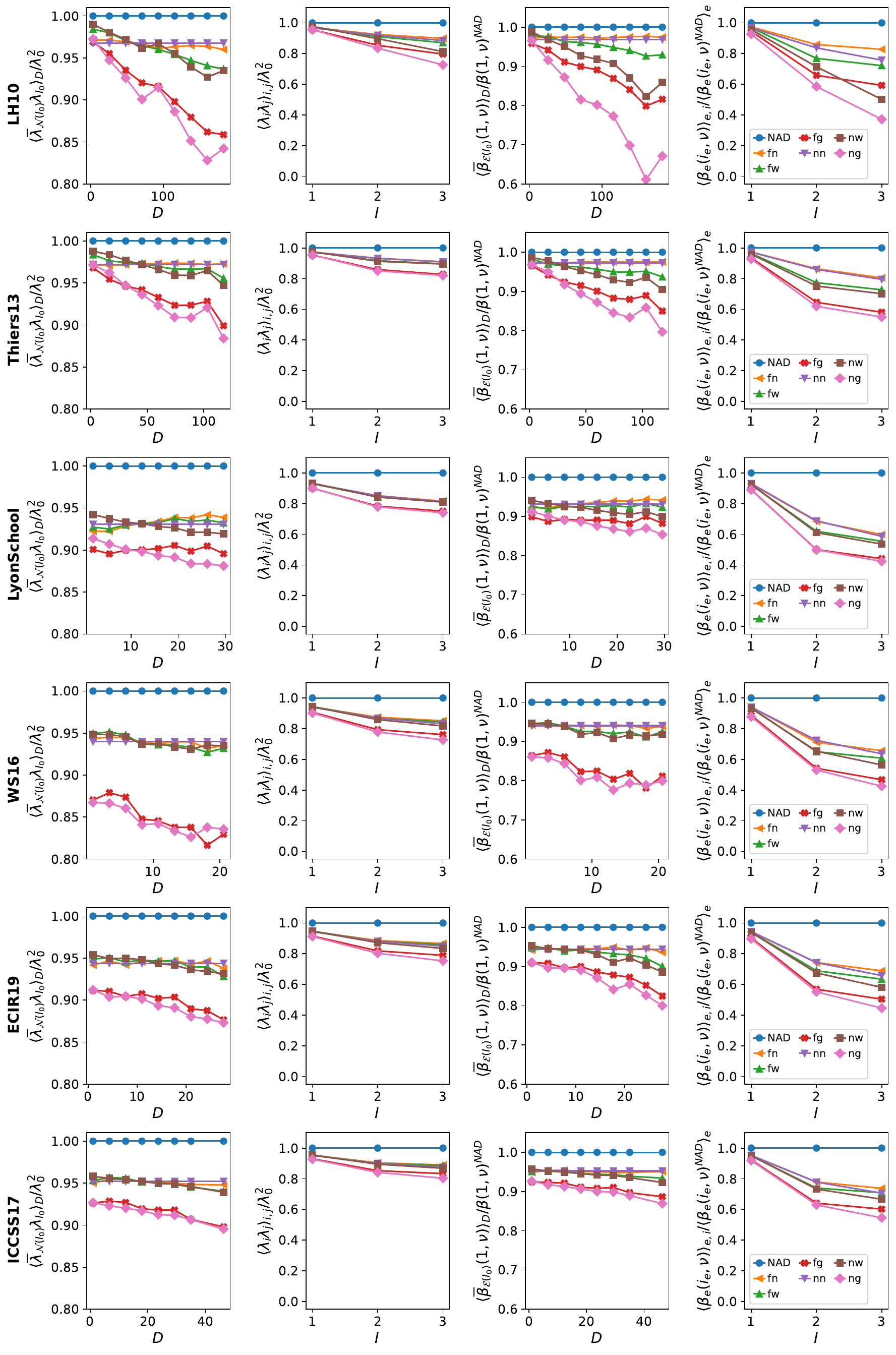}
\caption{\textbf{Local effects of adaptive behaviors.} Each row corresponds to a different dataset (see labels). For each of them the first two columns focus on the pairwise contagion, while the third and fourth columns focus on the higher-order contagion. We divide nodes into hyperdegree classes $D$ and for each class we estimate: (i) the average reduction in the pairwise infection probability $\langle \overline{\lambda}_{\mathcal{N}(I_0)} \lambda_{I_0} \rangle_D/\lambda_0^2$ (first column) and (iii) the average reduction in the higher-order contagion rate $\langle \overline{\beta}_{\mathcal{E}(I_0)}(1,\nu) \rangle_D/\beta(1,\nu)^{NAD}$ (third column), both evaluated in the neighborhood of a node $I_0$, when it is the only infected node in the population and it belongs to the class $D$. In the second and fourth columns, we show respectively (ii) the average reduction in contagion rates $\langle \lambda_i \lambda_j \rangle_{i,j}/\lambda_0^2$ and (iv) $\langle \beta_e(i_e,\nu) \rangle_{e,i}/\beta_e(i_e,\nu)^{NAD}$, in the neighbourhood of $I=[1,2,3]$ infected nodes (averaged over all the configuration of $I$ infected nodes and on their neighbourhood, assuming that if there are two or three infected they are connected). In all panels we consider all the adaptive strategies, $\nu=4$ and $\theta=0.3$ for the $ng$ and $fg$ strategies.}
\label{fig:figure2}
\end{figure*}

\clearpage
\begin{figure*}[ht!]
\includegraphics[width=\textwidth]{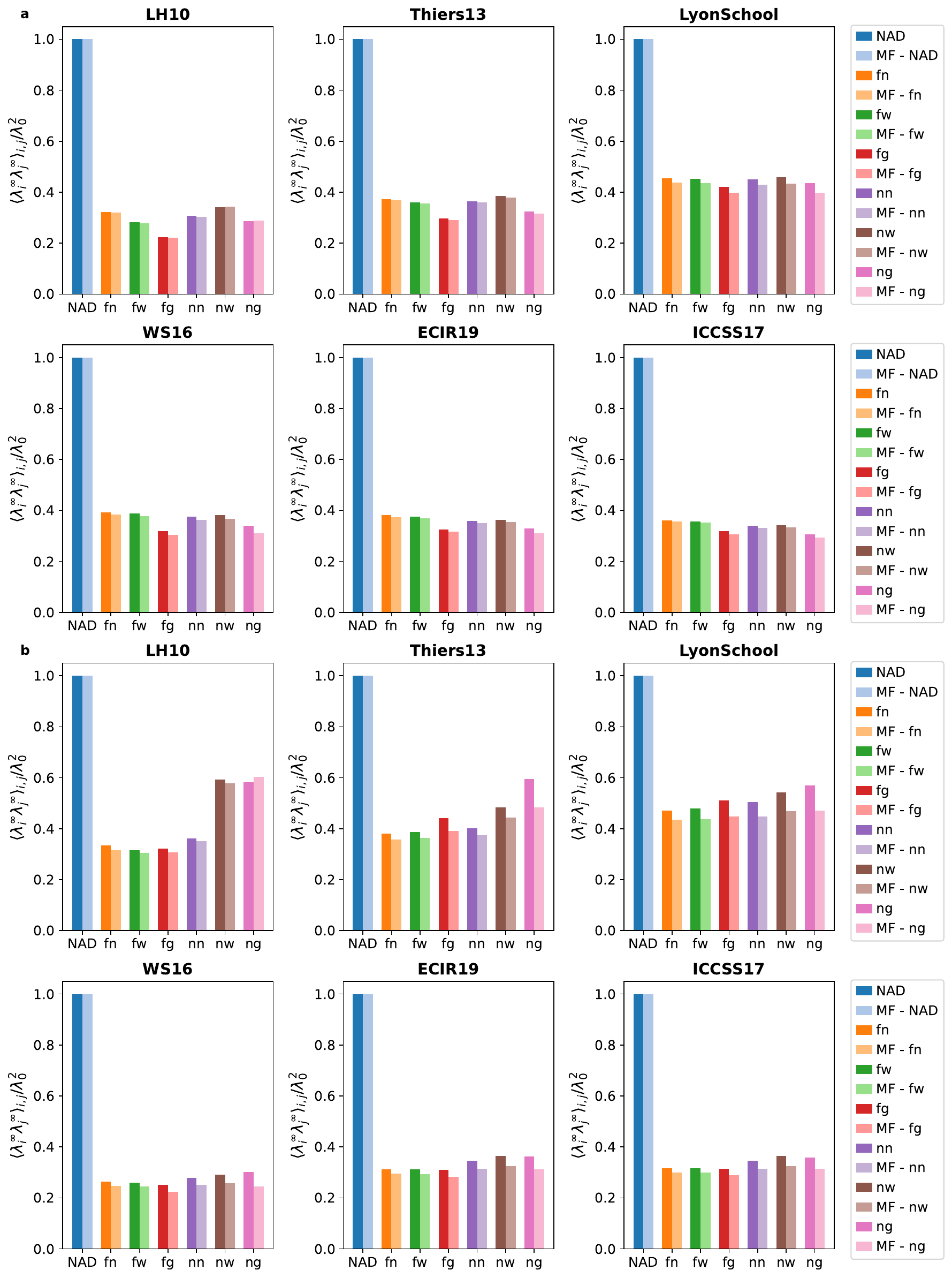}
\caption{\textbf{Alternative social cost evaluation - Empirical.} For each dataset (see labels) we show (through a bar-plot) the average reduction $\langle \lambda_i^{\infty} \lambda_j^{\infty} \rangle_{i,j} /\lambda_0^2$ of links in the asymptotic steady state of the pairwise contagion process (panel \textbf{a}) and of the higher-order contagion process (panel \textbf{b}), for the NAD case and for all the adaptive strategies. The dark-colored bars correspond to the average of 300 numerical simulations while light-colored bars are obtained through numerical integration of the mean-field equations (see legend). We consider $\theta=0.3$, the pairwise contagion with $r=0.05$ for the Thiers13 and LH10 datasets and $r=0.2$ for all the others, the higher-order contagion with $r=0.01$ for the LH10 dataset, $r=0.02$ for the Thiers13 dataset and $r=0.2$ for all the others.}
\label{fig:figureSM03}
\end{figure*}

\clearpage
\begin{figure*}[ht!]
\includegraphics[width=\textwidth]{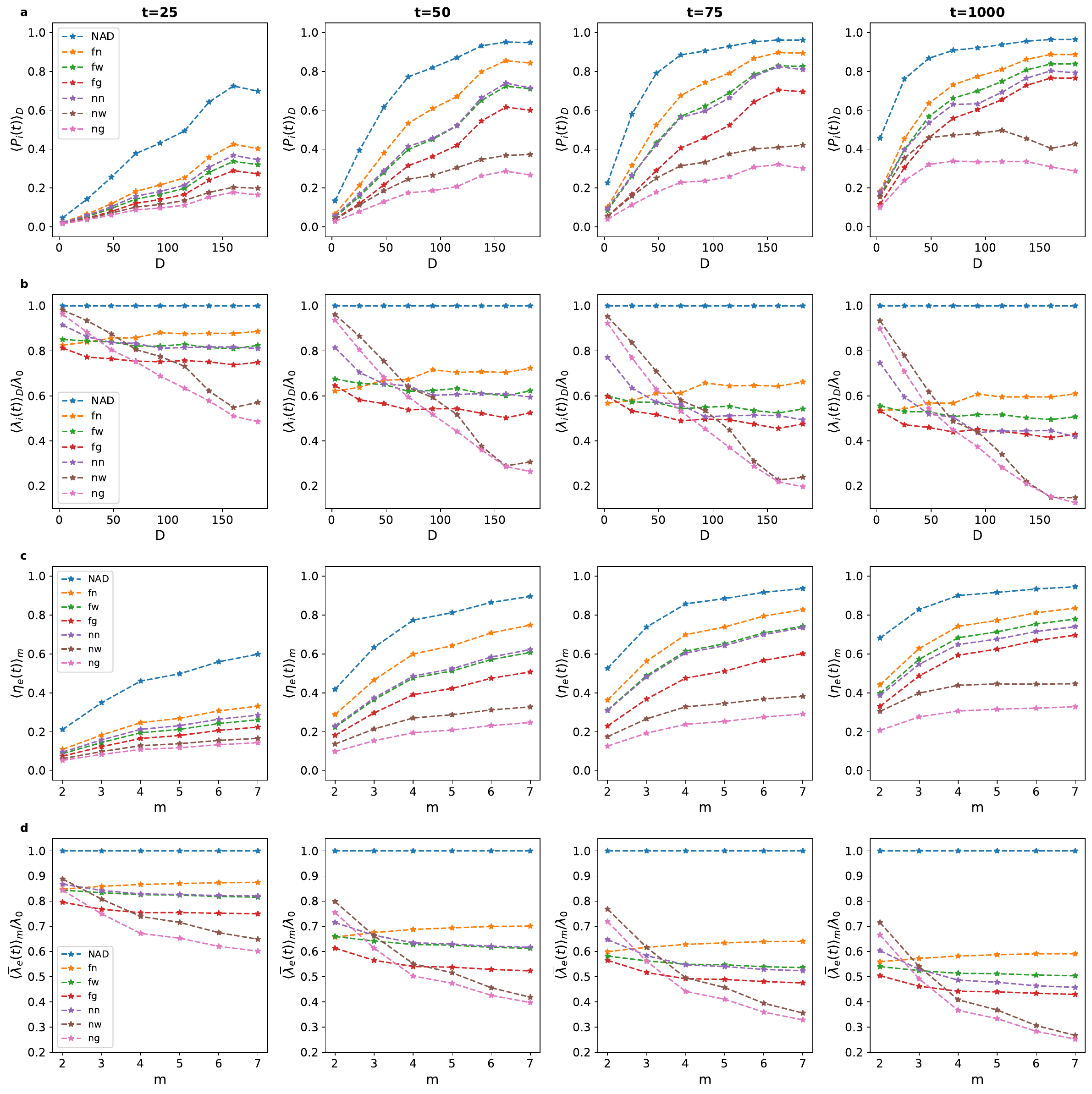}
\caption{\textbf{Temporal behavior of nodes and groups - LH10 - Pairwise contagion.} We consider the pairwise contagion process on the hospital dataset, $r=0.05$, $\theta=0.3$. We divide nodes into hyperdegree classes $D$ and for each class we estimate at each time $t$: \textbf{a}: the average probability that a node is infected $\langle P_i(t) \rangle_D$; \textbf{b}: the mean reduction in the risk parameter $\langle \lambda_i(t) \rangle_D/\lambda_0$. For each hyperedge size $m$ we estimate at each time $t$: \textbf{c}: the average fraction of infected nodes in a group $\langle \eta_e(t) \rangle_m$; \textbf{d}: the mean reduction in the risk parameter within a group $\langle \overline{\lambda}_e(t) \rangle_m/\lambda_0$. In all panels, we consider the NAD case and the six adaptive strategies and we show the results of numerical integrations of the mean-field equations.}
\label{fig:figureSM01}
\end{figure*}

\clearpage
\begin{figure*}[ht!]
\includegraphics[width=\textwidth]{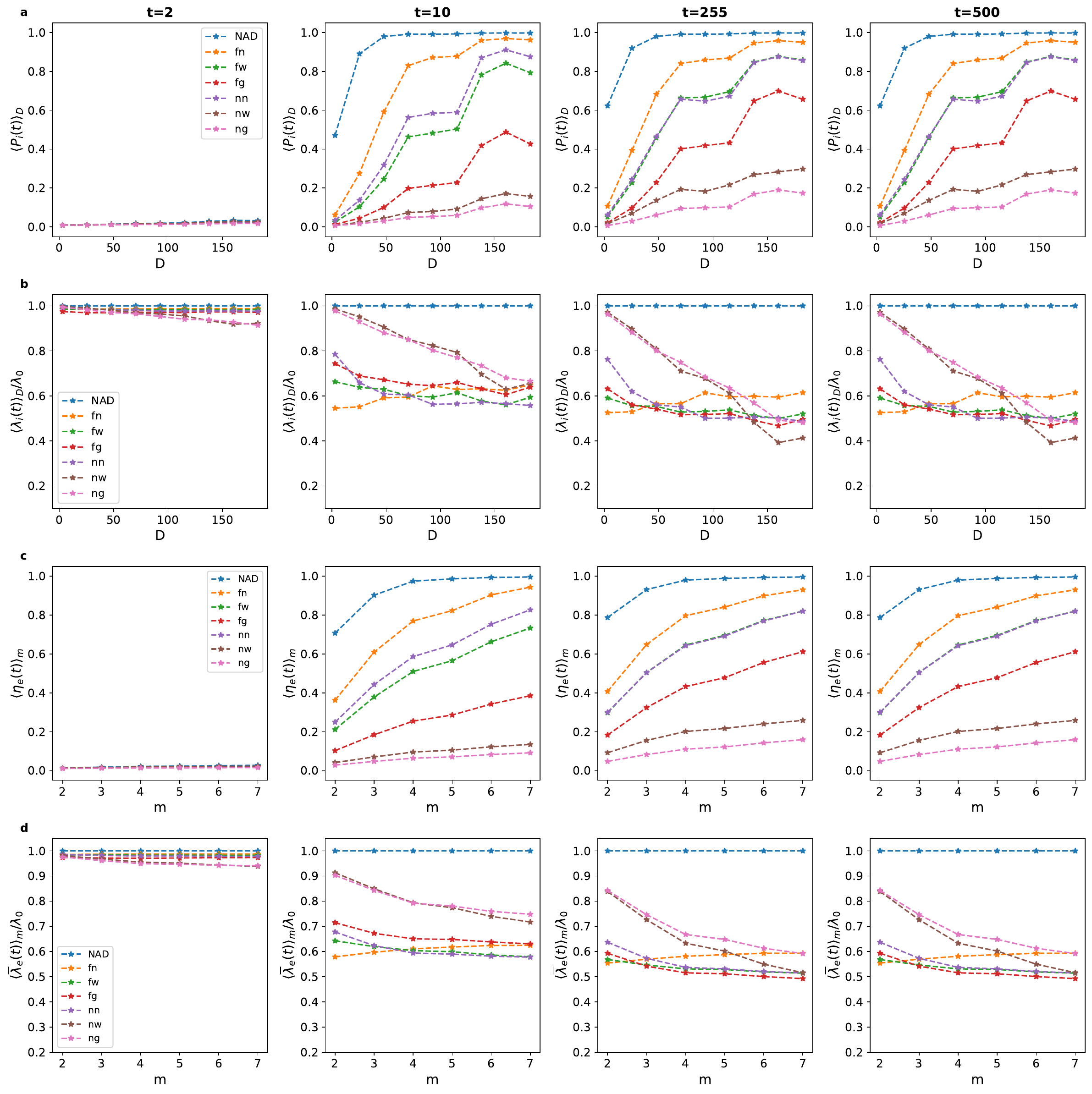}
\caption{\textbf{Temporal behavior of nodes and groups - LH10 - Higher-order contagion.} Analogous to Supplementary Fig. \ref{fig:figureSM01}, but here we consider the higher-order contagion with $\theta=0.3$, $r=0.01$. In all panels we consider the NAD case and the six adaptive strategies.}
\label{fig:figureSM02}
\end{figure*}

\clearpage
\section{Synthetic hypergraphs}
\label{sez:section3}
We consider three randomization procedures to build synthetic hypergraphs starting from an empirical one $\mathcal{H}$:
\begin{itemize}
    \item \textbf{G}: we construct a hypergraph $\mathcal{H}'$ by projecting each hyperedge $e \in \mathcal{H}$ onto its clique (fully connected) and without considering link weights. This procedure generates an unweighted pairwise graph, with the same number of nodes $N$ and same degree distribution $P(k)$ as the original hypergraph, but destroys any higher-order properties of $\mathcal{H}$, including the hyperedge size distribution $\Psi(m)$;
    \item \textbf{ER}: we build an Erdős–Rényi-like randomization of the empirical hypergraph, $\mathcal{H}'$ \cite{agostinelli2025higher}. The generated hypergraph has the same number $N$ of nodes and $E_m$ of hyperedges of size $m$ 
    as the empirical hypergraph $\mathcal{H}$: each hyperedge in $\mathcal{H}'$ is created by selecting its nodes uniformly at random within the population. This randomization preserves the number of nodes and the number of hyperedges of each size (therefore also the hyperedges size distribution $\Psi(m)$), but destroys the distribution of hyperdegrees $P(D)$ (overall and for each size) and any macro- and meso-structure (e.g. communities, hypercores).
    \item \textbf{k}: we construct a randomization of the empirical hypergraph, $\mathcal{H}'$, in the same spirit of the configuration model for graphs \cite{Mancastroppa2023,agostinelli2025higher}. Specifically, $\mathcal{H}'$ has the same number of nodes $N$ and hyperedges $E_m$ for each size $m$ as the original hypergraph $\mathcal{H}$: each hyperedge in $\mathcal{H}'$ is created by selecting nodes with probability proportional to their hyperdegree $D$ in the original hypergraph $\mathcal{H}$. This randomization preserves the number of nodes, the number of hyperedges of each size (therefore also the hyperedges size distribution $\Psi(m)$), and the hyperdegree distribution $P(D)$, but randomizes how different nodes participate in the different sizes (hyperdegrees by order) and any macro- and meso-structure (e.g. communities, hypercores).
\end{itemize}

We also consider two generation procedures, based on imposing specific properties for the synthetic hypergraphs, instead of randomizing interactions in the empirical datasets. We generate a hypergraph $\mathcal{H}'$ with $N$ nodes and $E$ hyperedges, whose sizes are drawn from the distribution of hyperedge sizes $\Psi(m)$ (hence $N$, $E$ and $\Psi(m)$ are imposed initially): for simplicity and to allow comparison with the other synthetic hypergraphs, we fix the number of nodes $N$ and the number of hyperedges $E_m$ of each size $m$ equal to those of an empirical hypergraph $\mathcal{H}$, thus reproducing its size distribution $\Psi(m)$. The two generation procedures work as follow:
\begin{itemize}
    \item \textbf{PD}: this model generates a hypergraph $\mathcal{H}'$ with a specific hyperdegree distribution $P(D)$. Given the hyperdegree distribution $P(D)$ desired, we assign a parameter $x$ to each node, extracted from a distribution $P(x)$ with the same functional form as $P(D)$; then we generate the hyperedges of $\mathcal{H}'$ by selecting nodes with a probability proportional to $x$. We consider three cases: a Dirac Delta $P(x) \sim \delta(x-1)$, for which all nodes have the same probability of participating in hyperedges; an exponential $P(x) \sim e^{-x/\beta}$ for various values of $\beta$ (see Supplementary Table \ref{tab:table2}); a power-law $P(x) \sim x^{\alpha-1}$ (with lower cut-off at $10^{-3}$ and upper cut-off at $1$) for various values of $\alpha$ (see Supplementary Table \ref{tab:table2}). The hypergraph $\mathcal{H}'$ has a hyperdegree distribution $P(D)$ with approximately the same shape as $P(x)$, whose average, minimum and maximum values depend on the number of hyperedges $E$ and of nodes $N$ chosen.
    \item \textbf{OV}: generates a hypergraph $\mathcal{H}'$ with a tunable level of hyperedges overlap, measured as $\langle w_l \rangle$, and with a homogeneous distribution of hyperdegree $P(D)$. We initially divide the population randomly into $n$ communities of approximately the same size. Each hyperedge $e$ of size $m$ (drawn from $\Psi(m)$) in $\mathcal{H}'$ is created as follows: (i) the first node in $e$ is selected uniformly at random in the whole population; (ii) the other $m-1$ nodes are selected with a probability proportional to $p$ if they belong to the same community, and with a probability proportional to $1-p$ if they belong to a different community. The value of $p$ tunes the overlap level: $p=0.5$ corresponds to a Erdős–Rényi-like hypergraph, since the nodes participating in an hyperedge are selected with the same probability regardless of their community; by increasing $p$ the overlap level is progressively increased, since interactions within the same community are favoured; $p \sim 1$ generates highly clustered hypergraphs with high overlap level.
\end{itemize}

We generate synthetic hypergraphs with these five methods starting from the hospital dataset (LH10). In the Supplementary Table \ref{tab:table2}, we report the main statistical properties of the synthetic hypergraphs generated (specifying the generation parameters if necessary). Moreover, in Supplementary Fig. \ref{fig:figure11} we show for selected synthetic hypergraphs: the hyperedge size distribution $\Psi(m)$, the distribution $P(k/\langle k \rangle)$ of the degree and $P(w_l/\langle w_l \rangle)$ of the link weights in $\mathcal{G}$, the distribution $P(D/\langle D \rangle)$ of the hyperdegree in $\mathcal{H}$.

\begin{table}[h!]
    \begin{tabular}{|c|c|c|c|c|c|c|c|}
    \hline
    \hline
           & $\langle m \rangle$ & $\langle D \rangle$ & $\langle D^2 \rangle/\langle D \rangle^2$ & $\langle k \rangle$ & $\langle k^2 \rangle/\langle k \rangle^2$ & $\langle w_l \rangle$ & $\langle w_l^2 \rangle/\langle w_l \rangle^2$ \\ \hline
         Data - LH10 & 3.4 & 50.0 & 2.0 & 30.4 & 1.3 & 4.6 & 3.4\\
         G & 2.0 & 30.4 & 1.3 & 30.4 & 1.3 & 1.0 & 1.0 \\
         ER & 3.4 & 50.0 & 1.0 & 62.9 & 1.0 & 2.2 & 1.3 \\
         k & 3.4 & 50.0 & 2.0 & 39.7 & 1.2 & 3.5 & 2.4 \\
         PD - Dirac delta & 3.4 & 50.0 & 1.0 & 63.4 & 1.0 & 2.2 & 1.3\\
         PD - Exp $\beta=0.5$ & 3.4 & 52.8 & 1.8 & 39.6 & 1.2 & 3.7 & 2.1 \\
         PD - Exp $\beta=1.5$ & 3.4 & 52.8 & 1.7 & 40.2 & 1.2 & 3.6 & 1.9 \\
         PD - Exp $\beta=3.5$ & 3.4 & 52.1 & 2.0 & 36.8 & 1.2 & 3.9 & 2.3\\
         PD - Power-law $\alpha=0.01$ & 3.4 & 52.8 & 2.8 & 27.4 & 1.4 & 5.4 & 3.3 \\
         PD - Power-law $\alpha=0.75$ & 3.4 & 53.5 & 1.4 & 43.3 & 1.2 & 3.4 & 1.6 \\
         PD - Power-law $\alpha=1.5$ & 3.4 & 52.0 & 1.2 & 54.5 & 1.0 & 2.7 & 1.4 \\
         PD - Power-law $\alpha=10$ & 3.4 & 52.0 & 1.0 & 61.7 & 1.0 & 2.3 & 1.3 \\
         OV - $n=7$, $p=0.5$ & 3.4  &  50.0  &  1.0  &  62.9  &  1.0  &  2.2  &  1.3 \\
         OV - $n=7$, $p=0.95$ & 3.4  &  50.0  &  1.0  &  47.4  &  1.0  &  2.9  &  2.2\\
         OV - $n=7$, $p=0.98$ & 3.4  &  50.0  &  1.0  &  34.1  &  1.0 &  4.1  &  2.4\\
    \hline
    \hline
    \end{tabular}
    \caption{\textbf{Properties of synthetic hypergraphs.} For each synthetic hypergraph generated we indicate: the parameters for its generation (if needed), the average hyperedge size $\langle m \rangle$, the average hyperdegree $\langle D \rangle$ and the heterogeneity of its distribution $\langle D^2 \rangle/\langle D \rangle^2$, the average degree $\langle k \rangle$ and the heterogeneity of its distribution $\langle k^2 \rangle/\langle k \rangle^2$, and the average weight $\langle w_l \rangle$ in the projected graph $\mathcal{G}$ and the heterogeneity of its distribution $\langle w_l^2 \rangle/\langle w_l \rangle^2$. Note that the statistics of the weights in $\mathcal{G}$ consider only existing links, so $\langle w_l \rangle \geq 1$. Moreover, for all the hypergraphs the numbers of nodes $N$ and of hyperedges $E_m$, for each size $m$, are as in the empirical LH10 dataset (see Supplementary Table \ref{tab:table1}).} 
    \label{tab:table2}
\end{table}

\clearpage
\begin{figure*}[ht!]
\includegraphics[width=\textwidth]{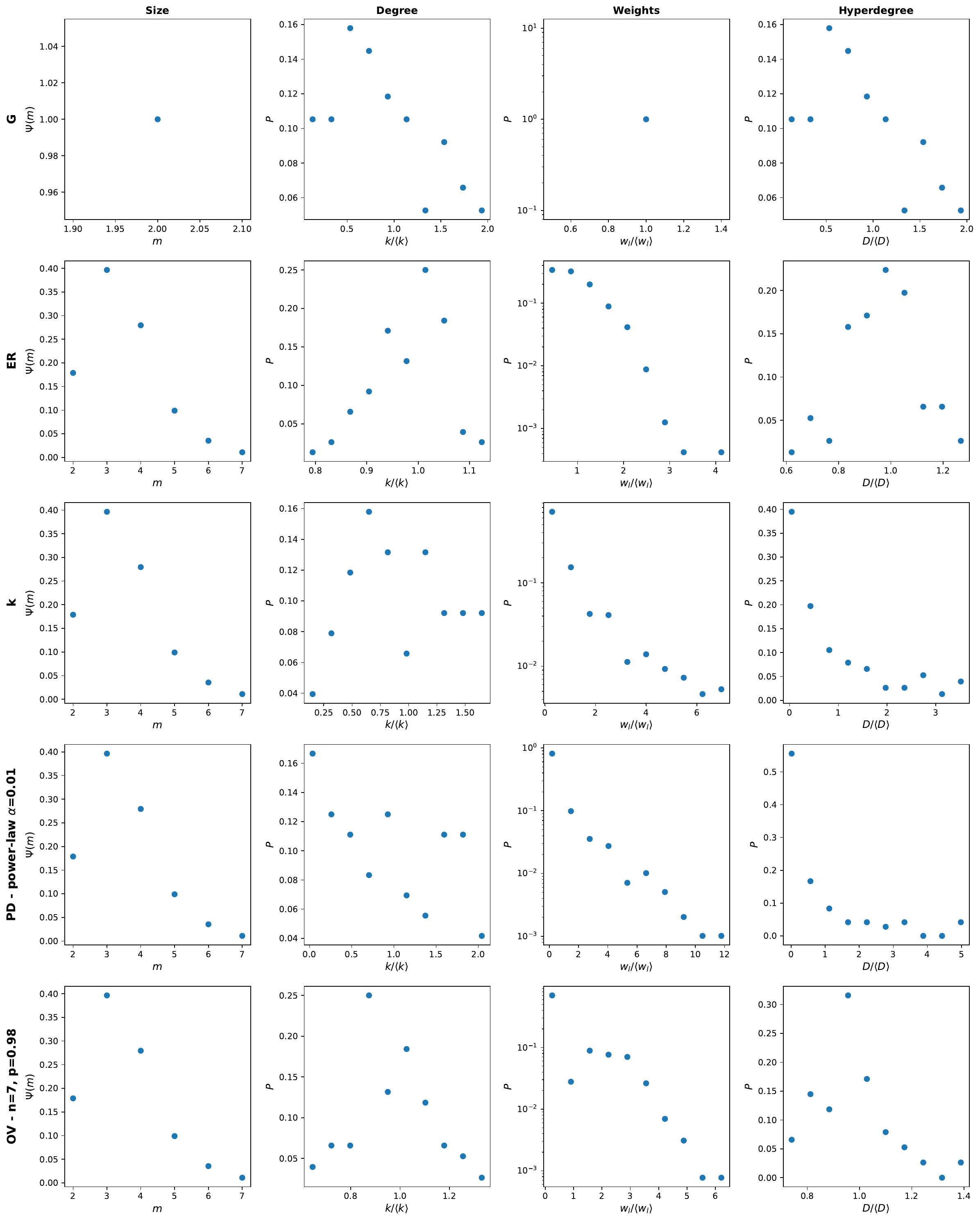}
\caption{\textbf{Properties of synthetic hypergraphs.} Each row corresponds to a different synthetic hypergraph (see labels): for each of them the first column show the hyperedge size distribution $\Psi(m)$; the second column the distribution $P(k/\langle k \rangle)$ of the degree in $\mathcal{G}$; the third column the distribution $P(w_l/\langle w_l \rangle)$ of the link weights in $\mathcal{G}$ (considering only non-zero $w_l$); the fourth column the distribution $P(D/\langle D \rangle)$ of the hyperdegree in $\mathcal{H}$.}
\label{fig:figure11}
\end{figure*}

\clearpage
\section{Results on synthetic hypergraphs}
\label{sez:section4}
Here we focus on the impact of all the adaptive strategies on pairwise and higher-order contagion processes, considering the generated synthetic hypergraphs as substrates for the spreading: we report results on the efficacy of the adaptive behaviors and on their social cost, comparing the NAD case with the different strategies. 
In Supplementary Fig.s \ref{fig:figure12}, \ref{fig:figure16} we report the temporal evolution of the fraction of infected nodes $I(t)$ and of the average risk parameter $\langle \lambda_i(t) \rangle/\lambda_0$, and we focus on the endemic state, showing the epidemic prevalence $I_{\infty}$ and the average  $\langle \lambda_i^{\infty} \rangle/\lambda_0$ in the asymptotic steady state.
Supplementary Fig.s \ref{fig:figure13}, \ref{fig:figure17} show the temporal evolution of the distribution $\rho(\lambda_i(t)/\lambda_0)$ of the risk parameter. 
In Supplementary Fig.s \ref{fig:figure14}, \ref{fig:figure18} we focus on how different nodes behave in the asymptotic steady state and in the epidemic transient: we divide nodes in classes of hyperdegree $D$ and for each class we estimate the average time of first infection $\langle o_i \rangle_D$, the average time spent infected in the asymptotic state $\langle \tau_i^{\infty} \rangle_D$ and the average asymptotic risk parameter $\langle \lambda_i^{\infty} \rangle_D/\lambda_0$.
Supplementary Fig.s \ref{fig:figure15}, \ref{fig:figure19} focus on how different groups behave in the asymptotic steady state and in the epidemic transient: we divide groups in classes of size $m$ and for each class we estimate the average time of first infection $\langle o_e \rangle_m$, the average fraction of infected nodes within them in the asymptotic state $\langle \eta_e^{\infty} \rangle_m$ and the average asymptotic risk parameter within them $\langle \overline{\lambda}_e^{\infty} \rangle_m/\lambda_0$.
In Supplementary Fig. \ref{fig:figureSM04} we report the social cost of the adaptive strategies with an alternative way to estimate it, as the reduction $\langle \lambda_i^{\infty} \lambda_j^{\infty} \rangle_{i,j}/\lambda_0^2$, i.e. as the change in the social intensity of the connection $i,j$, averaged over all the connections $i,j$ (see Section \ref{sez:section2}). \\

These results allow us to better understand the effect of the underlying higher-order structure on the performance of the different adaptive strategies (efficacy and cost), highlighting in particular the role played by hyperdegree heterogeneity and overlap.

If the underlying structure is a pairwise graph ($G$), higher-order strategies become equivalent to the pairwise ones, since there is no higher-order information. However, the differences between relative and absolute strategies are still present, with absolute strategies exploiting the graph's degree heterogeneity and producing better containment at a lower cost. In the ER-hypergraph, the phenomenology is opposite to the $G$-hypergraph: relative and absolute strategies are equivalent, since the network does not present topological heterogeneity, but differences remain between higher-order and pairwise strategies, due to the presence of group interactions and the ability of higher-order strategies to exploit them. Finally, in the k-hypergraph the impact of the different strategies is similar to the empirical case, including the hierarchy of strategy performance, since the generated hypergraph preserves many empirical higher-order topological properties. Note that all synthetic hypergraphs (including the k-hypergraph) do not reproduce the way in which strategies act on groups of different sizes: indeed, all the generated hypergraphs destroy the distributions of hyperdegree by order and the correlations in the behaviour of nodes at different orders.

Results on the generated synthetic hypergraphs show that the heterogeneity of the hyperdegree distribution plays an important role in both the pairwise and higher-order contagion processes, influencing the impact of the different adaptation strategies, while the overlap seems to have a more limited impact. Specifically, the more heterogeneous and with higher overlap level the underlying hypergraph is, the greater the differences between the various strategies are, especially between relative and absolute strategies (heterogeneity) and between higher-order and pairwise (heterogeneity and overlap). On the contrary, when this heterogeneity disappears, the strategies collapse and become equivalent.\\

To study this aspect in more detail, in Supplementary Fig.s \ref{fig:figure20}, \ref{fig:figure21} we show the efficacy of the different strategies and their social cost as a function of the hypergraph heterogeneity and overlap, considering the synthetic hypergraphs generated and both pairwise and higher-order contagion. Heterogeneity impacts differently the various strategies: pairwise and/or relative strategies exhibit almost a heterogeneity-independent efficacy, while absolute and higher-order strategies are significantly more effective in more heterogeneous systems. Hence, as the system's heterogeneity increases, differences between strategies are amplified. Similar results are observed for the social cost, with absolute and higher-order strategies having a cost that decreases significantly for increasing heterogeneity, while relative and/or pairwise strategies have a cost that increases slightly with heterogeneity. 
The impact of overlap is similar to that of heterogeneity, although its effect is smaller, since it is linked to heterogeneity. An increase in overlap increases the efficacy and reduces the social cost of absolute and higher-order strategies, while it has almost no effect on pairwise and/or relative strategies: this amplifies the differences between adaptive strategies.

Note that these comments concern the active phase of the epidemic: in \cite{mancastroppa2025_2} we studied in detail the impact of the underlying structure on the epidemic phase transition and its critical point (for both pairwise and higher-order processes), also showing the role of heterogeneity and overlap.

\clearpage
\begin{figure*}[ht!]
\includegraphics[width=0.95\textwidth]{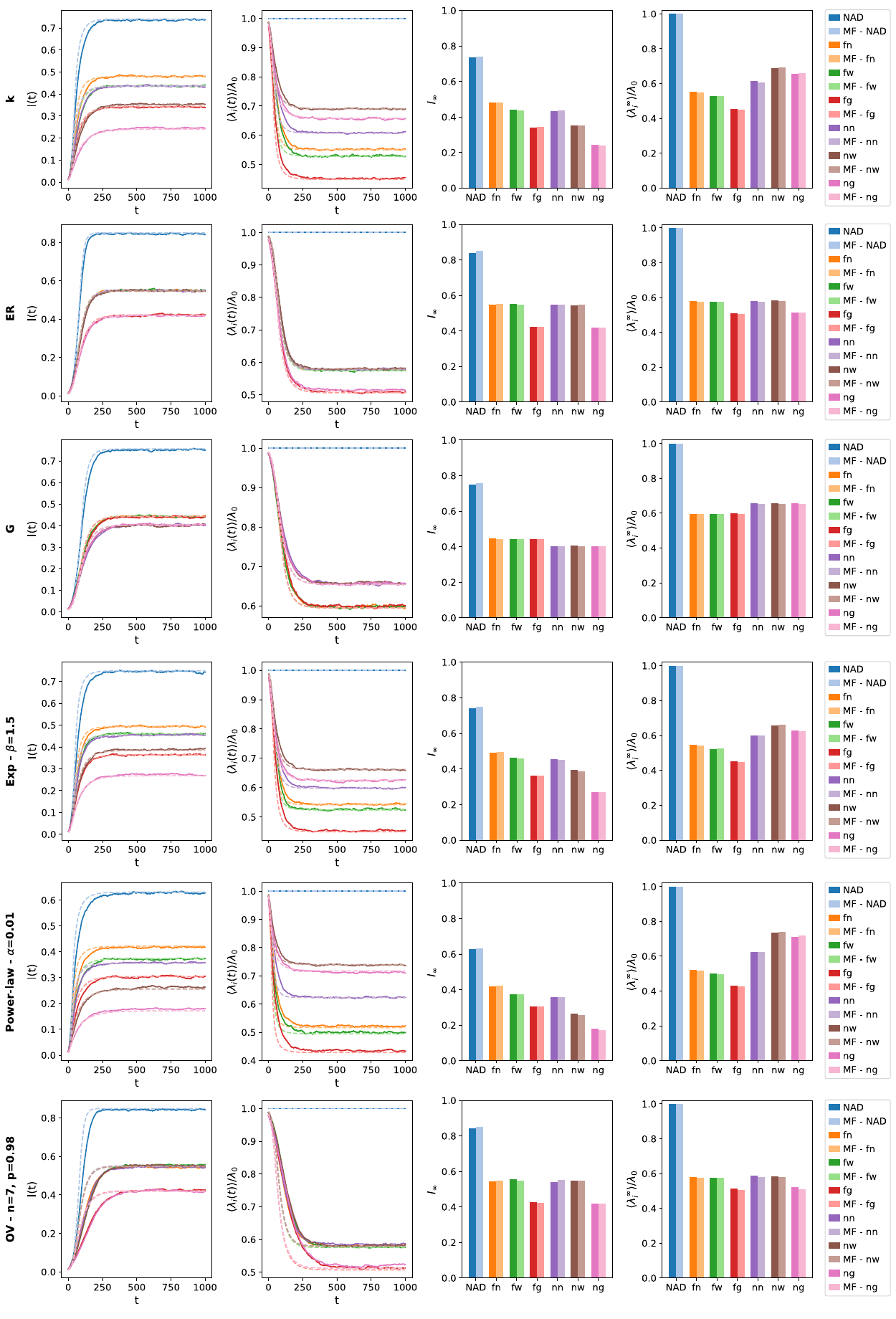}
\caption{\textbf{Containment efficacy and social cost - Synthetic - Pairwise contagion.} Analogous to Supplementary Fig. \ref{fig:figure3}, but here we consider the pairwise contagion on selected synthetic hypergraphs (see labels) with $\theta=0.3$, $r=0.016$ for the $G$ hypergraph and $r=0.05$ for all the others. In all panels we consider the NAD case and the six adaptive strategies.}
\label{fig:figure12}
\end{figure*}

\clearpage
\begin{figure*}[ht!]
\includegraphics[width=\textwidth]{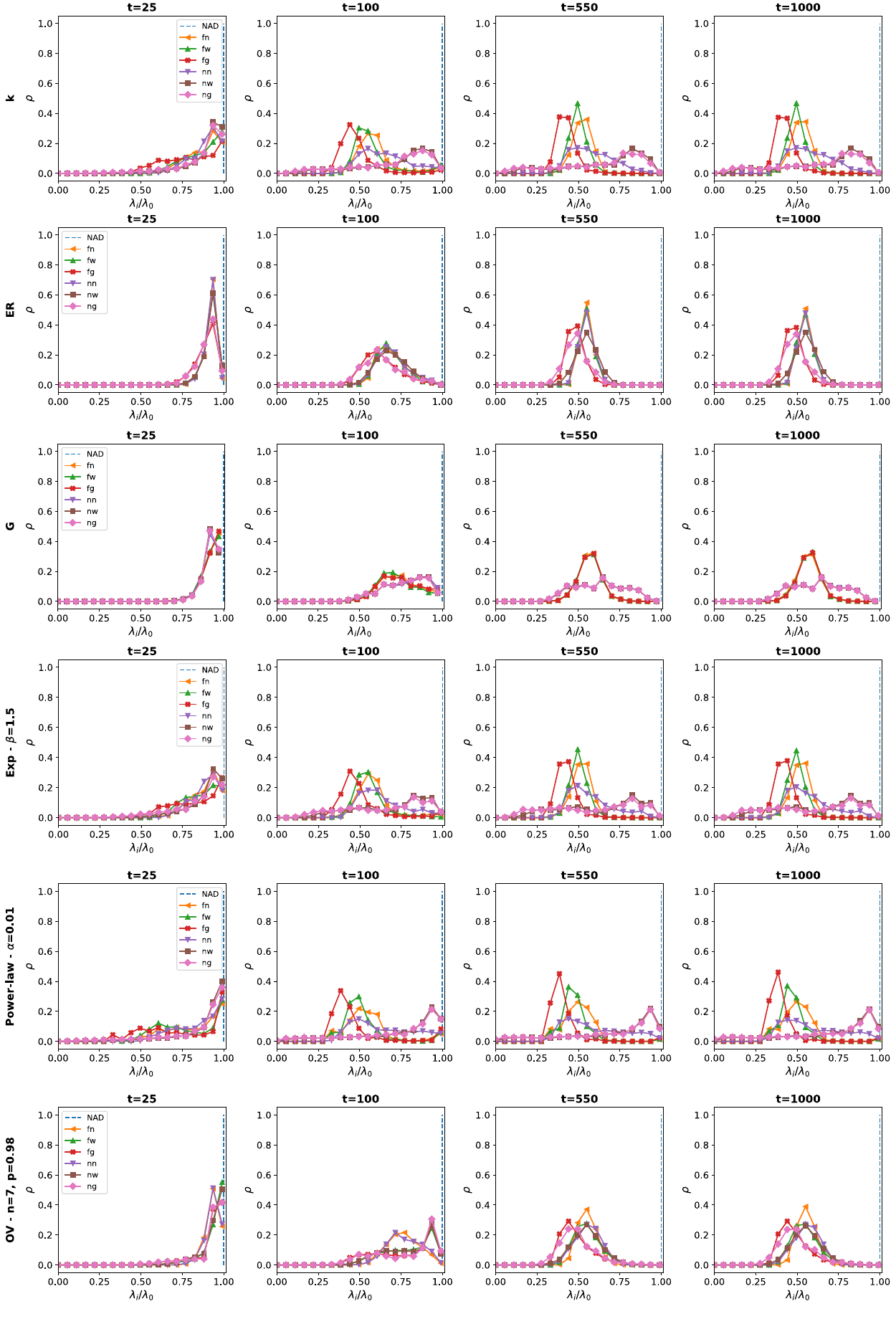}
\caption{\textbf{Evolution of risk parameter distribution - Synthetic - Pairwise contagion.} Analogous to Supplementary Fig. \ref{fig:figure4}, but here we consider the pairwise contagion on selected synthetic hypergraphs (see labels) with $\theta=0.3$, $r=0.016$ for the $G$ hypergraph and $r=0.05$ for all the others. In all panels we consider the NAD case and the six adaptive strategies.}
\label{fig:figure13}
\end{figure*}

\clearpage
\begin{figure*}[ht!]
\includegraphics[width=0.8\textwidth]{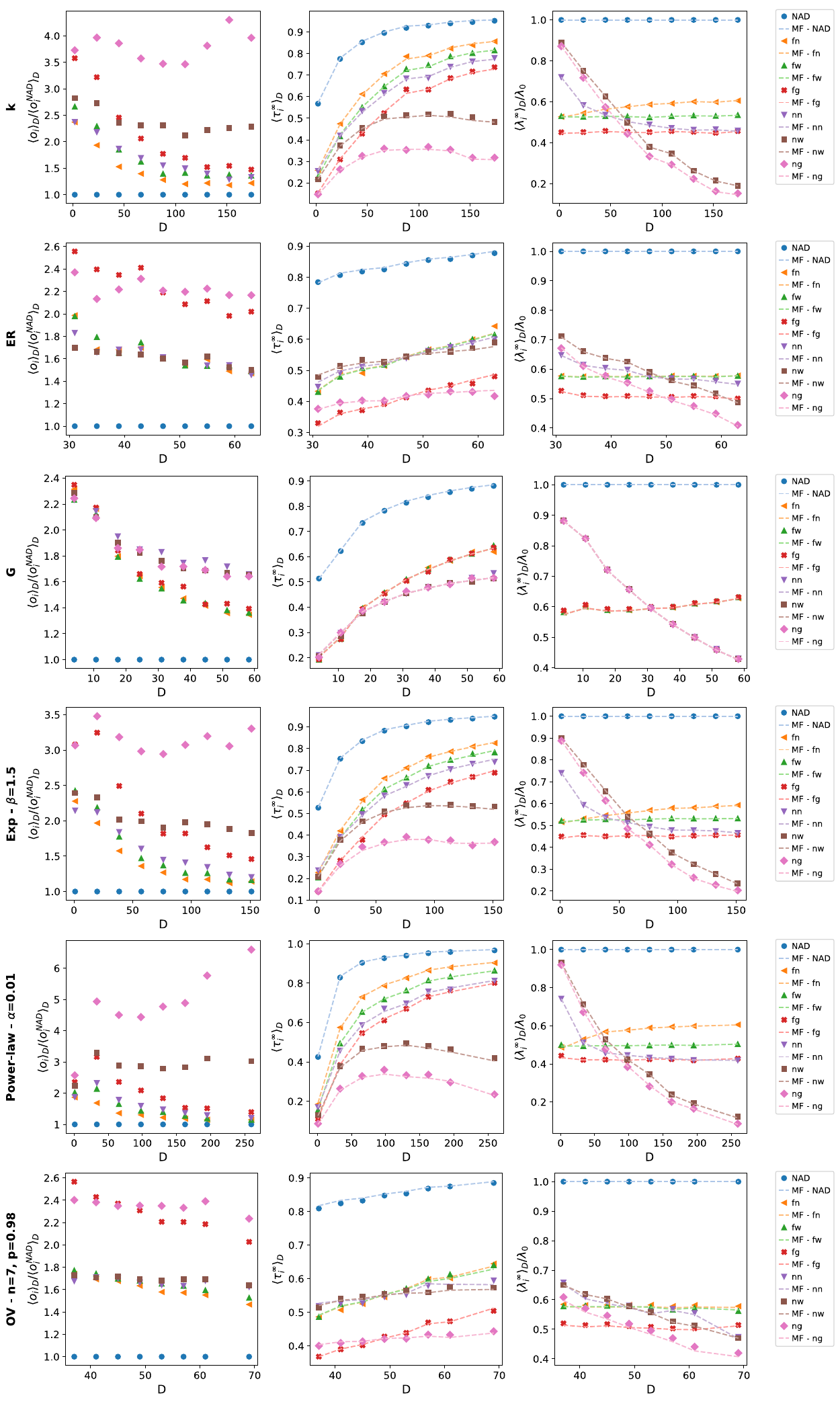}
\caption{\textbf{Behavior of single nodes - Synthetic - Pairwise contagion.} Analogous to Supplementary Fig. \ref{fig:figure5}, but here we consider the pairwise contagion on selected synthetic hypergraphs (see labels) with $\theta=0.3$, $r=0.016$ for the $G$ hypergraph and $r=0.05$ for all the others. In all panels we consider the NAD case and the six adaptive strategies.}
\label{fig:figure14}
\end{figure*}

\clearpage
\begin{figure*}[ht!]
\includegraphics[width=0.8\textwidth]{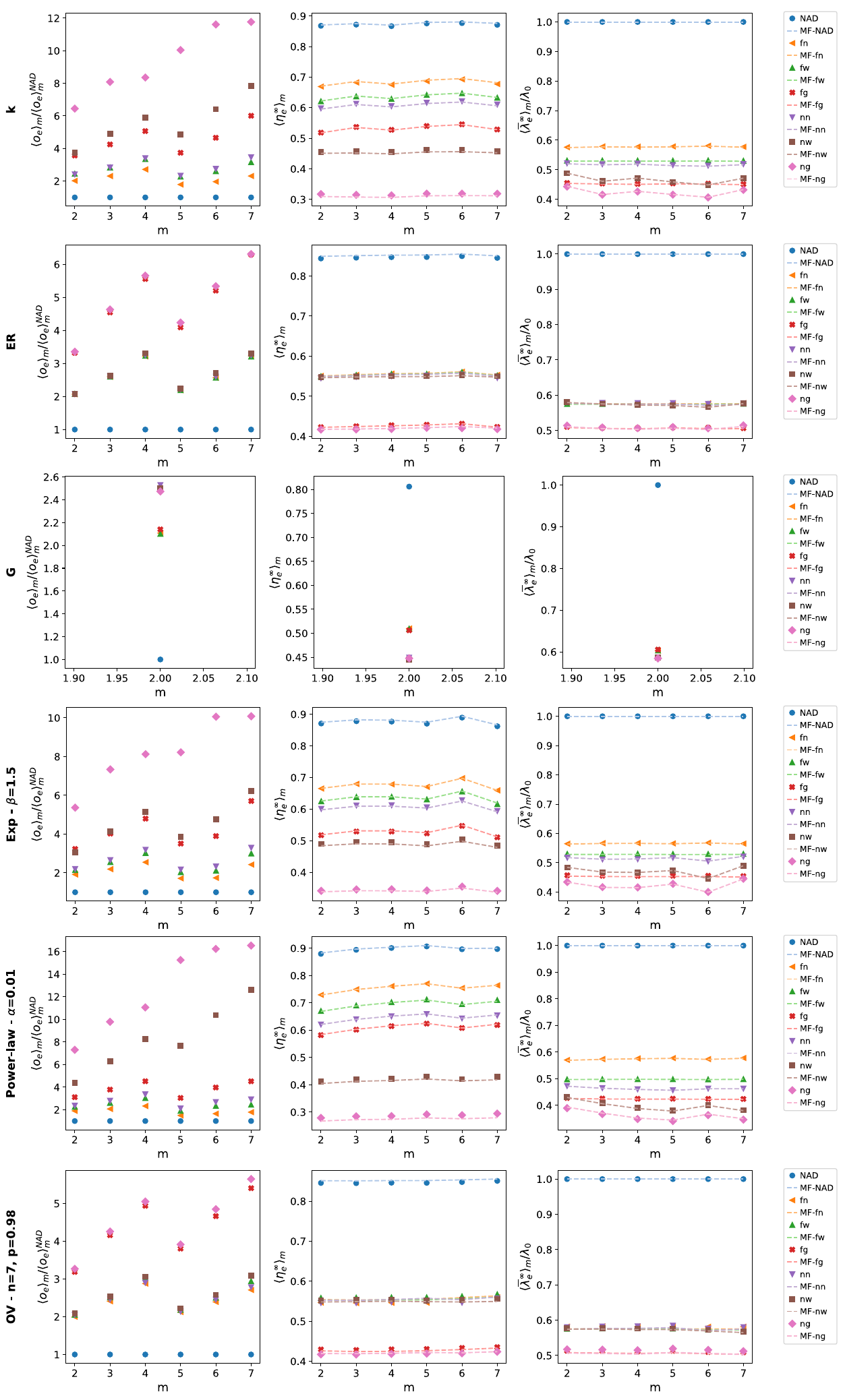}
\caption{\textbf{Behavior of single groups - Synthetic - Pairwise contagion.} Analogous to Supplementary Fig. \ref{fig:figure6}, but here we consider the pairwise contagion on selected synthetic hypergraphs (see labels) with $\theta=0.3$, $r=0.016$ for the $G$ hypergraph and $r=0.05$ for all the others. In all panels we consider the NAD case and the six adaptive strategies.}
\label{fig:figure15}
\end{figure*}

\clearpage
\begin{figure*}[ht!]
\includegraphics[width=0.95\textwidth]{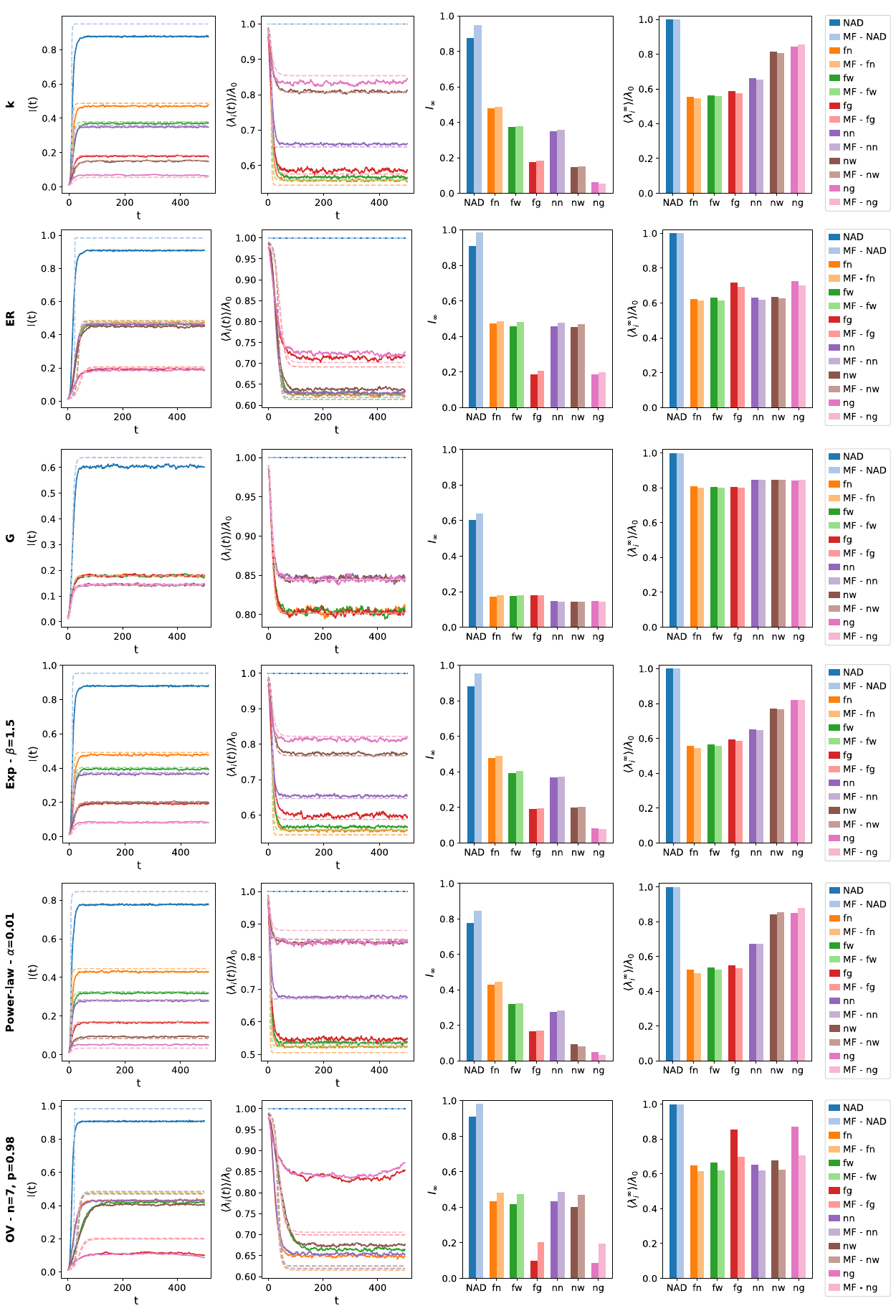}
\caption{\textbf{Containment efficacy and social cost - Synthetic - Higher-order contagion.} Analogous to Supplementary Fig. \ref{fig:figure3}, but here we consider the higher-order contagion on selected synthetic hypergraphs (see labels) with $\theta=0.3$, $r=0.1$ for the $G$ hypergraph and $r=0.01$ for all the others. In all panels we consider the NAD case and the six adaptive strategies.}
\label{fig:figure16}
\end{figure*}

\clearpage
\begin{figure*}[ht!]
\includegraphics[width=\textwidth]{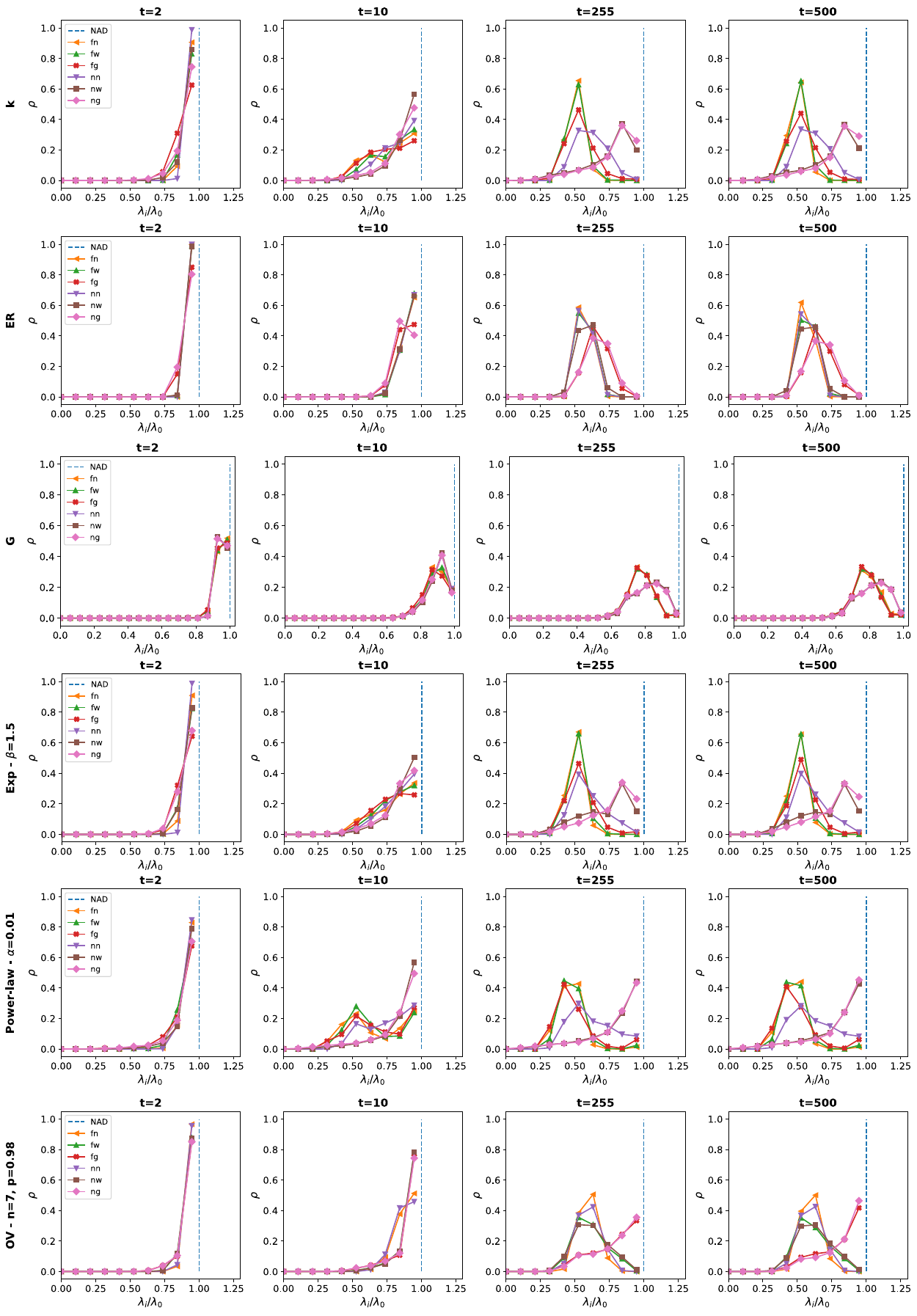}
\caption{\textbf{Evolution of risk parameter distribution - Synthetic - Higher-order contagion.} Analogous to Supplementary Fig. \ref{fig:figure4}, but here we consider the higher-order contagion on selected synthetic hypergraphs (see labels) with $\theta=0.3$, $r=0.1$ for the $G$ hypergraph and $r=0.01$ for all the others. In all panels we consider the NAD case and the six adaptive strategies.}
\label{fig:figure17}
\end{figure*}

\clearpage
\begin{figure*}[ht!]
\includegraphics[width=0.8\textwidth]{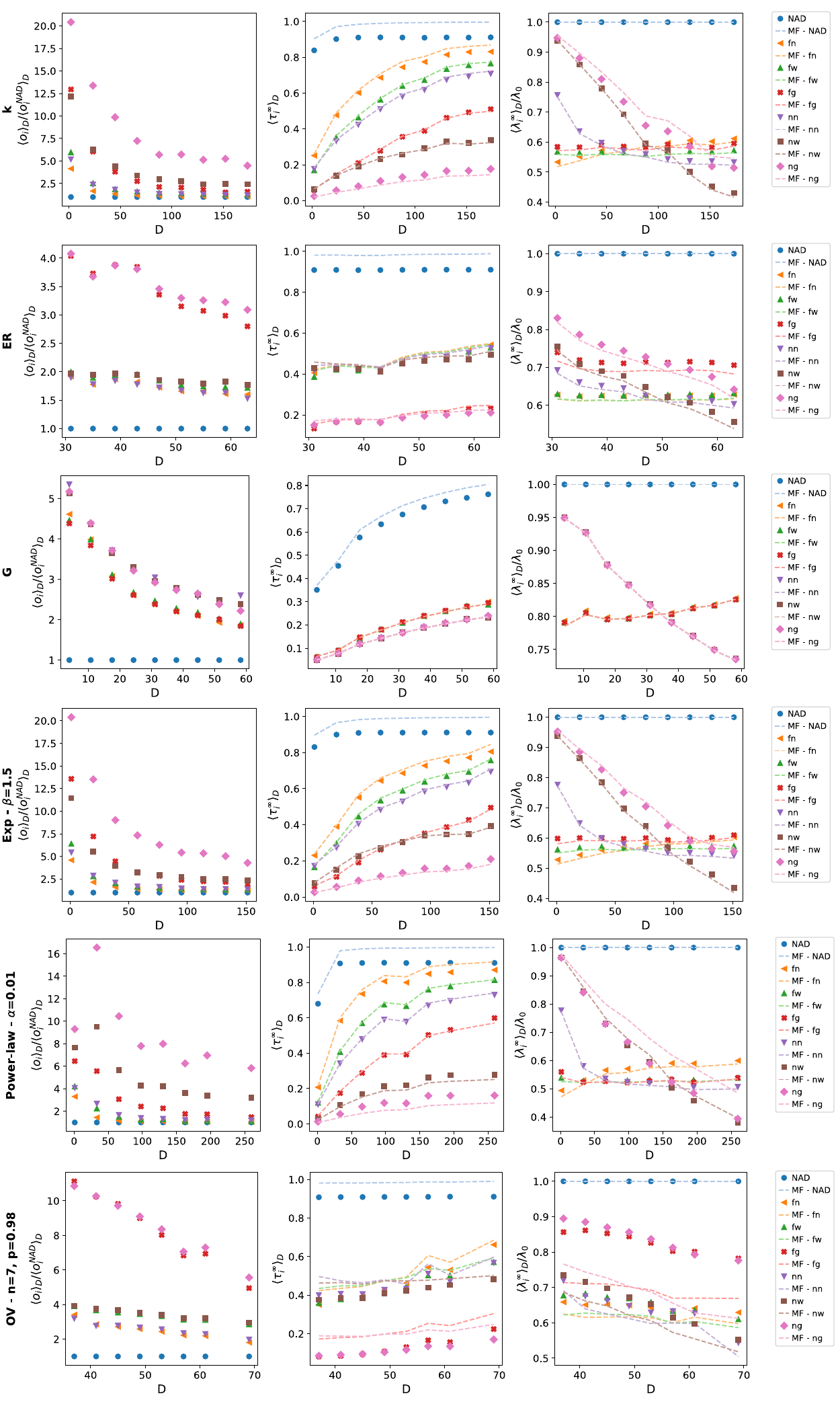}
\caption{\textbf{Behavior of single nodes - Synthetic - Higher-order contagion.} Analogous to Supplementary Fig. \ref{fig:figure5}, but here we consider the higher-order contagion on selected synthetic hypergraphs (see labels) with $\theta=0.3$, $r=0.1$ for the $G$ hypergraph and $r=0.01$ for all the others. In all panels we consider the NAD case and the six adaptive strategies.}
\label{fig:figure18}
\end{figure*}

\clearpage
\begin{figure*}[ht!]
\includegraphics[width=0.8\textwidth]{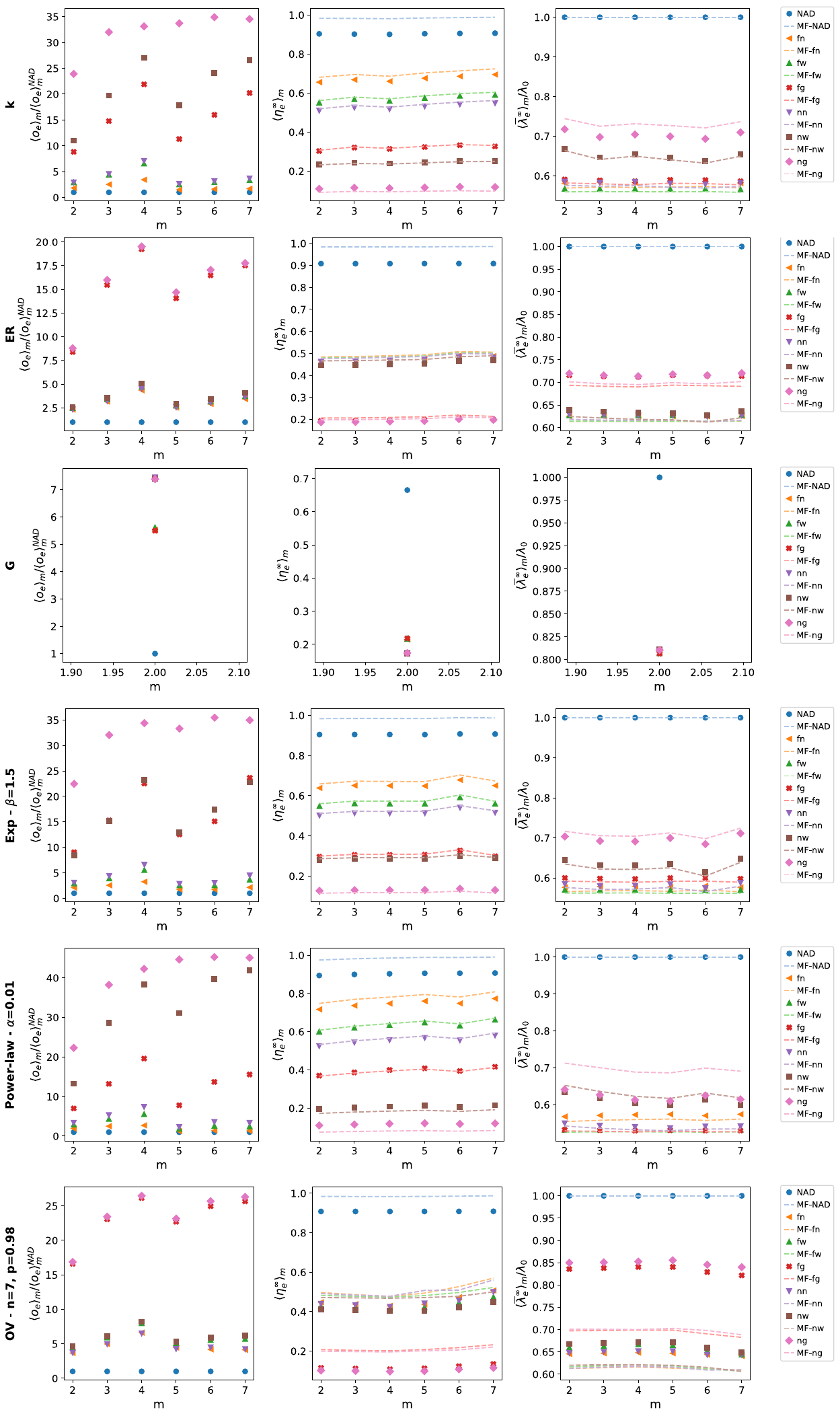}
\caption{\textbf{Behavior of single groups - Synthetic - Higher-order contagion.} Analogous to Supplementary Fig. \ref{fig:figure6}, but here we consider the higher-order contagion on selected synthetic hypergraphs (see labels) with $\theta=0.3$, $r=0.1$ for the $G$ hypergraph and $r=0.01$ for all the others. In all panels we consider the NAD case and the six adaptive strategies.}
\label{fig:figure19}
\end{figure*}

\clearpage
\begin{figure*}[ht!]
\includegraphics[width=\textwidth]{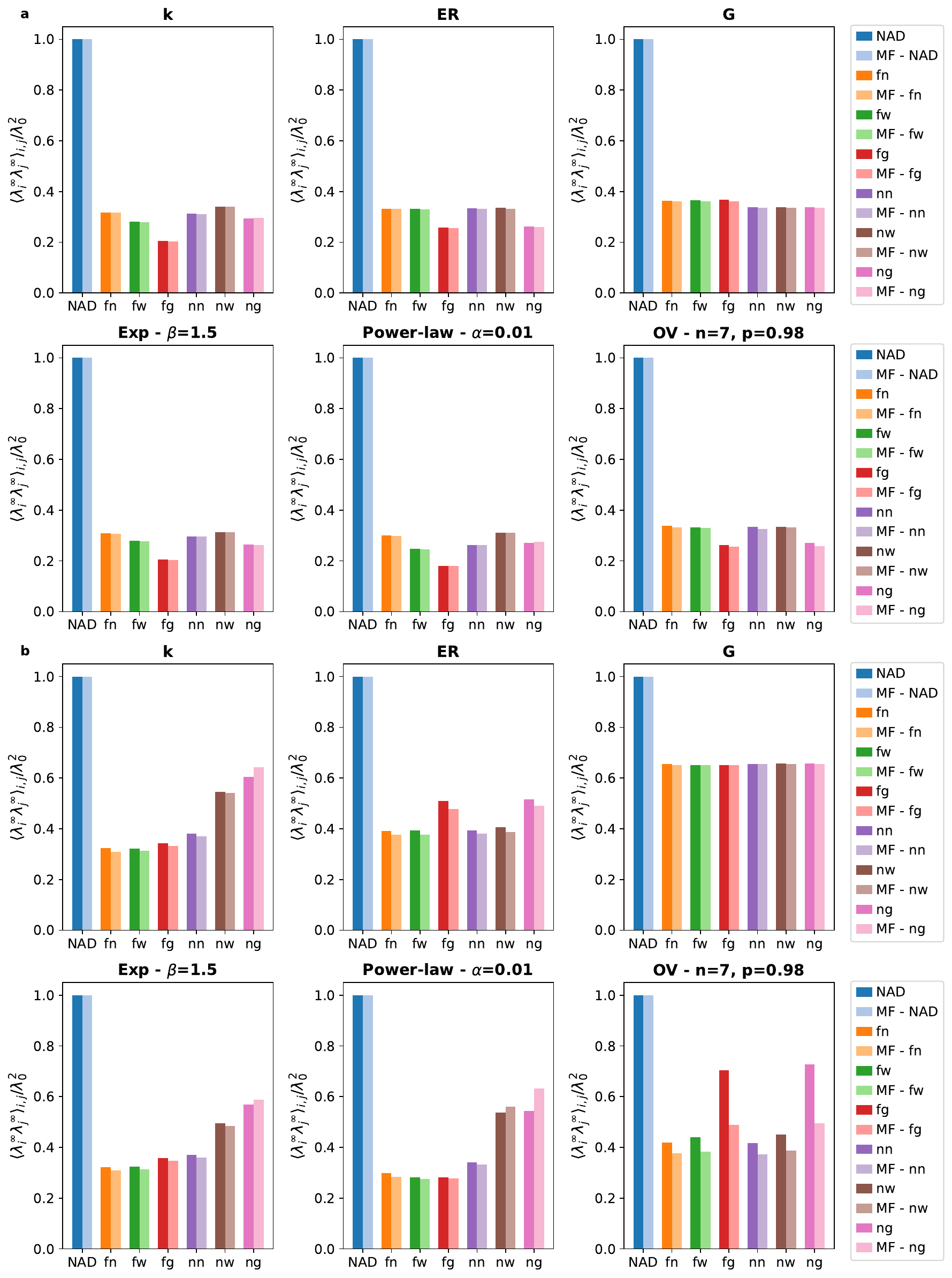}
\caption{\textbf{Alternative social cost evaluation - Synthetic.} Analogous to Supplementary Fig. \ref{fig:figureSM03}, but here we consider the contagion processes on selected synthetic hypergraphs (see labels) with $\theta=0.3$, the pairwise contagion with $r=0.016$ for the $G$ hypergraph and with $r=0.05$ for all the others, the higher-order contagion with $r=0.1$ for the $G$ hypergraph and $r=0.01$ for all the others. In all panels we consider the NAD case and the six adaptive strategies.}
\label{fig:figureSM04}
\end{figure*}

\clearpage
\begin{figure*}[ht!]
\includegraphics[width=0.8\textwidth]{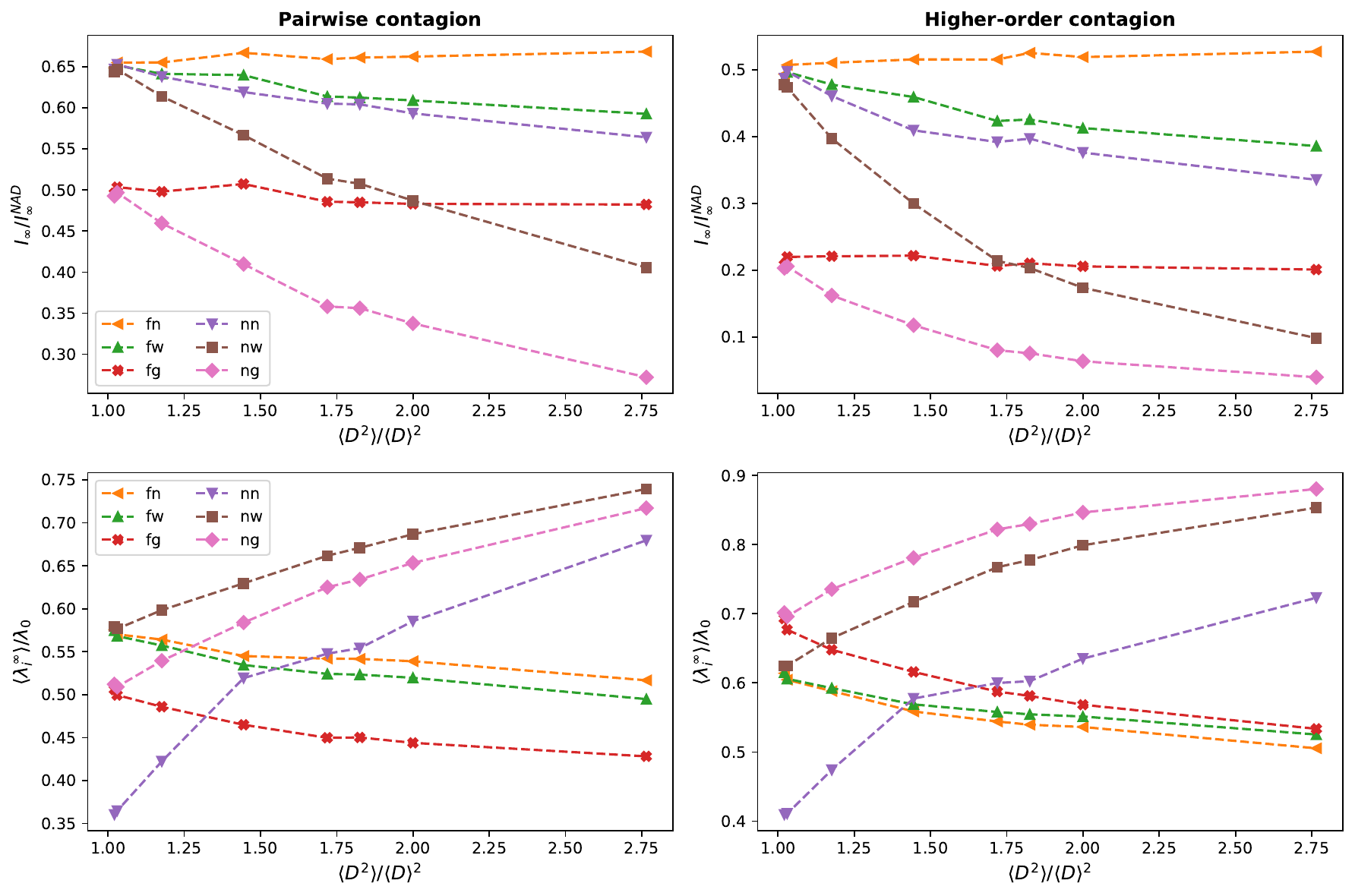}
\caption{\textbf{The role of hypergraph heterogeneity.} In the first and second row we show respectively the reduction in the epidemic prevalence $I_{\infty}/I_{\infty}^{NAD}$ and the average reduction in the risk parameter $\langle \lambda_i^{\infty} \rangle/\lambda_0$ in the asymptotic steady state, induced by adaptive strategies, as a function of the heterogeneity $\langle D^2 \rangle/\langle D \rangle^2$ of the hyperdegree distribution. The results are obtained through numerical integration of the mean-field equations. We consider the synthetic hypergraphs generated with the PD method, both for pairwise contagion (first column, with $r=0.05$) and higher-order contagion (second column, with $r=0.01$, $\nu=4$), and $\theta=0.3$.}
\label{fig:figure20}
\end{figure*}

\begin{figure*}[ht!]
\includegraphics[width=0.8\textwidth]{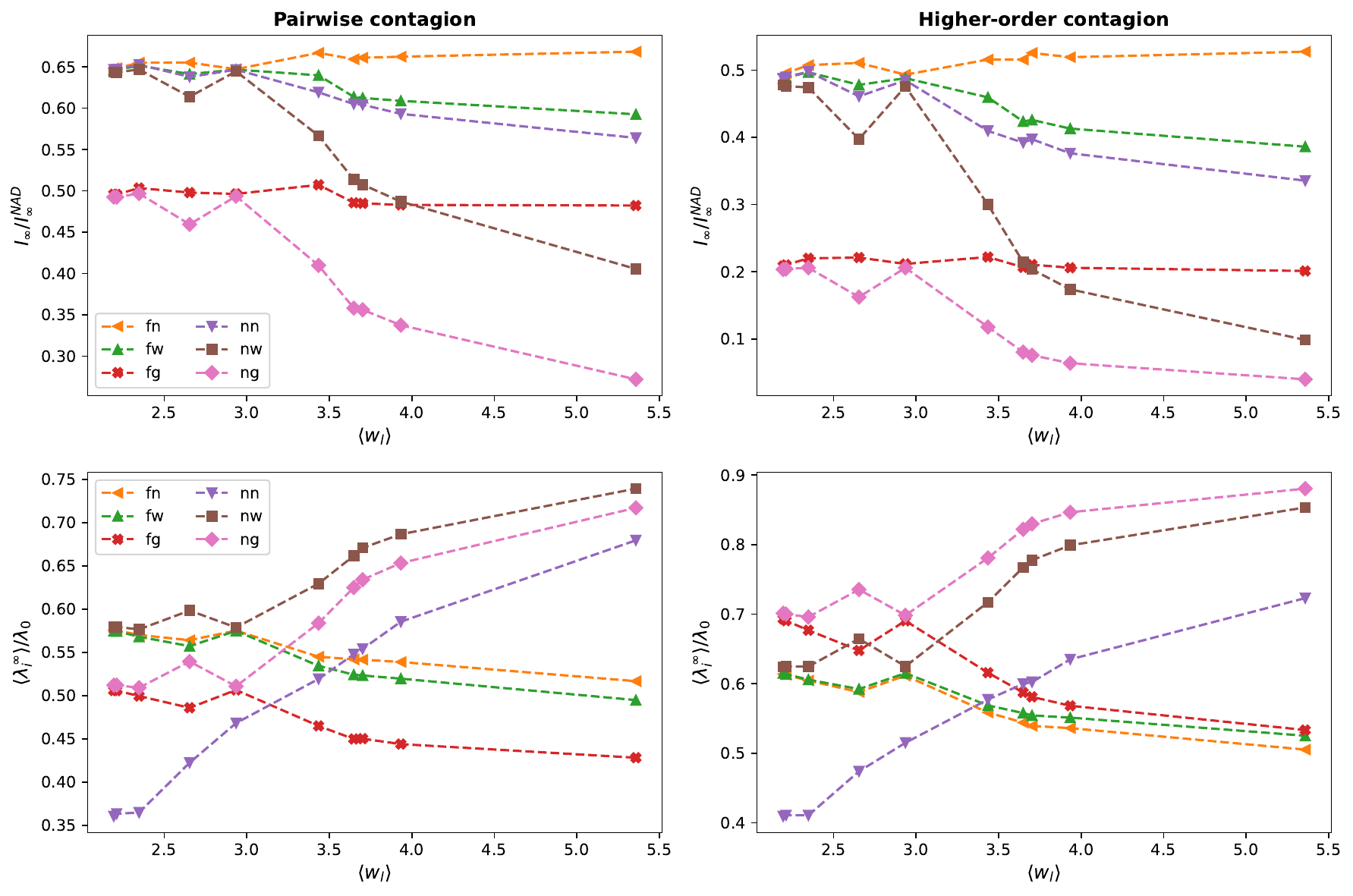}
\caption{\textbf{The role of hyperedge overlap.} Analogous to Supplementary Fig. \ref{fig:figure20}, but here we consider the efficacy and social cost  of the different adaptive strategies as a function of the hyperedge overlap $\langle w_l \rangle$ in $\mathcal{G}$. We consider the synthetic hypergraphs generated with the PD and OV methods, for both pairwise contagion (first column, with $r=0.05$) and higher-order contagion (second column, with $r=0.01$, $\nu=4$), and $\theta=0.3$.}
\label{fig:figure21}
\end{figure*}

\clearpage
\bibliographystyle{naturemag}
\bibliography{references_HO_adaptive_epi}